\documentclass[twocolumn]{aastex631}

\usepackage{savesym}
\savesymbol{tablenum}
\usepackage{siunitx}
\restoresymbol{SIX}{tablenum}

\usepackage[version=4]{mhchem}
\usepackage{graphicx,epsf}
\usepackage{subfigure}
\usepackage{graphicx}	
\usepackage{amsmath}	
\usepackage{amssymb}	
\usepackage{newtxtext,newtxmath}
\usepackage{siunitx}
\usepackage{footnote}

\newcommand{\Msun}{{\rm M}_\odot}
\newcommand{\Rsun}{{\rm R}_\odot}

\newcommand{\kms}{\textrm{km}\,\textrm{s}^{-1}}

\def\arcsec{\hbox{$^{\prime\prime}$}}

\newcommand{\mdot}{M$_{\odot}$~yr$^{-1}$}

\DeclareRobustCommand{\ion}[2]{\relax\ifmmode\ifx\testbx\f@series{\mathbf{#1\,\mathsc{#2}}}\else{\mathrm{#1\,\mathsc{#2}}}\fi\else\textup{#1\,{\mdseries\textsc{#2}}}\fi}

\newcommand{\code}[1]{\texttt{#1}}

\def\cmfgen{{\code{CMFGEN}}}

\usepackage[scale=0.94]{tgbonum}
\usepackage[mode=text]{siunitx}

\makeatletter
\DeclareTextCompositeCommand{\r}{OT1}{A}{%
  \leavevmode\vbox{%
    \offinterlineskip
    \ialign{\hfil##\hfil\cr\char23\cr\noalign{\kern-1.15ex}A\cr}%
  }%
}
\makeatother

\shorttitle{Flash Ultraviolet Spectroscopy}
\shortauthors{Jacobson-Gal\'an et al.}

\begin{document}

\title{Flash UV Spectroscopy: Decoding the Ultraviolet Diversity of Young Core-Collapse Supernovae with the Hubble Space Telescope}

\correspondingauthor{Wynn Jacobson-Gal\'{a}n (he, him, his)}
\email{wynnjg@caltech.edu}

\author[0000-0002-3934-2644]{W.~V.~Jacobson-Gal\'{a}n}
\altaffiliation{NASA Hubble Fellow}
\affiliation{Cahill Center for Astrophysics, California Institute of Technology, MC 249-17, 1216 E California Boulevard, Pasadena, CA, 91125, USA}

\author[0009-0005-3569-9944]{R.~J.~Yete}
\affiliation{Department of Astronomy, University of Maryland, College Park, MD 20742, USA}

\author[0000-0003-0599-8407]{L.~Dessart}
\affiliation{French-Chilean Laboratory for Astronomy, IRL 3386, CNRS and Instituto de Astrofísica, Pontificia Universidad Católica de Chile, Casilla 306, Santiago, Chile}
\affiliation{Institut d'Astrophysique de Paris, CNRS-Sorbonne Universit\'e, 98 bis boulevard Arago, F-75014 Paris, France}

\author[0000-0003-1546-6615]{J.~Sollerman}
\affiliation{The Oskar Klein Centre, Department of Astronomy, Stockholm University, AlbaNova, SE-10691 Stockholm, Sweden}

\author[0000-0003-2445-3891]{M.~R.~Siebert}
\affiliation{Space Telescope Science Institute, Baltimore, MD 21218}

\author[0000-0002-5683-2389]{N.~Rehemtulla}
\affiliation{Center for Interdisciplinary Exploration and Research in Astrophysics (CIERA) and Department of Physics and Astronomy, Northwestern University, Evanston, IL 60208, USA}
\affiliation{NSF-Simons AI Institute for the Sky (SkAI), 172 E. Chestnut St., Chicago, IL 60611, USA}

\author[0000-0001-8372-997X]{K.~K.~Das}
\affiliation{Center for Astrophysics $|$ Harvard \& Smithsonian, 60 Garden Street, Cambridge, MA 02138--1516, USA}

\author[0000-0003-4768-7586]{R.~Margutti}
\affiliation{Department of Astronomy, University of California, Berkeley, CA 94720-3411, USA}
\affiliation{Department of Physics, University of California, Berkeley, CA 94720, USA}
\affiliation{Berkeley Center for Multi-messenger Research on Astrophysical Transients and Outreach (Multi-RAPTOR), University of California, Berkeley, CA 94720-3411, USA}

\author[0000-0002-7706-5668]{R.~Chornock}
\affiliation{Department of Astronomy, University of California, Berkeley, CA 94720-3411, USA}
\affiliation{Berkeley Center for Multi-messenger Research on Astrophysical Transients and Outreach (Multi-RAPTOR), University of California, Berkeley, CA 94720-3411, USA}

\author[0000-0002-2445-5275]{R.~J.~Foley}
\affiliation{Department of Astronomy and Astrophysics, University of California, Santa Cruz, CA 95064, USA}

\author[0000-0002-5619-4938]{M.~M.~Kasliwal}
\affiliation{Cahill Center for Astrophysics, California Institute of Technology, MC 249-17, 1216 E California Boulevard, Pasadena, CA, 91125, USA}

\author[0000-0003-2238-1572]{O.~D.~Fox}
\affiliation{Space Telescope Science Institute, Baltimore, MD 21218}

\author[0000-0002-6652-9279]{A.~S.~Fruchter}
\affiliation{Space Telescope Science Institute, Baltimore, MD 21218}

\author[0000-0001-8018-5348]{E.~C.~Bellm}
\affiliation{DIRAC Institute, Department of Astronomy, University of Washington, 3910 15th Avenue NE, Seattle, WA 98195, USA}

\author[0009-0008-2714-2507]{A.~Bochenek}
\affiliation{Astrophysics Research Institute, Liverpool John Moores University, 146 Brownlow Hill, Liverpool L3 5RF, UK}

\author[0000-0001-8544-584X]{J.~D.~Carney}
\affiliation{Department of Physics and Astronomy, University of North Carolina at Chapel Hill, Chapel Hill, NC 27599-3255, USA}

\author[0000-0002-0987-3372]{J.~Casta\~neda Jaimes}
\affiliation{Division of Physics, Mathematics and Astronomy, California Institute of Technology, 1200 E. California Blvd, Pasadena, CA 91125, USA}

\author{E.~M.~Cotta}
\affiliation{Department of Electrical Engineering, University of California Los Angeles, 475 Portola Plaza, Los Angeles, CA 90095-1547}

\author[0000-0003-4263-2228]{D.~A.~Coulter}
\affiliation{Space Telescope Science Institute, Baltimore, MD 21218, USA}
\affiliation{Department of Physics and Astronomy, The Johns Hopkins University, Baltimore, MD 21218, USA}

\author[0000-0002-5680-4660]{K.~W.~Davis}
\affiliation{Department of Astronomy and Astrophysics, University of California, Santa Cruz, CA 95064, USA}

\author[0000-0002-4223-103X]{C.~Fremling}
\affiliation{Caltech Optical Observatories, California Institute of Technology, Pasadena, CA 91125, USA}
\affiliation{Division of Physics, Mathematics and Astronomy, California Institute of Technology, Pasadena, CA 91125, USA}

\author[0000-0002-3653-5598]{A.~Gal-Yam}
\affiliation{Department of Particle Physics and Astrophysics, Weizmann Institute of Science, 234 Herzl St, 76100 Rehovot, Israel}

\author[0000-0002-3884-5637]{A.~Gangopadhyay}
\affiliation{The Oskar Klein Centre, Department of Astronomy, Stockholm University, AlbaNova, SE-10691 Stockholm, Sweden}

\author[0000-0003-3703-5154]{S.~Gezari}
\affiliation{Department of Astronomy, University of Maryland, College Park, MD 20742-2421}

\author[0009-0002-9727-8326]{X.~Guo}
\affiliation{Center for Astrophysics $|$ Harvard \& Smithsonian, 60 Garden Street, Cambridge, MA 02138--1516, USA}

\author[0000-0002-0129-806X]{K-R.~Hinds}
\affiliation{Cahill Center for Astrophysics, California Institute of Technology, MC 249-17, 1216 E California Boulevard, Pasadena, CA, 91125, USA}

\author[0009-0005-3156-6386]{A.~P.~Huddleston}
\affiliation{Department of Astronomy and Astrophysics, University of California, Santa Cruz, CA 95064, USA}

\author[0000-0002-5740-7747]{C.~D.~Kilpatrick}
\affiliation{Center for Interdisciplinary Exploration and Research in Astrophysics (CIERA), Northwestern University, Evanston, IL 60202, USA}

\author[0000-0001-5390-8563]{S.~R.~Kulkarni}
\affiliation{Cahill Center for Astrophysics, California Institute of Technology, MC 249-17, 1216 E California Boulevard, Pasadena, CA, 91125, USA}

\author[0000-0002-2249-0595]{N.~LeBaron}
\affiliation{Department of Astronomy, University of California, Berkeley, CA 94720-3411, USA}

\author[0000-0003-4285-4453]{Z.~Li}
\affiliation{DIRAC Institute, Department of Astronomy, University of Washington, 3910 15th Avenue NE, Seattle, WA 98195, USA}

\author[0000-0002-8532-9395]{F.~J.~Masci}
\affiliation{IPAC, California Institute of Technology, 1200 E. California Blvd, Pasadena, CA 91125, USA}

\author[0000-0001-9515-478X]{A.~A.~Miller}
\affiliation{Center for Interdisciplinary Exploration and Research in Astrophysics (CIERA) and Department of Physics and Astronomy, Northwestern University, Evanston, IL 60208, USA}
\affiliation{NSF-Simons AI Institute for the Sky (SkAI), 172 E. Chestnut St., Chicago, IL 60611, USA}

\author[0000-0002-1092-6806]{K.~C.~Patra}
\affiliation{Department of Astronomy and Astrophysics, University of California, Santa Cruz, CA 95064, USA}

\author[0000-0003-1227-3738]{J.~Purdum}
\affiliation{Caltech Optical Observatories, California Institute of Technology, Pasadena, CA 91125, USA}

\author[0000-0002-4410-5387]{A.~Rest}
\affiliation{Space Telescope Science Institute, Baltimore, MD 21218}
\affiliation{Department of Physics and Astronomy, The Johns Hopkins University, Baltimore, MD 21218, USA}

\author[0000-0003-0427-8387]{R.~M.~Rich}
\affiliation{Department of Physics and Astronomy, University of California Los Angeles, 475 Portola Plaza, Los Angeles, CA 90095-1547}

\author[0000-0002-0387-370X]{R.~L.~Riddle}
\affiliation{Caltech Optical Observatories, California Institute of Technology, Pasadena, CA 91125, USA}

\author[0000-0001-6797-1889]{S.~Schulze}
\affiliation{Department of Particle Physics and Astrophysics, Weizmann Institute of Science, 234 Herzl St, 76100 Rehovot, Israel}

\author[0000-0001-8023-4912]{H.~Sears}
\affiliation{Department of Physics and Astronomy, Rutgers, the State University of New Jersey, 136 Frelinghuysen Road, Piscataway, NJ 08854-8019, USA}

\author[0000-0003-2091-622X]{A.~Singh}
\affiliation{The Oskar Klein Centre, Department of Astronomy, Stockholm University, AlbaNova, SE-10691 Stockholm, Sweden}

\author[0000-0001-7062-9726]{R.~M.~Smith}
\affiliation{Caltech Optical Observatories, California Institute of Technology, Pasadena, CA 91125, USA}

\author[0000-0002-5748-4558]{K.~Taggart}
\affiliation{Science and Technology Institute, Universities Space Research Association, Huntsville, AL 35805, USA}

\author[0000-0001-9834-3439]{J.~H.~Terwel}
\affiliation{School of Physics, Trinity College Dublin, The University of Dublin, Dublin 2, Ireland}

\author[0000-0001-5233-6989]{Q.~Wang}
\affiliation{Department of Physics and Kavli Institute for Astrophysics and Space Research, Massachusetts Institute of Technology, 77 Massachusetts Avenue, Cambridge, MA 02139, USA}
\affiliation{Department of Particle Physics and Astrophysics, Weizmann Institute of Science, 234 Herzl St, 76100 Rehovot, Israel}

\author[0000-0003-1710-9339]{L.~Yan}
\affiliation{Caltech Optical Observatories, California Institute of Technology, Pasadena, CA 91125, USA}

\author[0000-0001-6747-8509]{Y.~Yao}
\affiliation{Kavli Institute for Astronomy and Astrophysics, Peking University, Beijing 100871, China}

\author[0000-0001-8985-2493]{E.~A.~Zimmerman}
\affiliation{Department of Particle Physics and Astrophysics, Weizmann Institute of Science, 234 Herzl St, 76100 Rehovot, Israel}

\begin{abstract}

We present multi-wavelength observations and modeling of the largest sample to date of core-collapse supernovae (SNe) with Hubble Space Telescope (HST) ultraviolet (UV) spectroscopy within $\lesssim 2$~weeks of first light. Our total sample contains early-time UV spectroscopy from 12 Type II SNe (SNe II), 3 Type IIb SNe (SNe IIb), 3 Type IIn SNe (SNe IIn) and 1 Type I super-luminous SN (SLSN-I). This sample includes 11 objects with unpublished UV spectroscopy from HST Cycles 31 and 32 ``Flexible Thursdays" target of opportunity programs, increasing the number of core-collapse SNe with early-time HST UV spectra by a factor of 2.4. The early-time near-UV spectra of most sample SNe~II show similar Doppler-broadened P-Cygni profiles of metals (Mg, Fe, Ti, Cr), though with notable diversity in line velocities and absorption strengths. Comparison to non-Local Thermodynamic Equilibrium (nLTE) radiative transfer model grids at similar phases indicates that most sample SNe~II are consistent with low circumstellar material (CSM) densities at the time of UV observations (e.g., $\dot M = 10^{-6}$~\mdot, $v_w = 50~\kms$) and standard kinetic energies (e.g., $E_k = 0.6-1.2$~B), while two objects, SNe~2024faq and 2023ixf, show UV/optical evidence for interaction with more confined, higher-density CSM. SN~2023ixf and sample SNe~IIn exhibit relatively featureless early-time UV spectra despite displaying prominent optical Lorentzian emission lines from electron scattering in optically thick CSM -- this being attributed to differing properties of UV resonance versus optical recombination transitions as well as intrinsic CSM asymmetries. These results demonstrate the potential of early-time UV spectroscopy to constrain the outermost SN ejecta structure, explosion energetics/composition, and progenitor mass-loss histories of core-collapse SNe.

\end{abstract}


\section{Introduction} \label{sec:intro}

Type II supernovae (SNe~II) are the ideal stellar explosions for ultraviolet (UV) spectroscopy within their first 1--2 weeks post-explosion. Following shock breakout (SBO) from the stellar surface, the hot, ionized SN ejecta cool, radiating the majority of the flux at far-UV wavelengths ($\lambda < 2000$~\AA) during 1--2~days post-explosion for red supergiant (RSG) progenitor stars \citep{Gezari08, rabinak11, nakar10, sapir17, waxman17}. The UV emission in young SNe~II can be significantly boosted and lengthened if the SN ejecta interact with high-density circumstellar material (CSM) within the local progenitor environment that was created through mass-loss episodes in the final years before core-collapse \citep{dessart17, Haynie21, dessart22, Irani24, wjg24}. Confined CSM can extend the SBO emission by $\sim$days and manifest as a dramatic early-time color evolution while the first SN photons diffuse out of the dense material enshrouding the progenitor star before explosion \citep{gezari15, Li24, Zimmerman24, Shrestha24ggi, wjg24b, Morag26}. 

As predicted by non-LTE radiative transfer simulations, the interaction of SN ejecta with CSM leads to the formation of a plethora of narrow, high-ionization emission lines, the majority of which are located at UV wavelengths and are direct tracers of CSM properties as this gas is continuously photo-ionized by the SN shockwave \citep{dessart17}. At present, these ``IIn-like'' or ``flash'' features from early-time CSM interaction have been primarily detected at optical wavelengths \citep{Khazov16, galyam14, yaron17, terreran22, bruch21-1, bruch23, wjg22, wjg24}, with only one SN~II to date, SN 2023ixf, showing these narrow emission lines at near-UV wavelengths in Hubble Space Telescope (HST) spectra within the first week since first light \citep{Zimmerman24}. Although not a SN II, the stripped-envelope Type IIb SN (SN IIb) 1993J did show far-UV narrow line emission within the first $\sim$2--3~days after first light from photo-ionization of confined CSM by the SN radiation \citep{Benetti94, Garnavich94}. At later phases, shock power injection into the ``cold dense shell'' (CDS) between forward and reverse shocks from on-going ejecta-CSM interaction manifests as persistent UV emission lines of Ly-$\alpha$ and \ion{Mg}{ii} $\lambda$2803.5, which can respectively radiate $\sim 65\%$ and $\sim 10\%$ of the thermalized emission \citep{fransson05, Dessart23b, Bostroem24, Bostroem25, wjg25LT, Dessart26} -- this emission being dependent on a number of factors such as CDS mass, shock power and time since explosion.

Presently, the landscape of SN II UV spectroscopy observations remains sparse and unconstrained at early phases ($\sim$days to a week after explosion). The earliest UV spectra of core-collapse SNe are still the International Ultraviolet Explorer (IUE) spectral series of SN~1987A beginning at 1.7~days \citep{Kirshner87, Wamsteker87, Pun95} and SN~1993J beginning at $\sim 2.6$~days \citep{deBoer93} since first light. The current best UV spectrograph for SN observations is the Space Telescope Imaging Spectrograph (STIS) on the HST, but it is limited by the $>48$~hr target of opportunity response time. Consequently, the earliest such HST/STIS UV observation is of Type II SN 2020fqv at 3.4~days since first light but the extreme host galaxy reddening resulted in no detectable far-UV flux and a low signal-to-noise ratio near-UV spectrum \citep{tinyanont22}. SN~2023ixf was observed at a phase of 3.6~days since first light, making it the second youngest HST-observed SN~II \citep{Zimmerman24}. SN~2022acko became the first SN~II with a complete far- and near-UV HST spectral sequence beginning at 5.2~days \citep{bostroem22acko}. Besides these SNe, other efforts for HST UV spectra have yielded some limited observations of SNe~II beginning at $\sim 10$~days, such as SNe~2021yja and 2022wsp \citep{vasylyev22, vasylyev23}. 

In this study, we present observations and modeling of the largest sample to date of core-collapse SNe with HST UV spectroscopy at $\lesssim$2~weeks since first light. In Section \ref{sec:obs}, we define the sample and present the spectroscopic and photometric observations. Section \ref{sec:photanalysis} presents an analysis of the multi-band light curves for all sample objects. In Section \ref{sec:specanalysis}, we perform spectroscopic analysis and model-matching for each SN sub-type in our sample. Our results are discussed in Section \ref{sec:discussion} and our conclusions are presented in Section \ref{sec:conclusion}. All phases reported in this paper are with respect to the adopted time of first light and are in rest-frame days. The time since first light ($\delta t$) and its uncertainty are calculated from the average phase between the last deep non-detection and the first detection using forced photometry from the survey that initially imaged the SN. When possible, we use redshift-independent host-galaxy distances and adopt standard $\Lambda$CDM cosmology ($H_{0}$ = 70~km~s$^{-1}$~Mpc$^{-1}$, $\Omega_M = 0.27$, $\Omega_{\Lambda} = 0.73$) if only redshift information is available for a given object.


\section{Observations} \label{sec:obs}

\subsection{Ultraviolet Spectroscopic Sample}\label{SubSec:sample}

\begin{figure*}
\centering
\includegraphics[width=\textwidth]{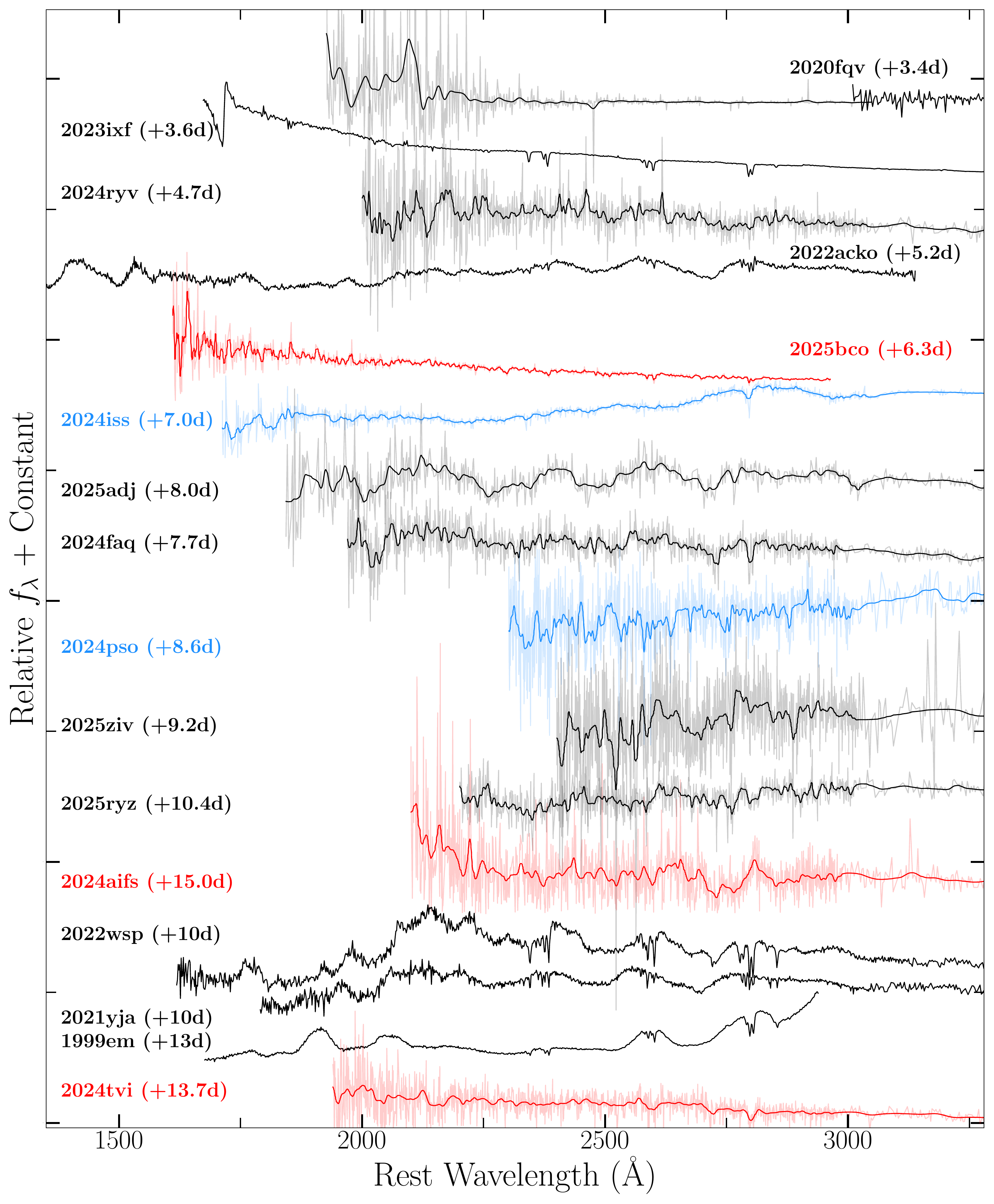}
\caption{Early-time ($\delta t \lesssim 14$~days) UV spectra of core-collapse SNe by HST. Type II-P/L SNe are shown in black, Type IIn SNe in red, and Type IIb SNe in blue. Smoothed spectra from convolution with a Gaussian kernel shown as higher opacity lines; raw spectra shown as lower opacity lines.
    \label{fig:spec_all} }
\end{figure*}

The SN sample analyzed in this work consists of core-collapse SNe with UV spectroscopic observations from HST within $\lesssim$2 weeks after first light. In total, our sample contains 11 objects with unpublished HST UV spectroscopy as well as 8 objects already published in the literature. Published objects with early-time HST spectra are SNe~1999em  \citep{Baron00}, 1998S  \citep{fransson05}, 2013df \citep{benami15}, 2020fqv \citep{tinyanont22}, 2021yja \citep{vasylyev22}, 2022wsp \citep{vasylyev23}, 2022acko \citep{bostroem22acko}, and 2023ixf \citep{Zimmerman24, Bostroem24}. The complete HST sample contains 12 SNe II, 3 SNe~IIb, 3 SNe~IIn and 1 Type I super-luminous SN (SLSN-I). Additionally, we include early-time IUE UV spectroscopy of SNe~1987A and 1993J for comparison analysis. Sample SNe~IIn are distinct from SN~II with early-time CSM interaction (i.e., ``23ixf-like'' or SNe~IIP-CSM; \citealt{wjg24}) events because they show IIn-like emission line profiles for $>$10~days.  We note that early-time UV spectra of other core-collapse SNe have been obtained with the Swift/UVOT grism (e.g., see \citealt{Immler07, Bufano09, margutti14,  Dhungana16}) or GALEX (e.g., \citealt{galyam08}), but we do not include them in our sample analysis given their lower spectral resolution and signal-to-noise (SNR). We do include Swift/UVOT grism spectra of SN~2021yja at $\delta t = 4.5 \ \& \ 7$~days for comparison.  Properties of each sample object are presented in Table \ref{tab:sample_specs}, and the earliest HST UV spectra for the sample objects are shown in Figure \ref{fig:spec_all}.

New HST UV spectra for our sample were obtained through the ``Flexible Thursdays'' target of opportunity (ToO) program, which began in Cycle 31 (Nov. 2023). This program allows for up to 2 ``FlexDay'' ToO activations on one Thursday each month of the cycle. ToO triggers must be submitted by 10:00 UT on the preceding Tuesday so that observations can be scheduled during a 24--36 hour window beginning on the Thursday. We utilized this ToO initiative through Cycles 31--32 (HST GO Programs 17507 \& 17807) in order to obtain up to 14 UV/optical STIS/CCD ($\sim$1600--10000~\AA) spectra (7 per cycle) of core-collapse SNe within their first $\sim$1--2~weeks post-explosion. In Cycle 31, we triggered our program on 7 SNe but only 5 successful UV spectra were obtained due to loss of guiding for two of the activations. In Cycle 32, we triggered our program on 7 SNe and only 1 target had a guiding failure resulting in an unsuccessful UV spectral observation. Two ToO activations during Cycle 32 had overlapping targets (SNe~2025bco and 2025adj) with another Flexible Thursday program (GO-17788) and were observed through that program. 

\begin{figure*}
\centering
\includegraphics[width=\textwidth]{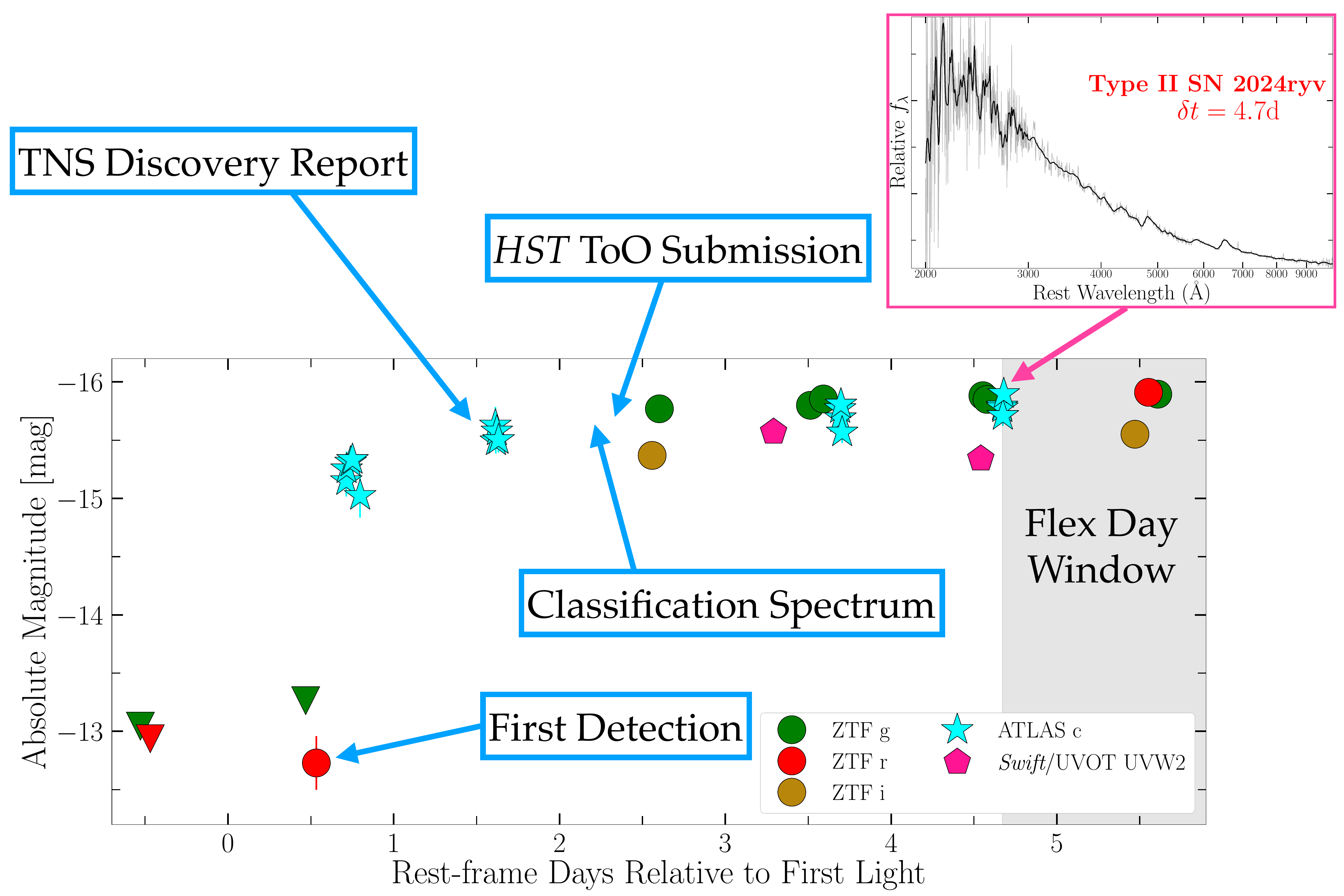}
\caption{Example timeline of ``Flex Day'' target of opportunity trigger for Type II SN 2024ryv, which was observed in the UV by HST STIS/CCD at $\delta t = 4.7$~days since first light. Multi-color light curve of SN~2024ryv with respect to time since first light (MJD $60532.82 \pm 0.47$) from ZTF (circles), Swift (polygons), and ATLAS (stars). ZTF upper limits shown as downward triangles. Observed photometry is presented in the AB magnitude system and has been corrected for host galaxy and MW extinction.\label{fig:ToO} }
\end{figure*}

In order to facilitate the discovery of young SNe for HST spectroscopy, we employed a specialized observation strategy with the ZTF 48-in telescope during part of Cycle 31 and all of Cycle 32. Here, the ``ZTF Flexible Weekends'' program performed a targeted observation strategy in the $g$ and $r$ bands of ZTF fields that would be visible to HST. Every visible field was observed $\sim$1--2 times per night from the Friday -- Sunday preceding a FlexDay opportunity, guaranteeing that a new SN discovered during this time window would be within $\sim$1~week of explosion at the time of the HST observation on the following Thursday -- this is visualized in Figure \ref{fig:ToO}. Details of the ZTF discovery efforts are discussed in \cite{Rehemtullaflex} and \cite{wjgflex}. All promising targets were classified using various spectroscopic facilities including Palomar/SEDM, LT/SPRAT, and Shane/Kast before the Tuesday ToO deadline. All sample objects received Swift UV imaging through community and GI ToO program (PID \#2023100) requests. 

\subsection{Photometric Observations}\label{SubSec:Phot}

The Ultraviolet Optical Telescope (UVOT; \citealt{roming_swift_2005}) on board the Neil Gehrels Swift Observatory \citep{gehrels_swift_2004} provided UV/optical light curves for all SNe in our sample, as shown in Figure \ref{fig:LCs}. We performed aperture photometry with a 5$\arcsec$ region radius with \texttt{uvotsource} within HEAsoft v6.33 \citep{HEAsoft}\footnote{We used the most recent calibration database (CALDB) version.}, following the standard guidelines from \cite{Brown14}\footnote{\url{https://github.com/gterreran/Swift_host_subtraction}}. In order to remove contamination from the host galaxy, we employed pre- and post-explosion template images to subtract the measured count rate at the location of the SN from the count rates in the SN images and corrected for point-spread-function (PSF) losses following the prescriptions of \cite{Brown14}.

We supplement Swift-UVOT light curves for all sample objects with $gri$-band Zwicky Transient Facility (ZTF; \citealt{bellm_zwicky_2019, graham19, dekany2020}) photometry from Public, Partnership and Caltech streams, which was generated through the ZTF forced-photometry service \citep{Masci19}. Additional $ugri$ band photometry was obtained with the Spectral Energy Distribution Machine (SEDM; \citealt{fremling16, blagorodnova_sed_2018, Kim22, Rigault19}) on the 60-in telescope at Palomar Observatory. Data storage and observation follow-up for ZTF and SEDM were performed through the SkyPortal instance of the Kowalski broker and Fritz marshal \citep{skyportal, Duev19, Coughlin23}.

For all SNe, the Milky Way (MW) $V$-band extinction and color excess along the SN line of sight are inferred using a standard \cite{fitzpatrick99} reddening law ($R_V = 3.1$). In addition to the MW color excess \citep{schlafly11}, we estimate the contribution of host-galaxy extinction in the local SN environment using \ion{Na}{i}~D absorption lines for all sample objects. To determine whether \ion{Na}{i}~D is detected, we fit the continuum in a region around the transition based on the spectral resolution and calculate the residuals between the continuum fit and the spectral data. We then integrate the residual flux and confirm that it is greater than or equal to three times the residual flux uncertainty in order to claim a ``detection.'' We calculate the \ion{Na}{i}~D equivalent width (EW) and use $A_V^{\rm host} = (0.78\pm0.15)~{\rm mag} \times ({\rm EW_{NaID}}$/\AA) from \cite{Stritzinger18} to convert these EWs to an intrinsic host-galaxy $E(B-V)$, also using the \cite{fitzpatrick99} reddening law. For non-detections, we calculate an upper limit on the EW and host reddening using the fitted continuum flux. All adopted extinction (MW and host), redshift, distance, and first light date values are reported for the sample objects in Table \ref{tab:sample_specs}. Complete, multi-band light curves are shown in Figure \ref{fig:LCs}. All photometric data/logs will be publicly available in an online data repository\footnote{\url{https://github.com/wynnjacobson-galan/UV_Spectra_Sample}}.

\subsection{Spectroscopic Observations}\label{SubSec:Spec}

Early-time, ``FlexDay'' UV spectra were obtained in the G230LB (1680 – 3060~\AA), G430L (2900 – 5700~\AA), and G750L (5236 – 10266~\AA) gratings of the HST STIS/CCD instrument. Each spectrum was reduced with the \texttt{stistools} package by performing cosmic ray rejection, matching the extraction aperture size across the gratings, increasing the local background size, and placing the local background closer to the spectral trace. We obtained optical spectra of sample objects with the Kast spectrograph on the 3-m Shane telescope at Lick Observatory \citep{KAST}. For all of these spectroscopic observations, standard CCD processing and spectrum extraction were accomplished with \textsc{IRAF}\footnote{\url{https://github.com/msiebert1/UCSC\_spectral\_pipeline}}. The data were extracted using the optimal algorithm of \citet{1986PASP...98..609H}. Low-order polynomial fits to calibration-lamp spectra were used to establish the wavelength scale and apply small adjustments derived from night-sky lines in the object frames. 

Additional optical spectra of sample objects were obtained with the Palomar Spectral Energy Distribution Machine (SEDM; \citealt{blagorodnova_sed_2018, Rigault19, Kim22}), the Nordic Optical Telescope Alhambra Faint Object Spectrograph and Camera (ALFOSC)\footnote{\href{http://www.not.iac.es/instruments/alfosc}{{http://www.not.iac.es/instruments/alfosc}}}, the Liverpool Telescope Spectrograph for the Rapid Acquisition of Transients (SPRAT; \citealt{piascik_sprat_2014}), the P200 Palomar Double Spectrograph (DBSP) and Next Generation Palomar Spectrograph (NGPS). Optical spectra obtained with the Low Resolution Imaging Spectrometer (LRIS; \citealt{oke_keck_1995}) on the 10-m Keck I telescope were reduced with {\tt Lpipe} \citep{Perley19}. Near-infrared (NIR) spectra were obtained with the $R\approx2700$ Near-Infrared Echelle Spectrograph (NIRES; \citealt{2004SPIE.5492.1295W}) located on the 10-m Keck II telescope. The data was reduced using a custom version of the IDL-based reduction package {\tt Spextool} \citep{2004PASP..116..362C} modified for use with NIRES. We used {\tt xtellcor} \citep{2003PASP..115..389V} to correct for telluric features in our spectrum using an A0 standard star observed close in airmass and time to our target. A complete log of spectroscopic observations is presented in Table \ref{tab:spec_all}. All spectroscopic observations will be released publicly on GitHub$^3$ and WISeREP.






\section{Photometric Analysis}\label{sec:photanalysis}

\begin{figure*}
\centering
\subfigure{\includegraphics[width=0.95\textwidth]{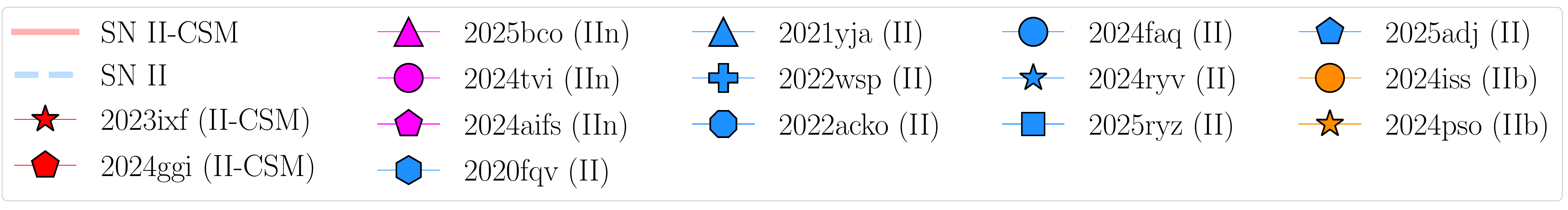}}\\ \vspace{-3mm}
\subfigure{\includegraphics[width=0.49\textwidth]{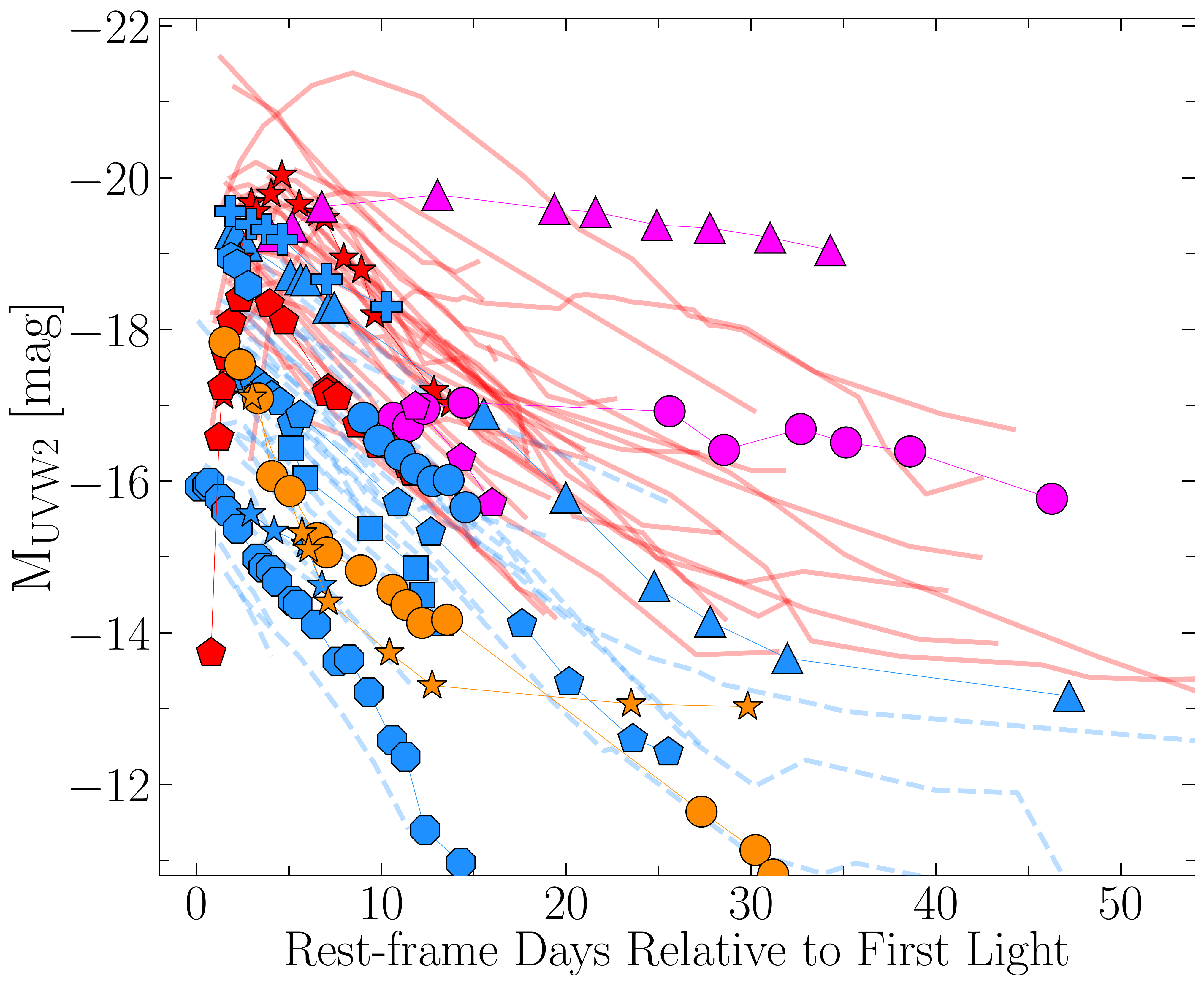}}
\subfigure{\includegraphics[width=0.49\textwidth]{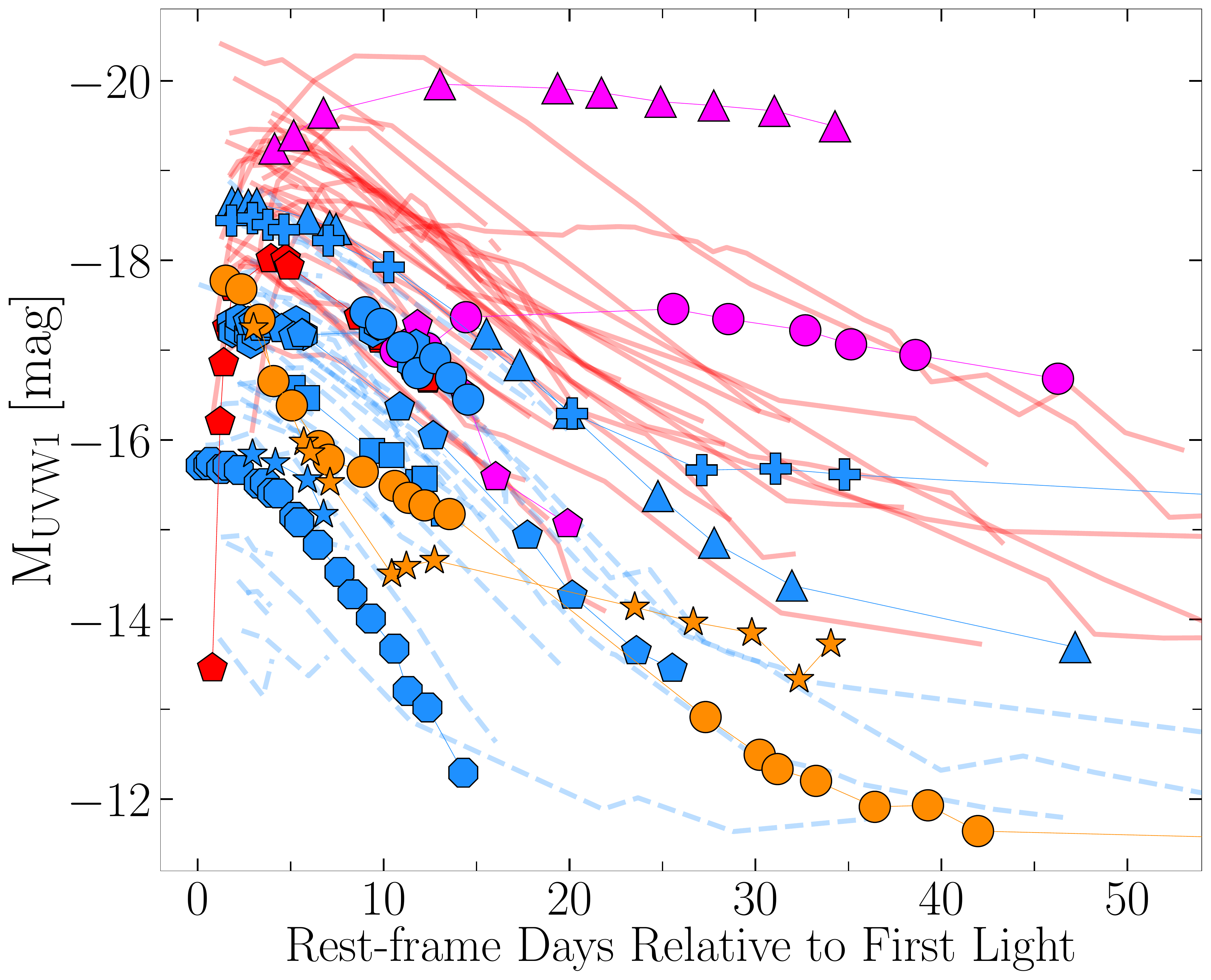}}\\
\subfigure{\includegraphics[width=0.49\textwidth]{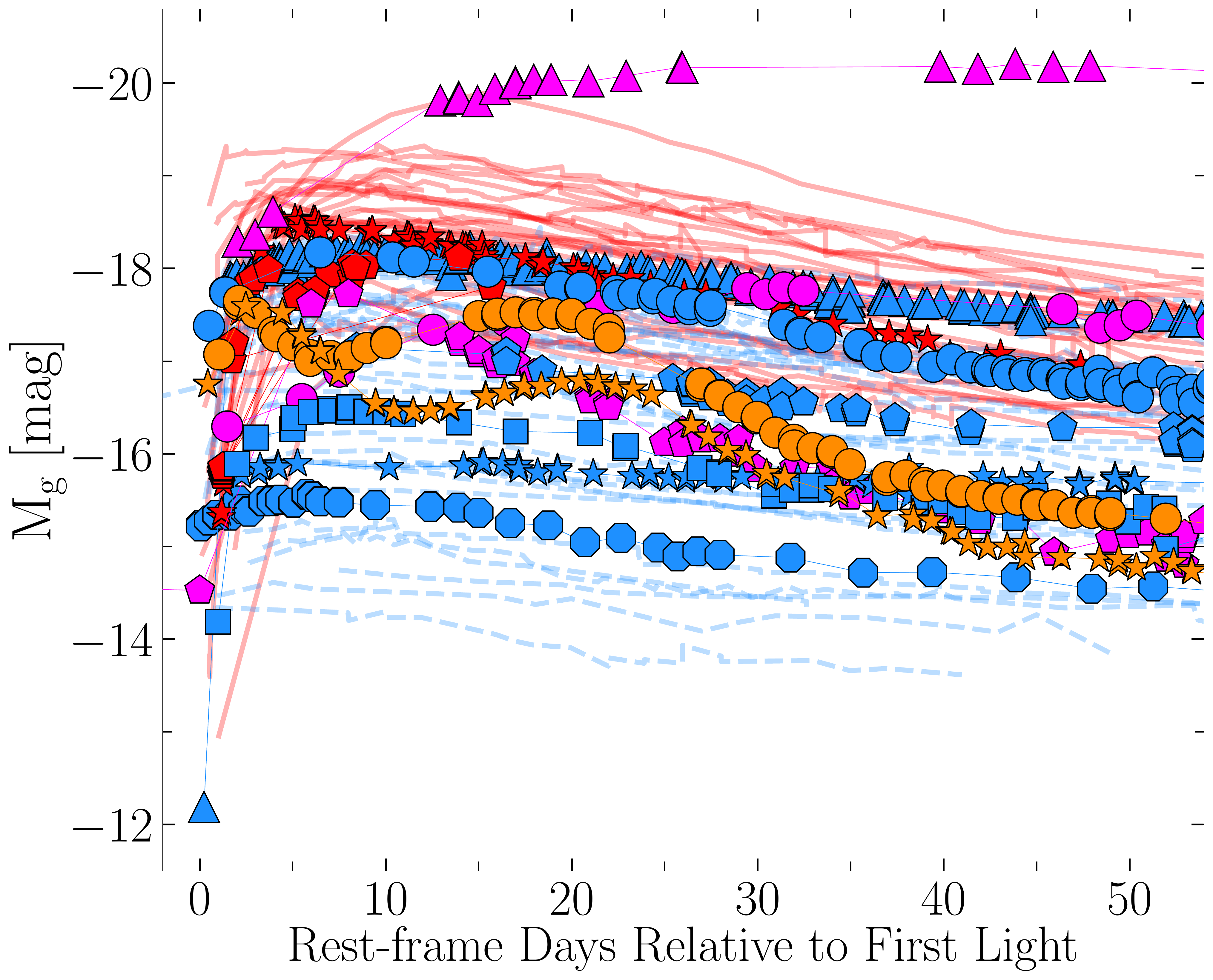}}
\subfigure{\includegraphics[width=0.49\textwidth]{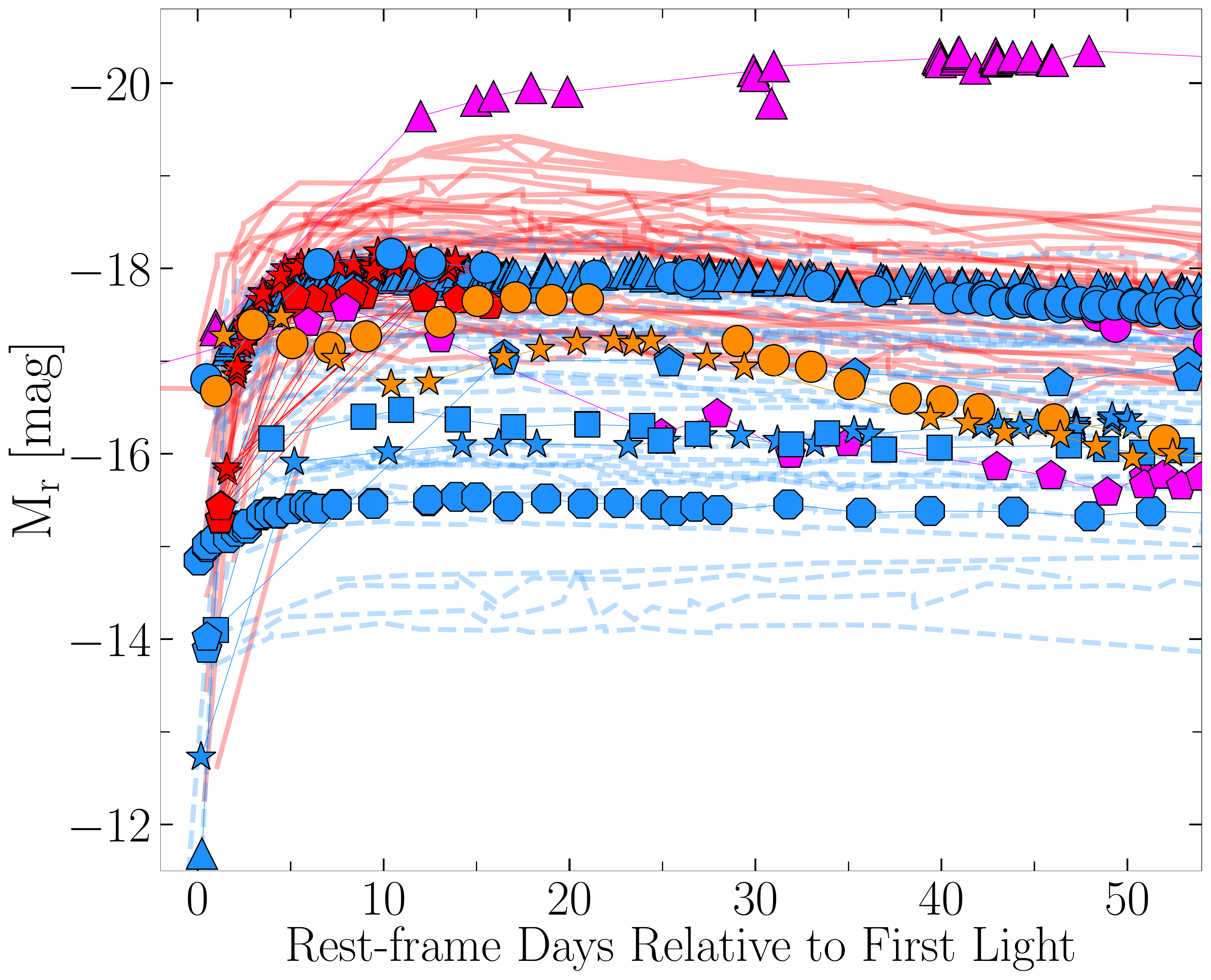}}
\caption{{\it Top Left:} {\it UVW2}-, {\it Top Right:} {\it UVW1}-, {\it Bottom Left:} {\it g}-, and {\it Bottom Right:} {\it r}-band light curves of SNe II/IIn with HST UV spectra compared to SNe II with (red solid lines) and without (blue dashed lines) ``IIn-like'' features in their early-time optical spectra \citep{wjg24, wjg25a}. SNe~II-P plotted as blue markers, SNe IIn as magenta markers and SNe IIb as orange markers. \label{fig:LCs} }
\end{figure*}

\begin{figure*}
\centering
\subfigure{\includegraphics[width=0.49\textwidth]{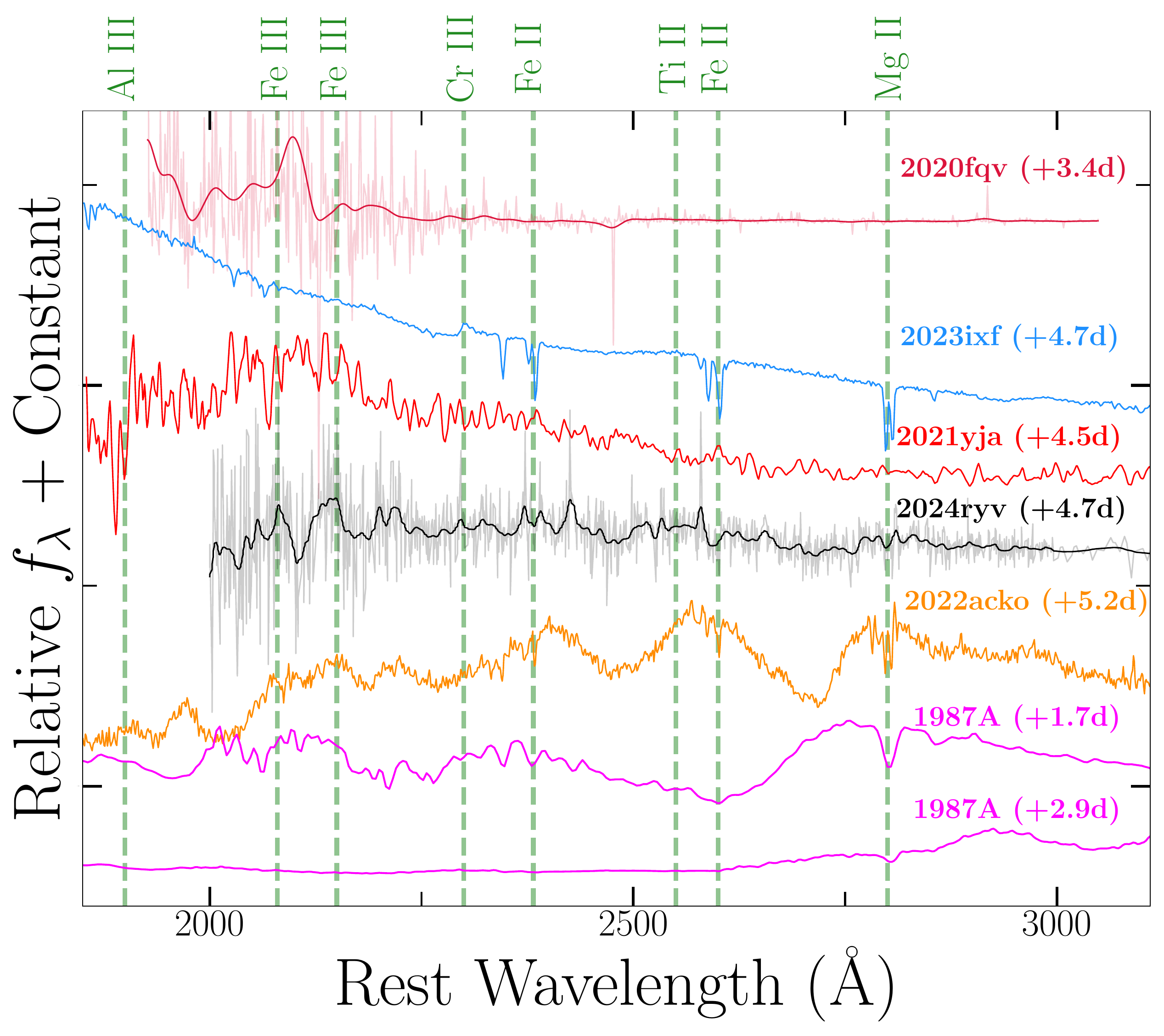}}
\subfigure{\includegraphics[width=0.49\textwidth]{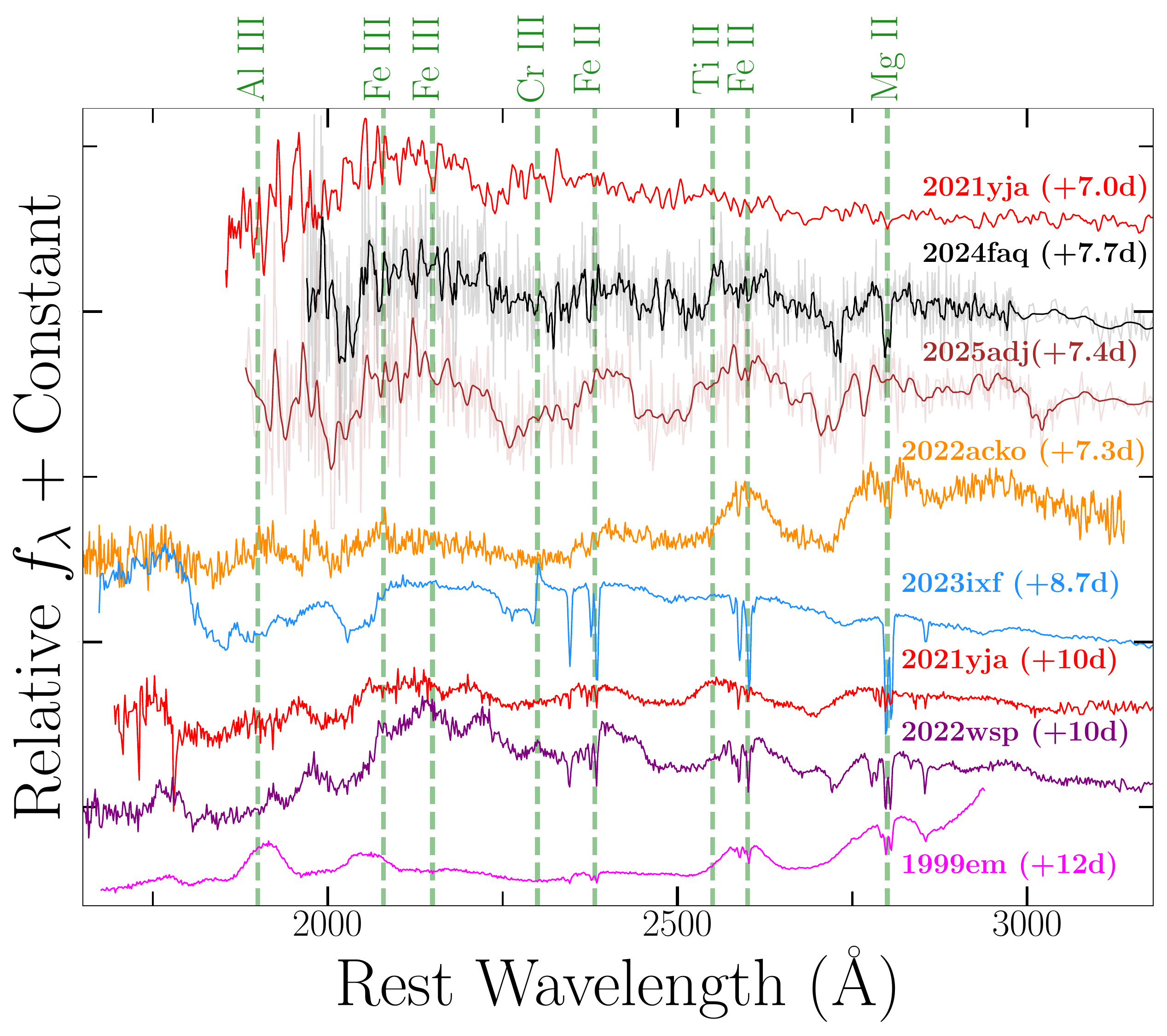}}\\
\caption{{\it Left:} Comparison of the youngest SNe II with UV spectroscopy at $\delta t < 6$~days. UV spectra of SNe~2023ixf, 2024ryv and 2022acko are from HST, while SN~1987A was observed with IUE and SN~2021yja was observed with Swift. Line IDs shown as vertical dashed lines. {\it Right:} UV spectra comparison of SNe~II observed at $7 < \delta t < 12$~days. All sample objects presented were observed with HST but the earliest UV spectra of SN~2021yja were obtained by Swift. \label{fig:SNIIUV} }
\end{figure*}

We present extinction-corrected Swift/UVOT $w2,w1$ and ZTF $g,r$-band light curves of all sample objects in Figure \ref{fig:LCs}. We compare object light curves to the sample of SNe II compiled by \cite{wjg24, wjg25a}, which all have early-time ($\delta t \lesssim 2$~days) optical spectra with (red solid lines) or without (blue dashed lines) IIn-like (``flash'') features indicative of ejecta interaction with confined CSM. With the exception of SN~2023ixf, all sample SNe~II have UV light curves that are consistent with the UV luminosity evolution of SNe~II with no IIn-like features, likely suggesting low CSM densities and progenitor mass-loss rates. This conclusion is further validated by the lack of IIn-like features in all early-time optical spectra obtained at $\delta t \approx 1.4 - 2.4$~days for the majority of sample SNe~II. We note that the first spectrum of Type II SN 2025ryz was obtained at $\delta t = 4.1$~days and it shows both Doppler-broadened Balmer series profiles as well as blue-shifted \ion{He}{ii} $\lambda$4686 emission. This profile is also seen in sample object SN~2021yja and can be explained by spectral models for SNe~II interacting with low density CSM formed from progenitor mass-loss rates as low as $\dot M = 10^{-6}$~\mdot \citep{Hosseinzadeh22, wjg24}. Unlike the other SNe~II in the sample, SN~2023ixf displays prominent IIn-like features and a luminous, longer-rising UV light curve, all of which are the product of ejecta interaction with high-density CSM (i.e., $\dot M \approx 10^{-2}$~\mdot, $v_w = 25~\kms$) confined to $<10^{15}$~cm \citep{wjg23, Bostroem23, Zimmerman24, Dickinson25}. 

As shown in Figure \ref{fig:LCs}, we also present UV/optical light curves of three sample SNe~IIn and two SNe~IIb with early-time UV spectroscopy. Similar to CSM-interacting SNe~II, the sample SNe~IIn display longer rise-times and remain UV-bright out to later phases due to on-going ejecta interaction with high-density CSM. The most luminous event in our sample is SN~2025bco, which has a UV rise-time of $\sim15$~days and peaks at $M_{\rm UV} \approx -20$~mag. Amongst the SN~IIn sample objects, we note that SN~2024aifs displays a luminous precursor event ($M_r \approx -16$~mag) in the final $\sim$40~days before the adopted explosion date (e.g., see Appendix Figure \ref{fig:24aifs}). Additionally, our sample includes Type IIb SNe 2024iss and 2024pso, both of which display a double-peaked light curve indicative of initial emission from shock cooling of an extended envelope which is then followed by emission from radioactive decay power. The maximum luminosity of the first light curve peak in both events is similar ($M_g \approx -17.8$~mag) but the duration of the shock cooling emission is longer in SN~2024pso ($\sim$10~days) compared to SN~2024iss ($\sim$6~days) suggestive of a larger extended envelope in the former. Modeling of the early-time, multi-band light curve of SN~2024iss suggests an extended H-rich progenitor envelope of $\sim100-320~\Rsun$ \citep{yete26, chen26} and the intra-night ZTF detection at $\sim$40~min after first light was shown to be consistent with shock breakout emission \citep{yete26}. 

In Appendix Figures \ref{fig:BB1} \& \ref{fig:BB2}, we show blackbody model fits to both the multi-band UV/optical photometry and HST UV/optical spectra at similar early-time epochs for our new sample SNe~II, IIb and IIn. We find that similar blackbody temperatures and radii can be derived from both photometry and spectra that extend from optical to near-UV wavelengths. However, line blanketing from the forest of metal lines in the UV as well as scattering does cause a deviation from a smooth, thermal continuum in some SNe~II and IIb \citep{dessart05}. This is seen visually in sample objects with wavelength coverage at $\lambda < 2000$~\AA~ (e.g., SNe~2024iss, 2022acko, 2021yja, 2022wsp), where metal-line blanketing (e.g., \ion{Fe}{iii}) causes significant deviation from a blackbody model \citep{D15_2n, dessart26a}. \cite{yete26} showed that the root mean square (RMS) fractional deviation between the observed spectrum of SN~IIb 2024iss and the best-fit blackbody model was as high as $\sim47\%$ between $\lambda = 1600-2200$~\AA. Furthermore, \cite{bostroem22acko} found that a suppression factor was required to fit a blackbody model to the early-time far-UV to optical spectra of SN~2022acko, and they also  noted that the blackbody temperatures were systematically higher than the temperatures applied in the non-LTE model spectra. 

\section{Spectroscopic Analysis}\label{sec:specanalysis}

\subsection{Type II Supernovae}\label{sec:specII}

\subsubsection{Comparisons and Line Profiles}\label{sec:SNIIcomp}

In Figure \ref{fig:SNIIUV}, we present comparisons of near-UV spectra of SNe~II at $\delta t < 6$~days (left panel) and $7 < \delta t < 12$~days (right panel). We supplement this sample comparison with Swift Grism spectra of SN~2021yja at $\delta t = 4.5$~days \citep{vasylyev22} and IUE spectra of SN~1987A at $\delta t = 1.7 - 2.9$~days \citep{Pun95}. To date, the earliest HST UV spectral observations are still those of SN~2020fqv at $\delta t = 3.4$~days \citep{tinyanont22}, followed by SN~2023ixf at $\delta t = 3.6$~days \citep{Zimmerman24}. However, through the ``FlexDay'' ToO program, SN~2024ryv was observed at $\delta t = 4.7$~days, making it the third youngest SN~II with UV spectroscopy from HST. Furthermore, the HST STIS/MAMA observations of SN~2022acko yielded one of the earliest and highest SNR far- and near-UV spectral sequences of a SN~II published to date \citep{bostroem22acko}. Nonetheless, there has been no earlier UV SN spectrum than SN~1987A, which was observed with the IUE at $\delta t = 1.7$~days \citep{Kirshner87, Wamsteker87, Pun95}. 

Despite the limited observations, UV spectra of SNe~II at $\delta t < 6$~days show some noticeable diversity. For example, SNe~2020fqv, 2021yja, 2023ixf and 2024ryv are relatively featureless compared to SNe~2022acko and 1987A, which show clear, broad features of Mg, Fe, Cr, and Ti from the fast-moving SN ejecta. For SN~2023ixf, there are narrow emission lines of \ion{C}{iii} and \ion{N}{iv} from photo-ionized and radiatively-accelerated gas ahead of the forward shock, which are attached to deeper blue-shifted absorption ``troughs'' that likely trace faster-moving material in the post-shock region. However, in SN~2024ryv, close examination of the UV spectra does show weak, broad absorption profiles consistent with \ion{Mg}{ii} as well as some other Fe and Ti blends. The lack of SN ejecta features at these phases is explained by on-going CSM interaction for SN~2023ixf, but it is more ambiguous in other objects such as SNe~2020fqv, 2021yja and 2024ryv. These objects do not show IIn-like (or any narrow) features in their optical spectra, suggesting that any confined CSM must be relatively low density compared to SNe~II with stronger CSM interaction signatures at early times. Overall, it is possible that ejecta features are suppressed at such early phases for a variety of reasons, e.g., a larger stellar envelope, larger expansion velocities, and/or some ejecta interaction with low-density, confined CSM.

\begin{figure*}
\centering
\subfigure{\includegraphics[width=0.35\textwidth]{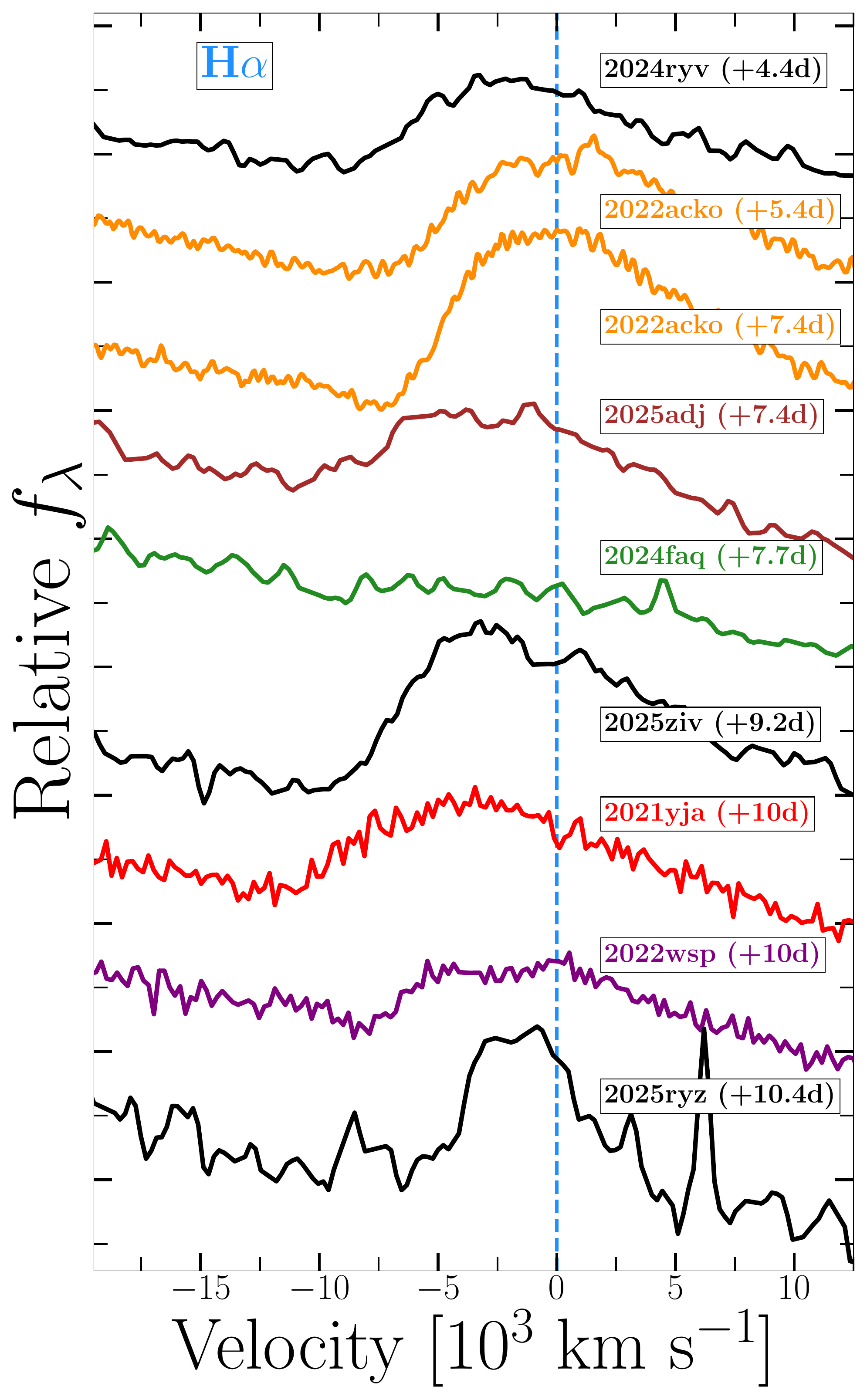}}
\subfigure{\includegraphics[width=0.32\textwidth]{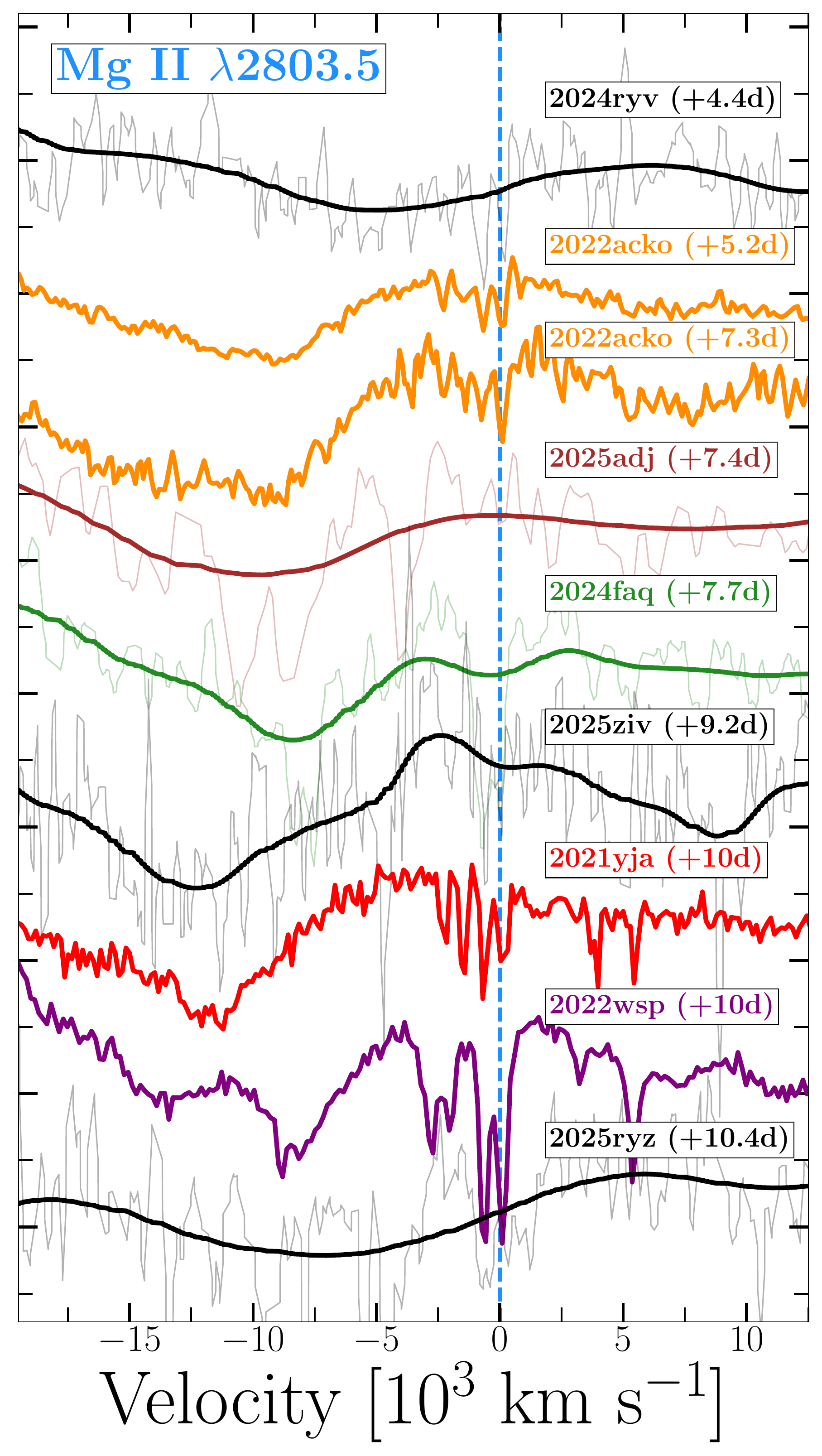}}
\subfigure{\includegraphics[width=0.32\textwidth]{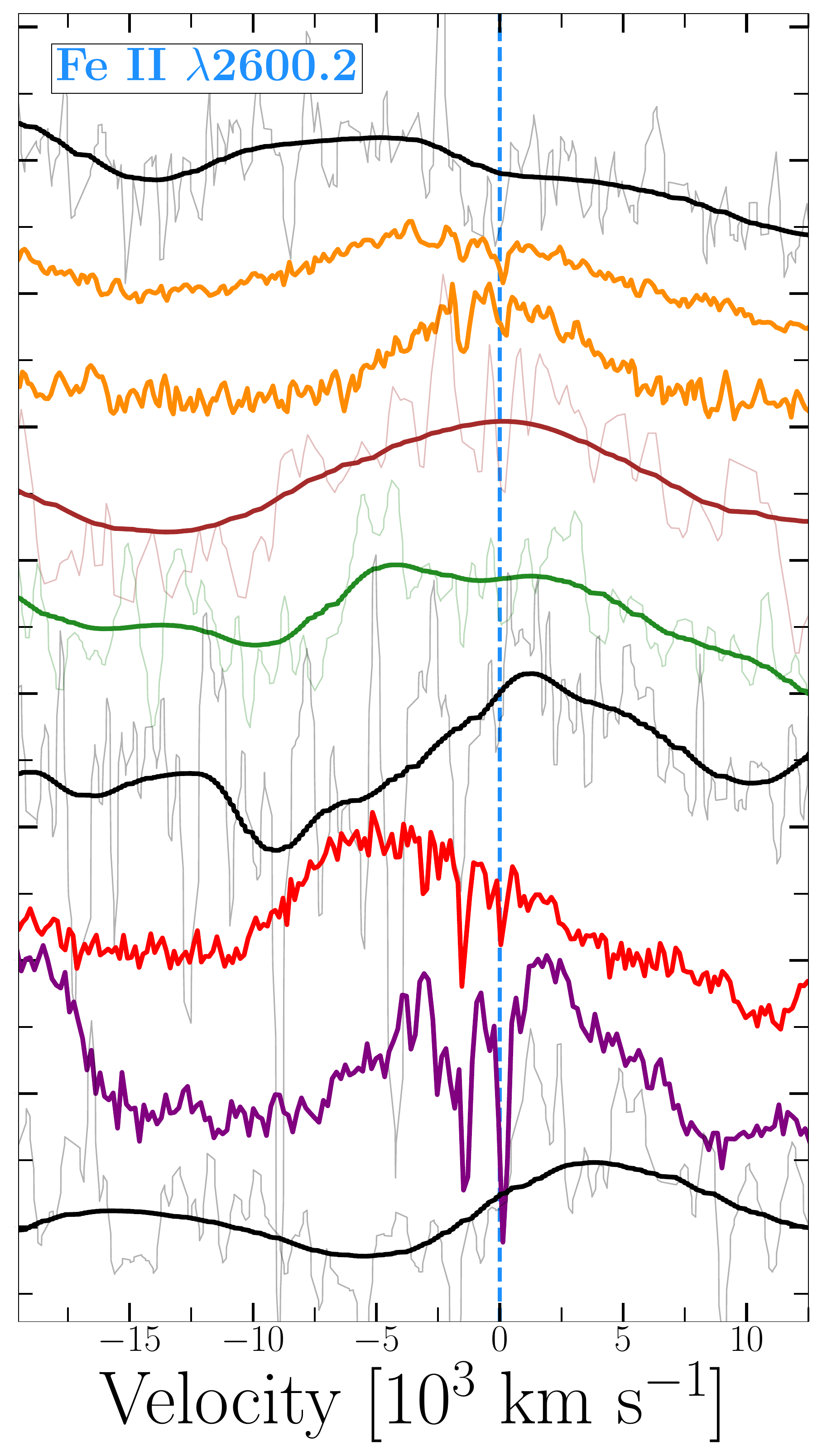}}
\caption{{\it Left:} H$\alpha$, {\it Middle:} \ion{Mg}{ii} $\lambda$2803.5, and {\it Right:} \ion{Fe}{ii} $\lambda$2600 velocities of sample SNe~II, ordered by increasing phase since first light. We note that precise measurement of \ion{Fe}{ii} velocities is challenging given the numerous blended lines from metals near this central wavelength. The systematic blueshift of the H$\alpha$ emission centroid is a known property of young SNe~II and linked to the outer ejecta density profile \citep{Anderson14b}. \label{fig:vels} }
\end{figure*}

\begin{figure*}
\centering
\subfigure{\includegraphics[width=0.48\textwidth]{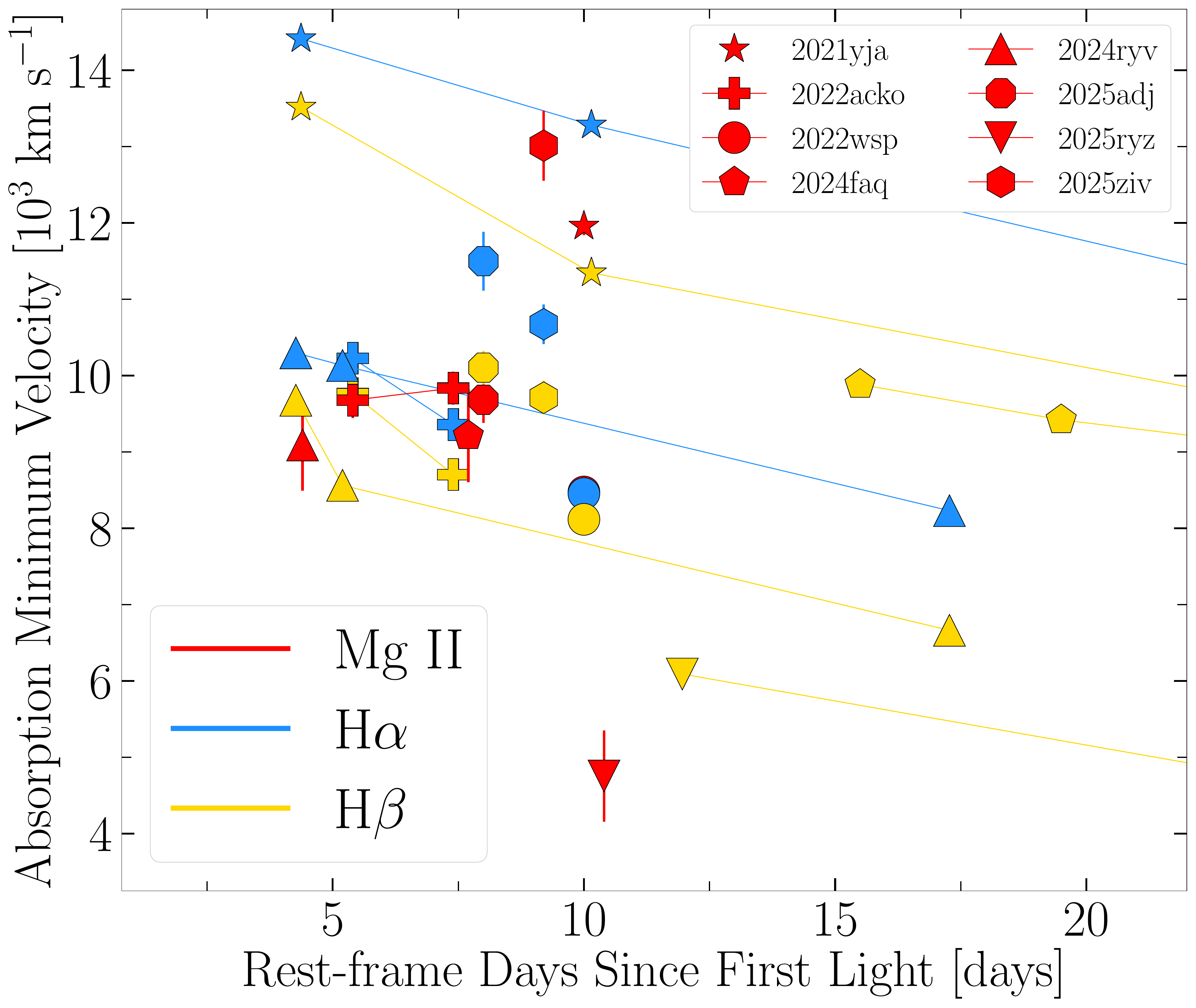}}
\subfigure{\includegraphics[width=0.515\textwidth]{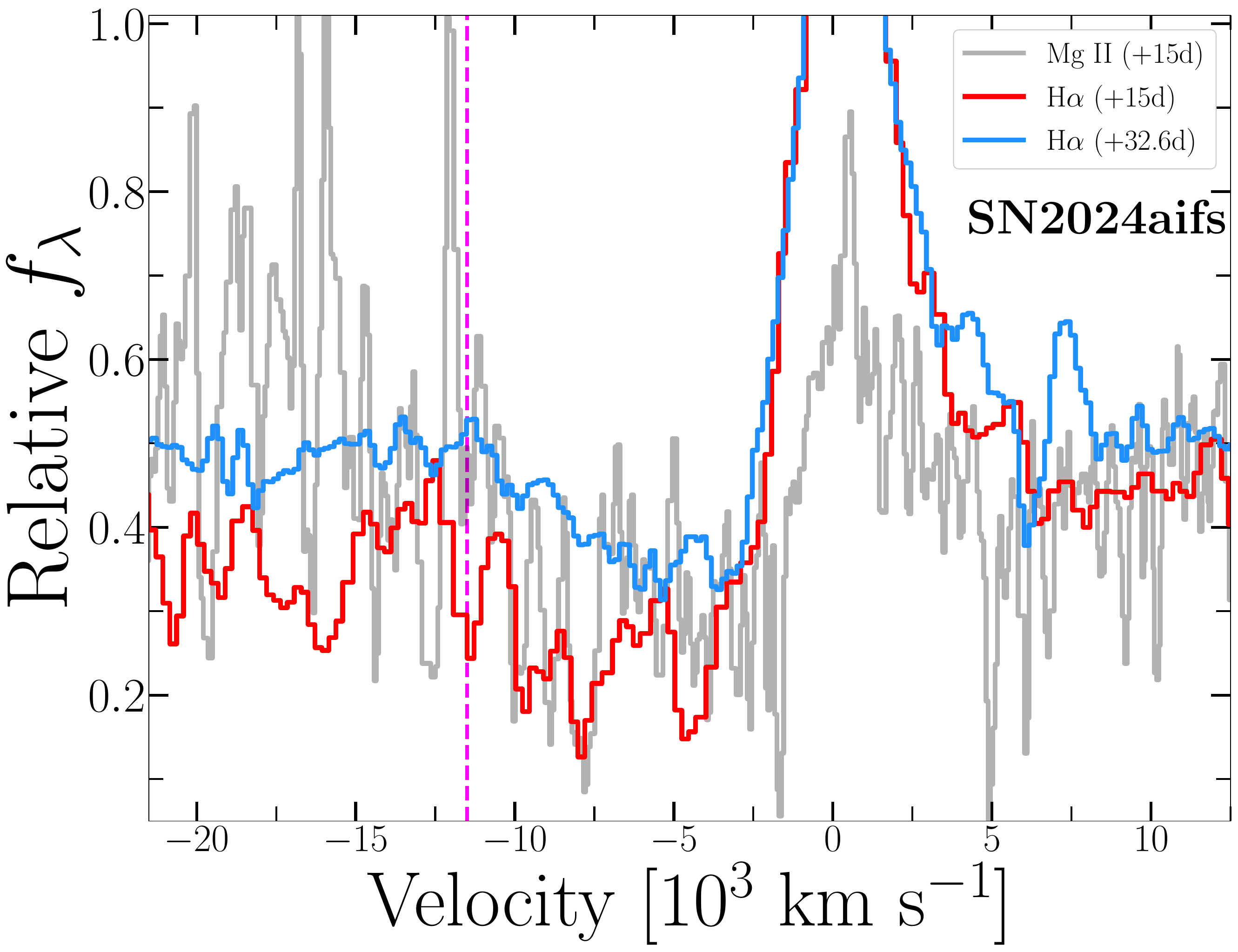}}
\caption{{\it Left:} H$\alpha$ (blue), H$\beta$ (yellow), and \ion{Mg}{ii} $\lambda$2803.5 (red) velocities of sample SNe~II 
relative to days since first light as measured from fitting of each absorption profile minimum. 
{\it Right:} \ion{Mg}{ii} $\lambda$2803.5 velocity profile in Type IIn SN 2024aifs at $\delta t = 15$~days compared to H$\alpha$ at the same phase and at $\delta t = 32.6$~days. Approximate maximum blueward velocity marked with dashed magenta line. \label{fig:mgIIvels} }
\end{figure*}

We also compare the near-UV spectra of sample SNe~II at $\delta t \approx 1-2$~weeks in the right panel of Figure \ref{fig:SNIIUV}. By these phases, Doppler-broadened profiles of metals such as Al, Fe, Ti, Cr and Mg located in the fast-moving outer SN ejecta are clearly visible. We label some of the prominent transitions for each element in Figure \ref{fig:SNIIUV}, but given the forest of lines in the UV, it is difficult to disentangle individual ion transitions and their corresponding line velocities like at optical and IR wavelengths. In Figure \ref{fig:vels}, we plot line velocities for all sample objects centered on H$\alpha$, \ion{Mg}{ii}~$\lambda2803.5$ and \ion{Fe}{ii}$~\lambda2600.2$. Additionally, we measure the absorption minimum of the \ion{Mg}{ii}$~\lambda2803.5$, H$\alpha$ and H$\beta$ profiles as a function of time since first light in Figure \ref{fig:mgIIvels}. We find that the \ion{Mg}{ii}~$\lambda2803.5$ line profile is consistent with H$\alpha$ in its shape and extent to the fastest velocities. As shown in Figure \ref{fig:mgIIvels}, the absorption minimum of \ion{Mg}{ii}~$\lambda2803.5$ is smaller than for  H$\alpha$ for most SNe~II at early-time phases, but can be comparable to the minimum velocity measured for H$\beta$. Overall, the near-UV spectra of the sample SNe~II without strong CSM interaction all show similar spectral profiles, with the primary distinguishing feature being the line velocities, which can be linked to differences in kinetic energy. However, it is important to note that ejecta velocities and line profiles can be impacted by the presence of confined CSM since that interaction will extract kinetic energy from the SN ejecta.

\subsubsection{Type II SN Model Matching}\label{sec:SNIImodels}

\begin{figure*}
\centering
\subfigure{\includegraphics[width=0.49\textwidth]{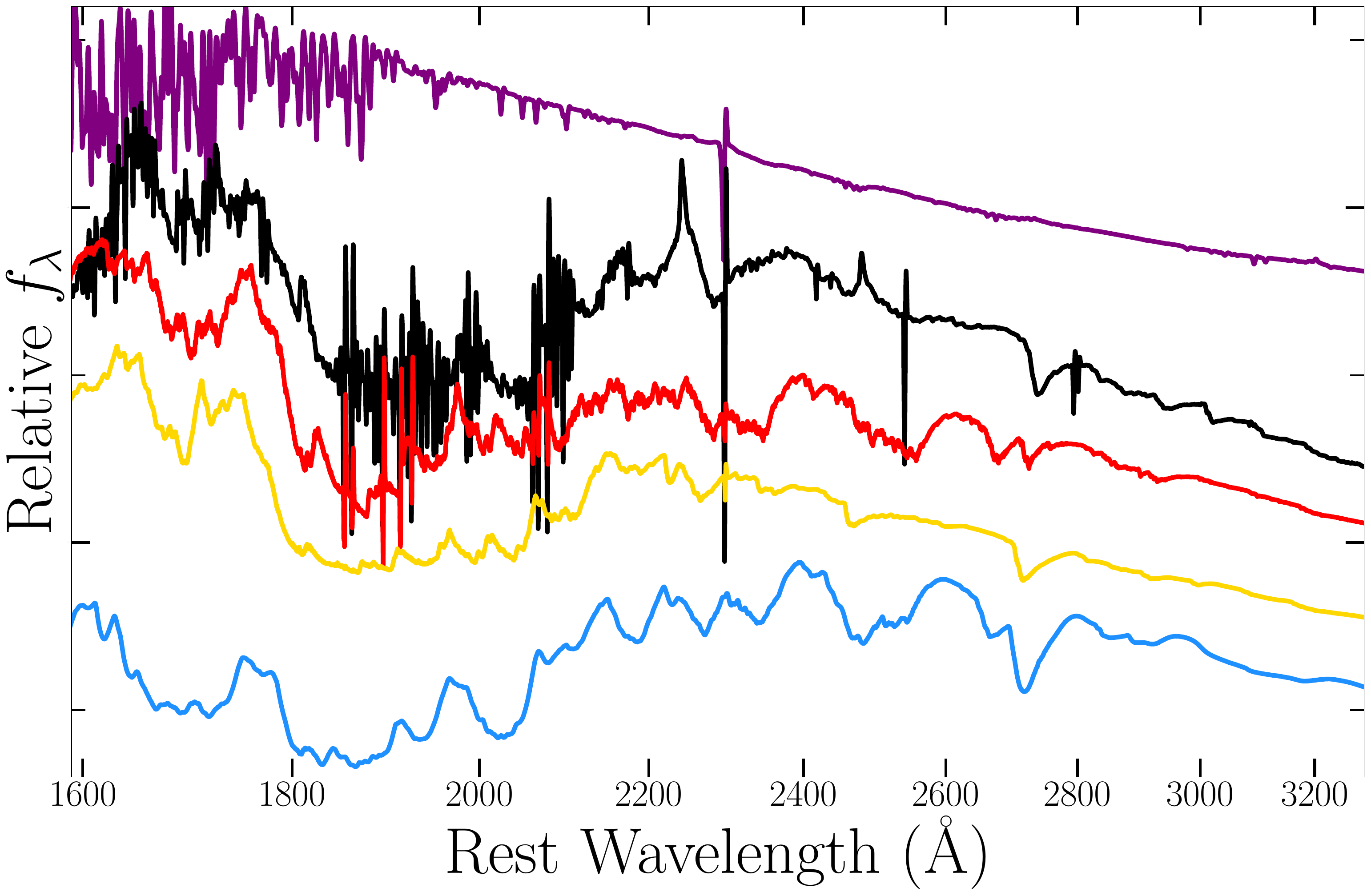}}
\subfigure{\includegraphics[width=0.49\textwidth]{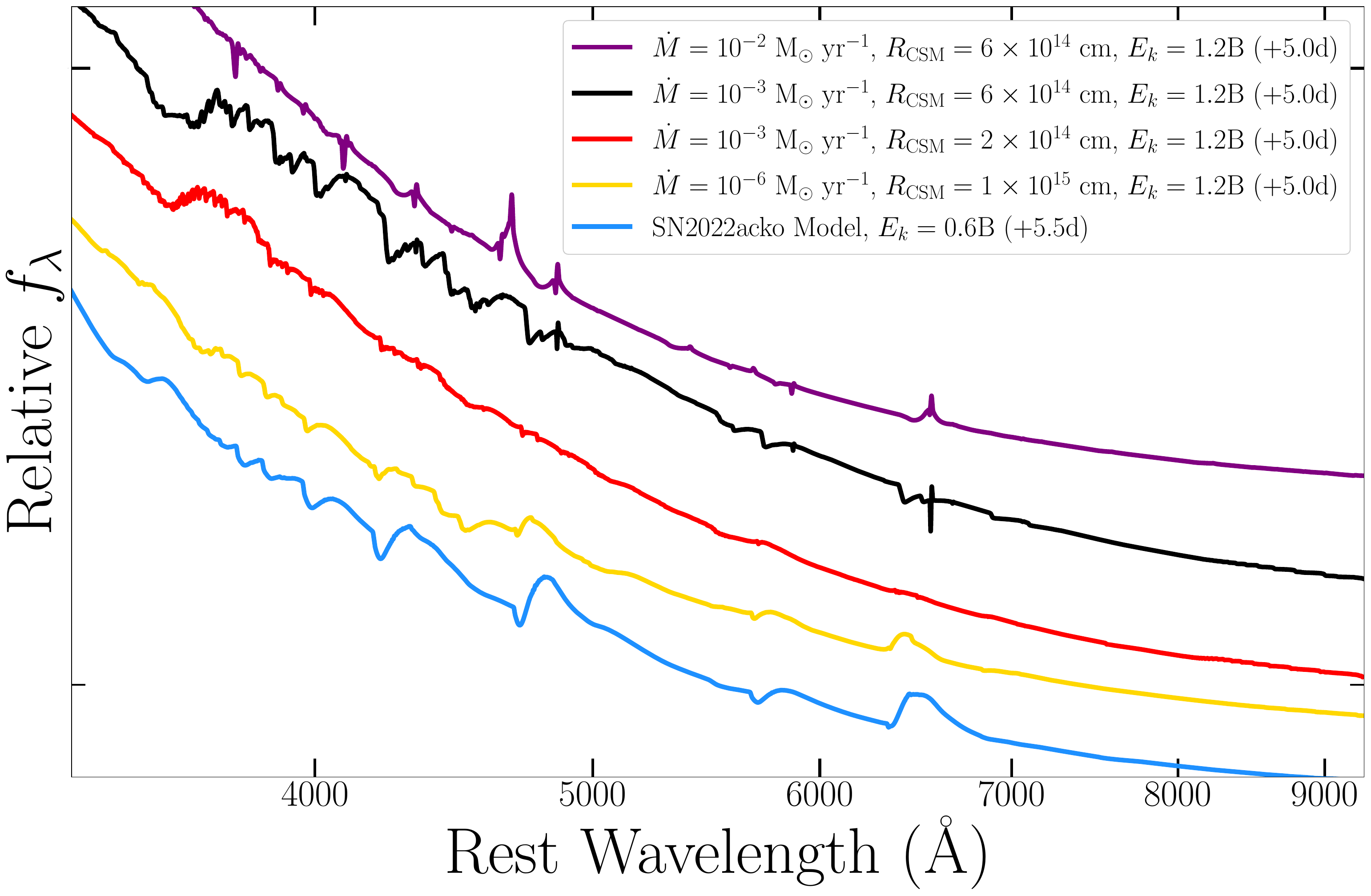}}
\caption{Example UV (left) and optical (right) spectra of \cmfgen\ models with varying mass-loss rates, CSM radii and kinetic energies from \cite{bostroem22acko} and \cite{dessart25CSMgrid} at $\delta t \approx 5$~days. Model spectra with high mass-loss rates ($\dot M = 10^{-3} - 10^{-2}$~\mdot) and extended CSM ($R_{\rm CSM} = 6\times 10^{14}$~cm) are shown in purple and black; compact CSM ($R_{\rm CSM} = 2\times 10^{14}$~cm) shown in red. Models with weak mass-loss rates ($\dot M = 10^{-6}$~\mdot) and kinetic energies of 1.2~B and 0.6~B are shown in yellow and blue. The presence of confined CSM increases the temperature of the continuum and suppresses emission from the outer SN ejecta at early-time phases.  \label{fig:models} }
\end{figure*}

To further understand the physical properties of the sample SNe II, we compare their  UV spectra to non-LTE radiative transfer simulation spectra created using \cmfgen\ \citep{hillier12, D15_2n}. For model matching, we utilize the grid of SN II models presented in \cite{dessart25CSMgrid}, which uniformly covers mass-loss rates of $10^{-3} - 10^{-1}$~\mdot, CSM radii of $(0.2-1)\times 10^{15}$~cm, and kinetic energies of 0.6, 1.2 and 1.8~B. This grid also includes one low mass-loss rate setup ($10^{-6}$~\mdot, $R_{\rm CSM} = 10^{15}$~cm) for all three kinetic energies e.g., the {\tt m1em6\_e1p2\_r1e15} model has $E_k = 1.2$~B. These models were generated using a $15~\Msun$ progenitor model with final mass and radius of $12.4~\Msun$ and $652~\Rsun$, respectively. Additionally, we compare observed UV/optical spectra to \cmfgen\ model spectra from \cite{bostroem22acko}, which were created with two kinetic energies (0.4 and 0.6 B) and low-density CSM (i.e., $\dot M = 10^{-6}$~\mdot). These models have an initial progenitor ZAMS mass of $12~\Msun$, a final mass of $9.7~\Msun$, and final radius of $500~\Rsun$. All models were created with an ejecta/CSM metallicity based on the surface abundances of the {\tt MESA} RSG progenitor model. The details of each non-LTE solver and model setups are described in the above papers. In Figure \ref{fig:models} we show example \cmfgen\ spectra at UV/optical wavelengths for varying mass-loss rates and CSM radii from both sets of models at $\sim$5 days. 

\begin{figure*}
\centering
\subfigure{\includegraphics[width=0.49\textwidth]{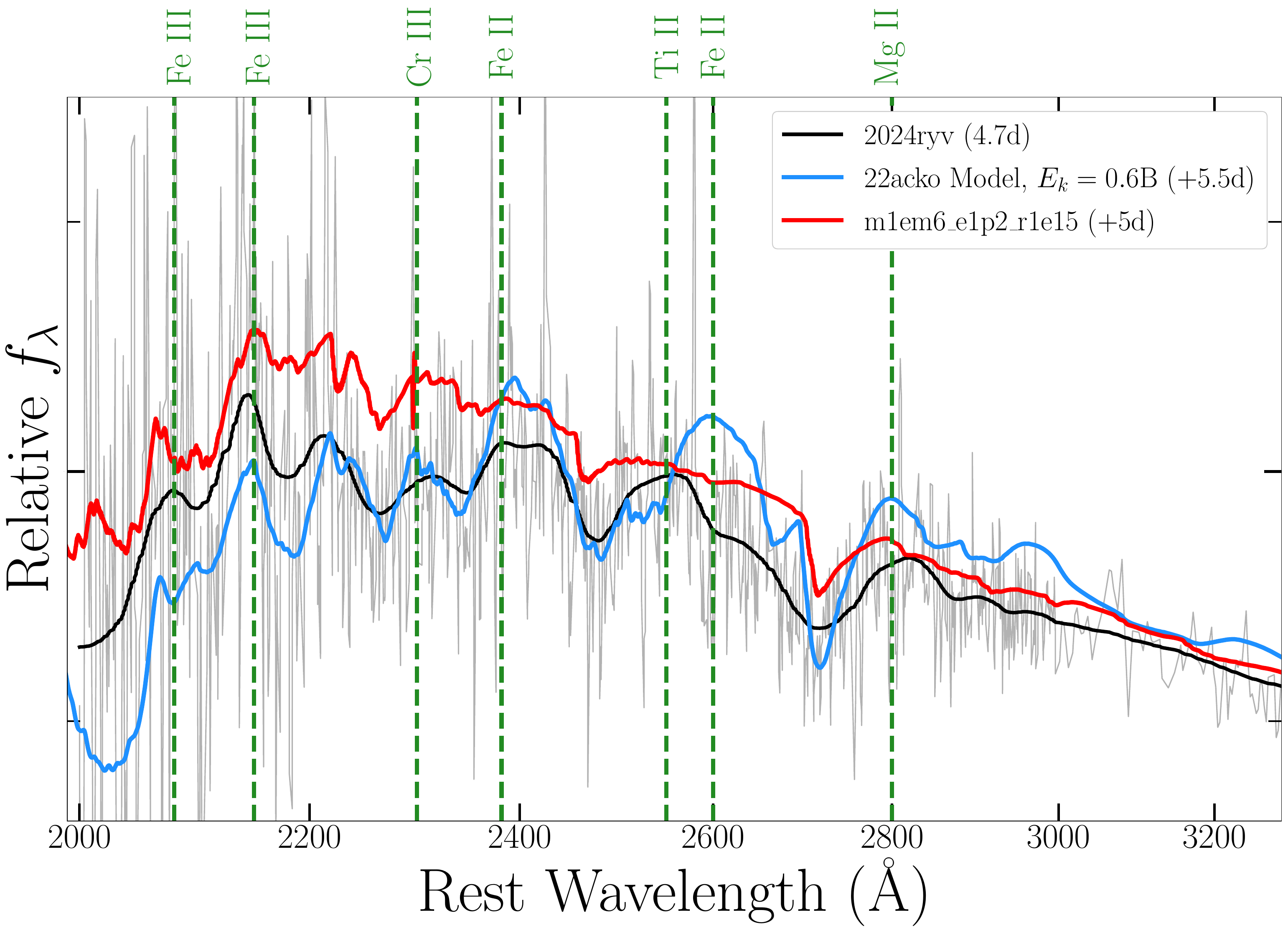}}
\subfigure{\includegraphics[width=0.49\textwidth]{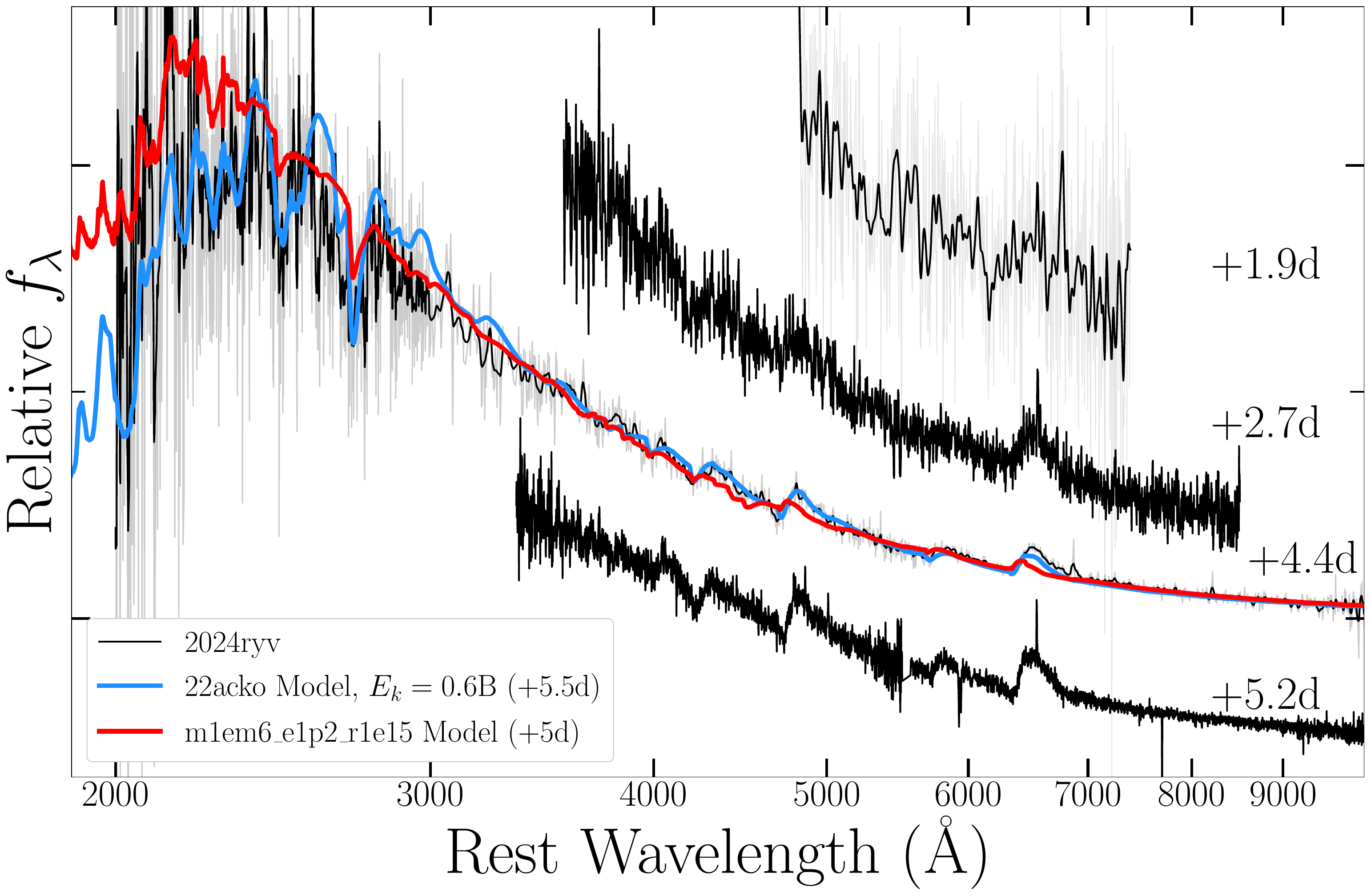}}\\
\subfigure{\includegraphics[width=0.49\textwidth]{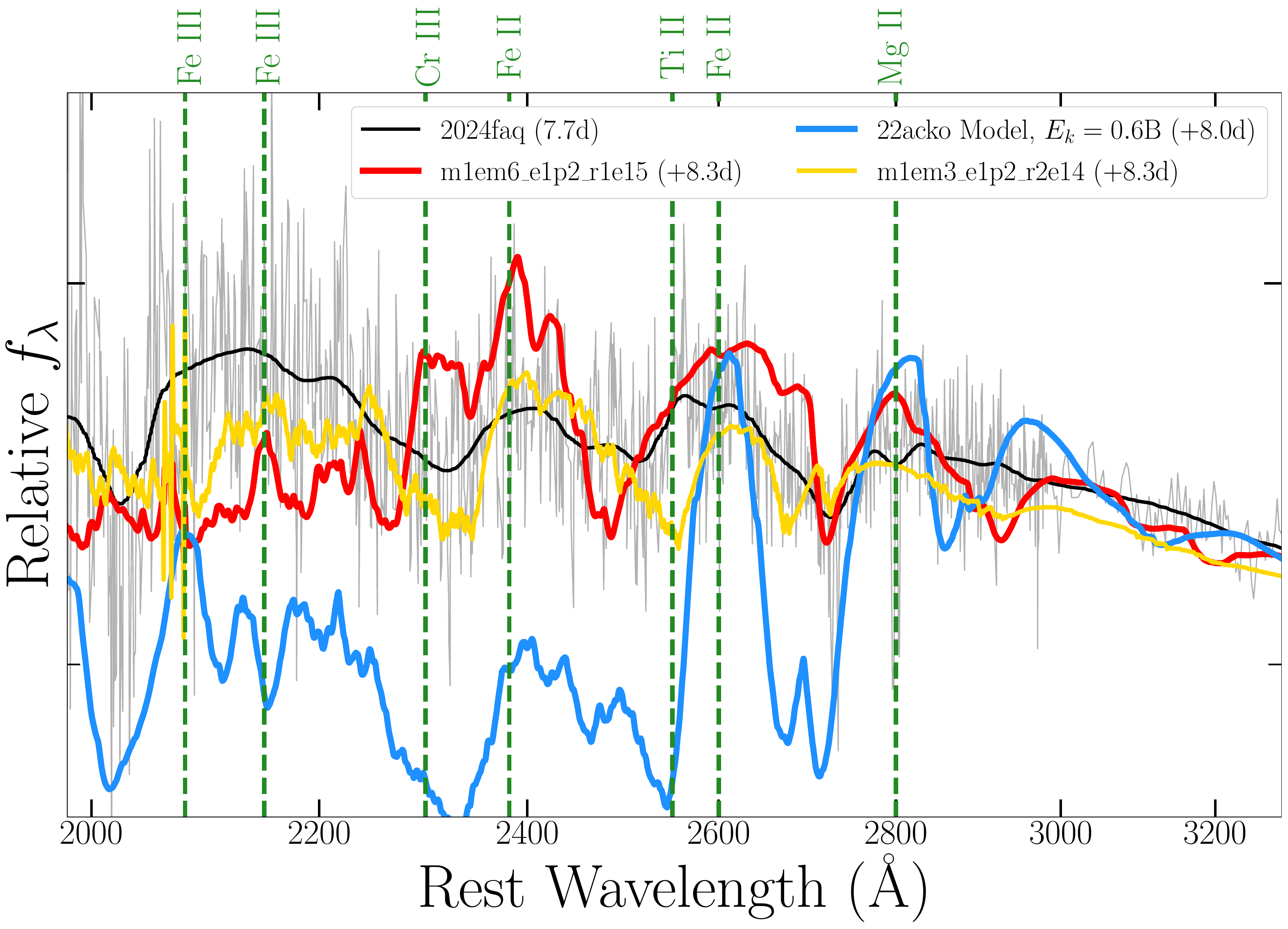}}
\subfigure{\includegraphics[width=0.49\textwidth]{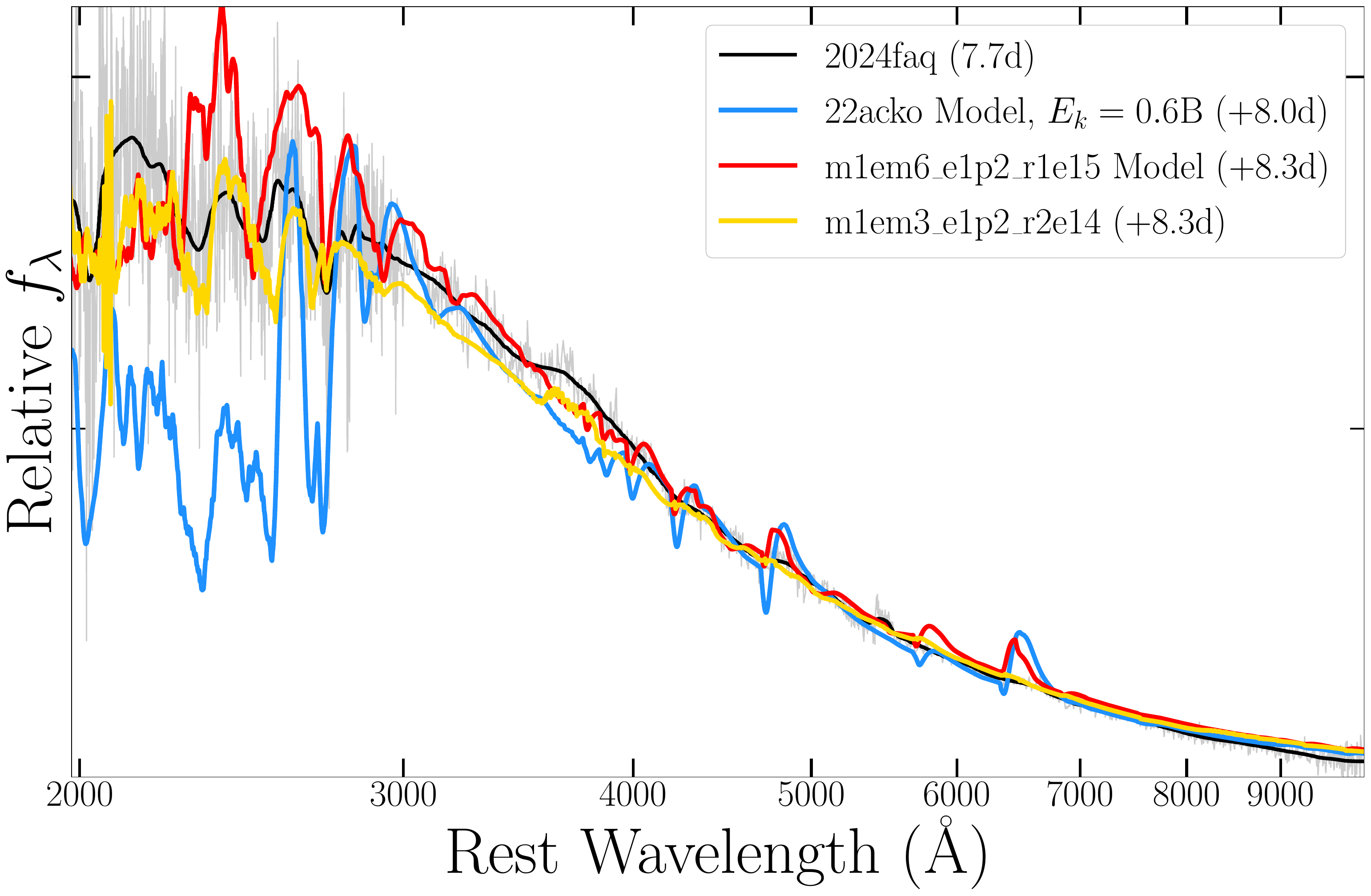}}\\
\subfigure{\includegraphics[width=0.49\textwidth]{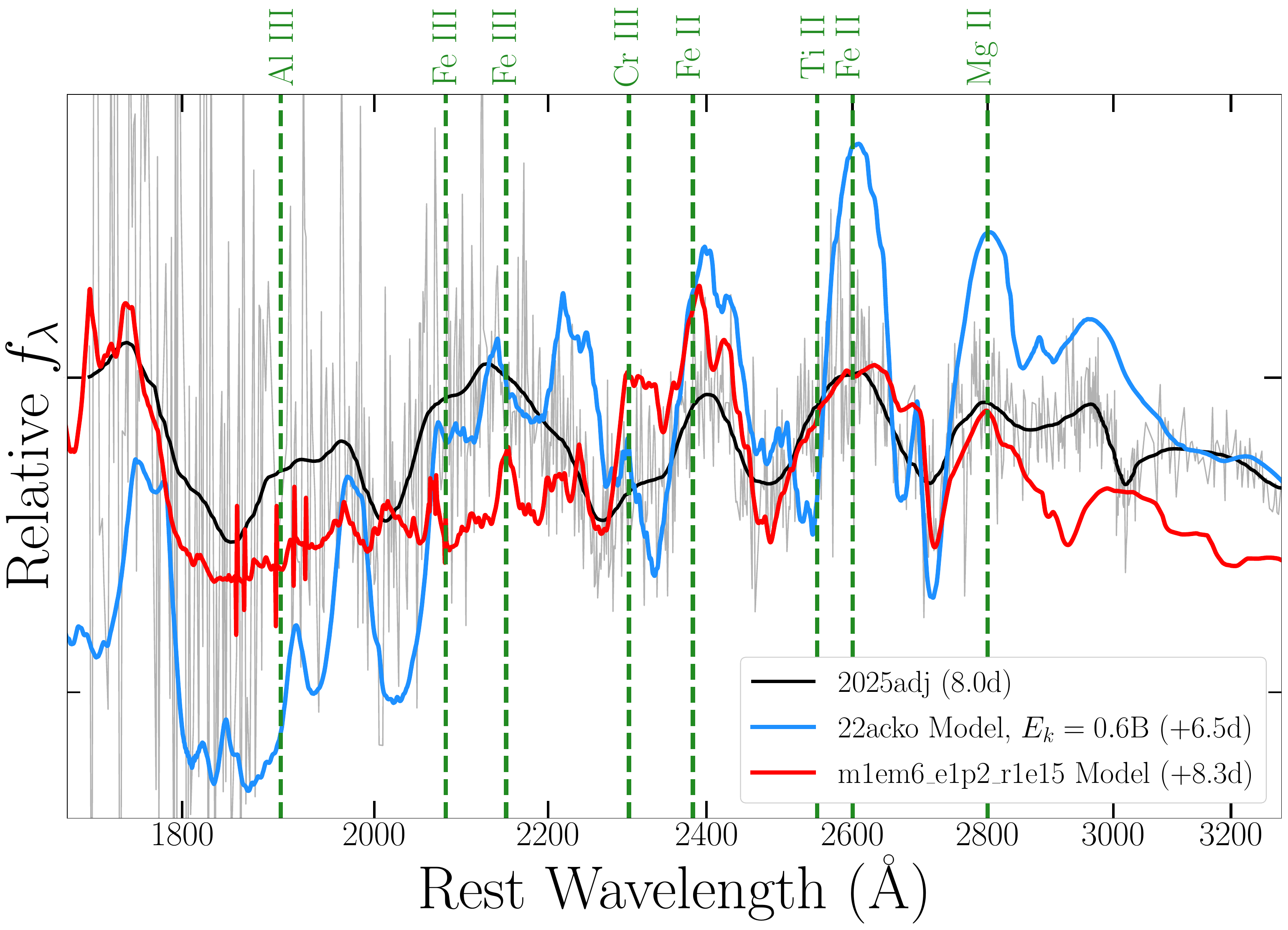}}
\subfigure{\includegraphics[width=0.49\textwidth]{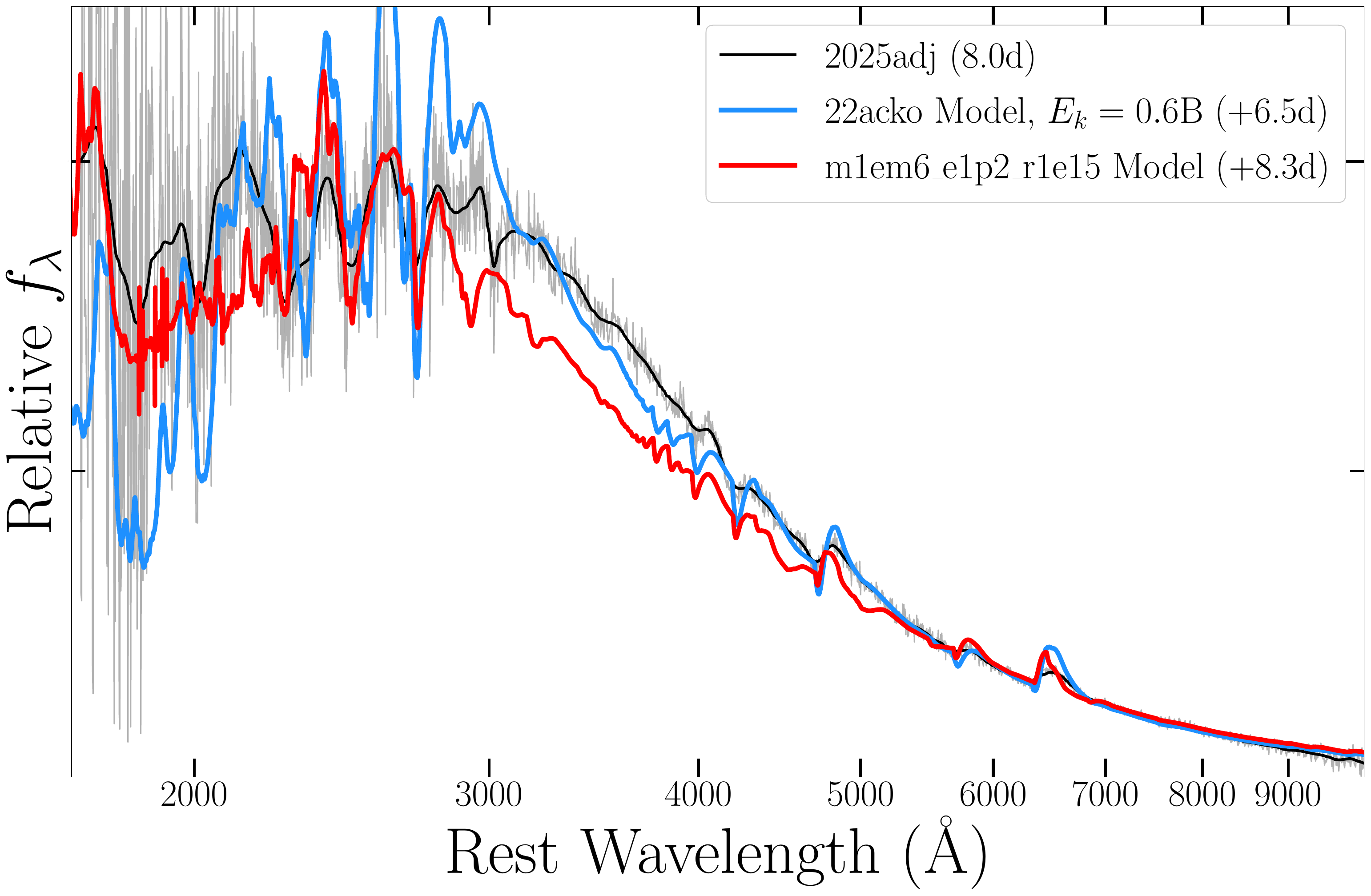}}\\
\caption{ Comparison of sample SNe~II (black) to the best-matched \cmfgen\ model spectra at varying phases as described in \S\ref{sec:SNIImodels}. Model spectra developed for SN~II 2022acko are shown in blue \citep{bostroem22acko}, low mass-loss model ($10^{-6}$~\mdot) shown in red, and higher mass-loss model ($10^{-3}$~\mdot) shown in yellow. Spectra are normalized to the optical continuum in order to determine the best-match to the UV continuum. Discrepant phases between models and data indicate that a larger progenitor model radius may be required to match the observations. \label{fig:snII_cmfgen1} }
\end{figure*}

\begin{figure*}
\centering
\subfigure{\includegraphics[width=0.49\textwidth]{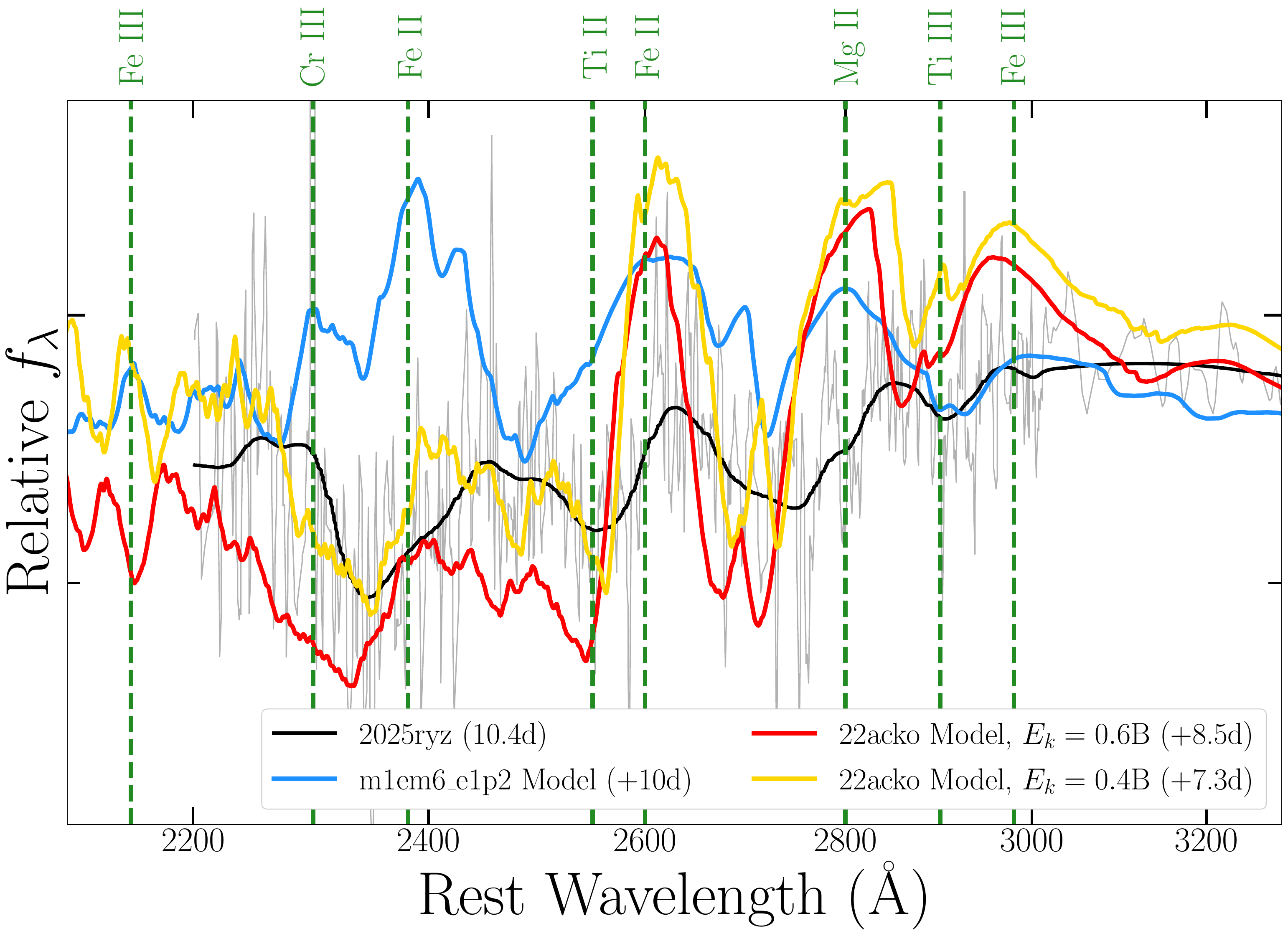}}
\subfigure{\includegraphics[width=0.49\textwidth]{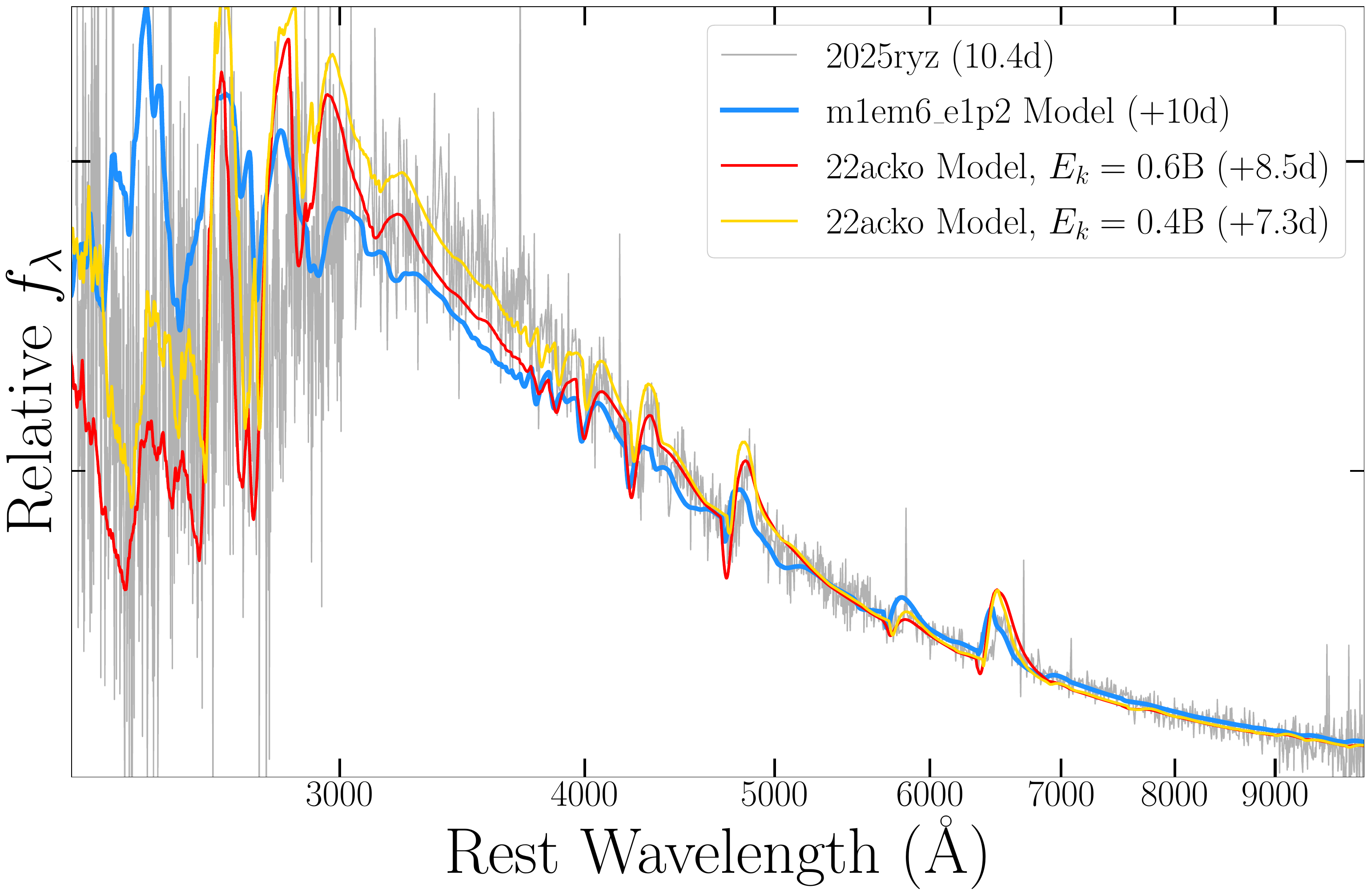}}\\
\subfigure{\includegraphics[width=0.49\textwidth]{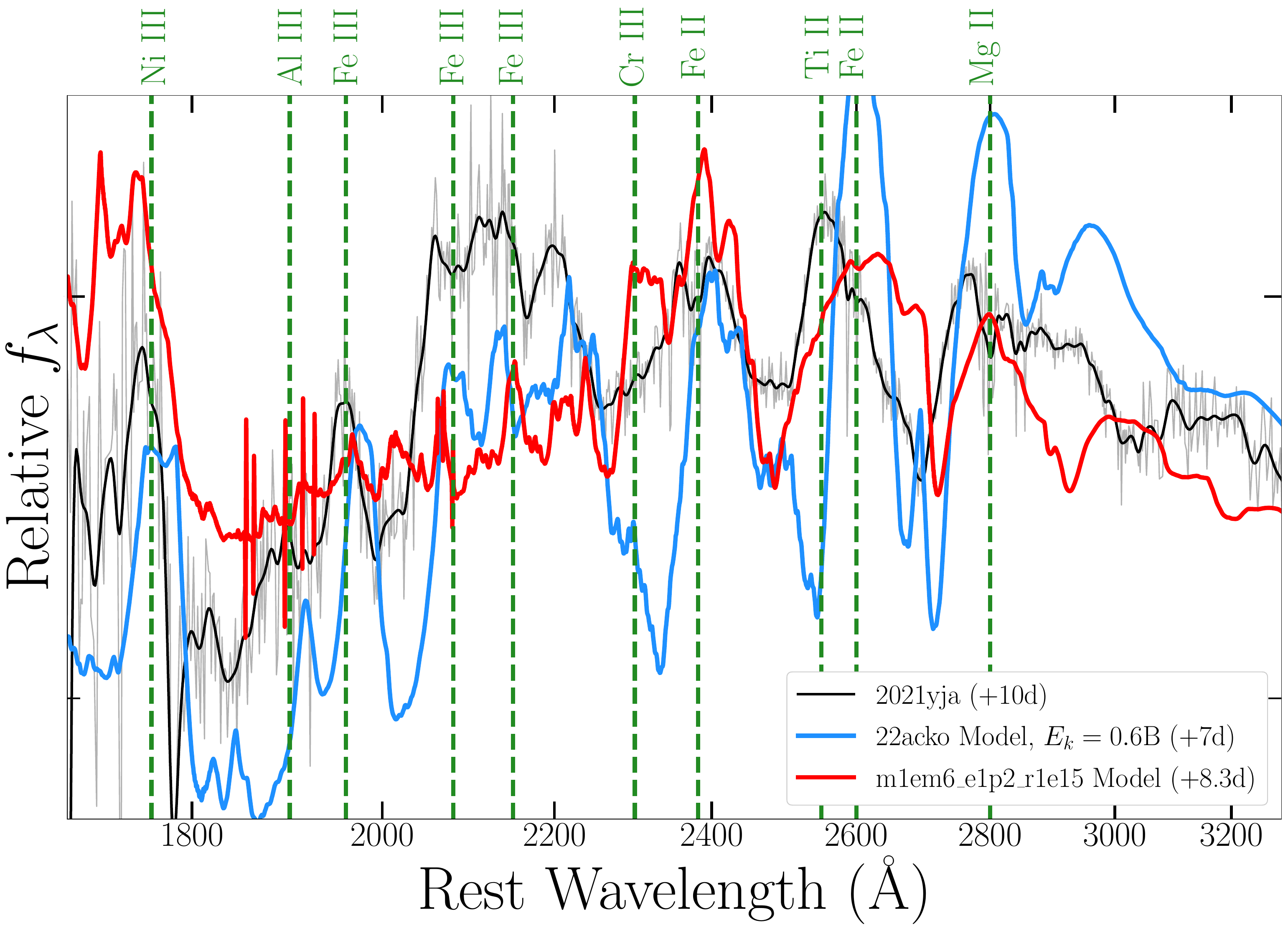}}
\subfigure{\includegraphics[width=0.49\textwidth]{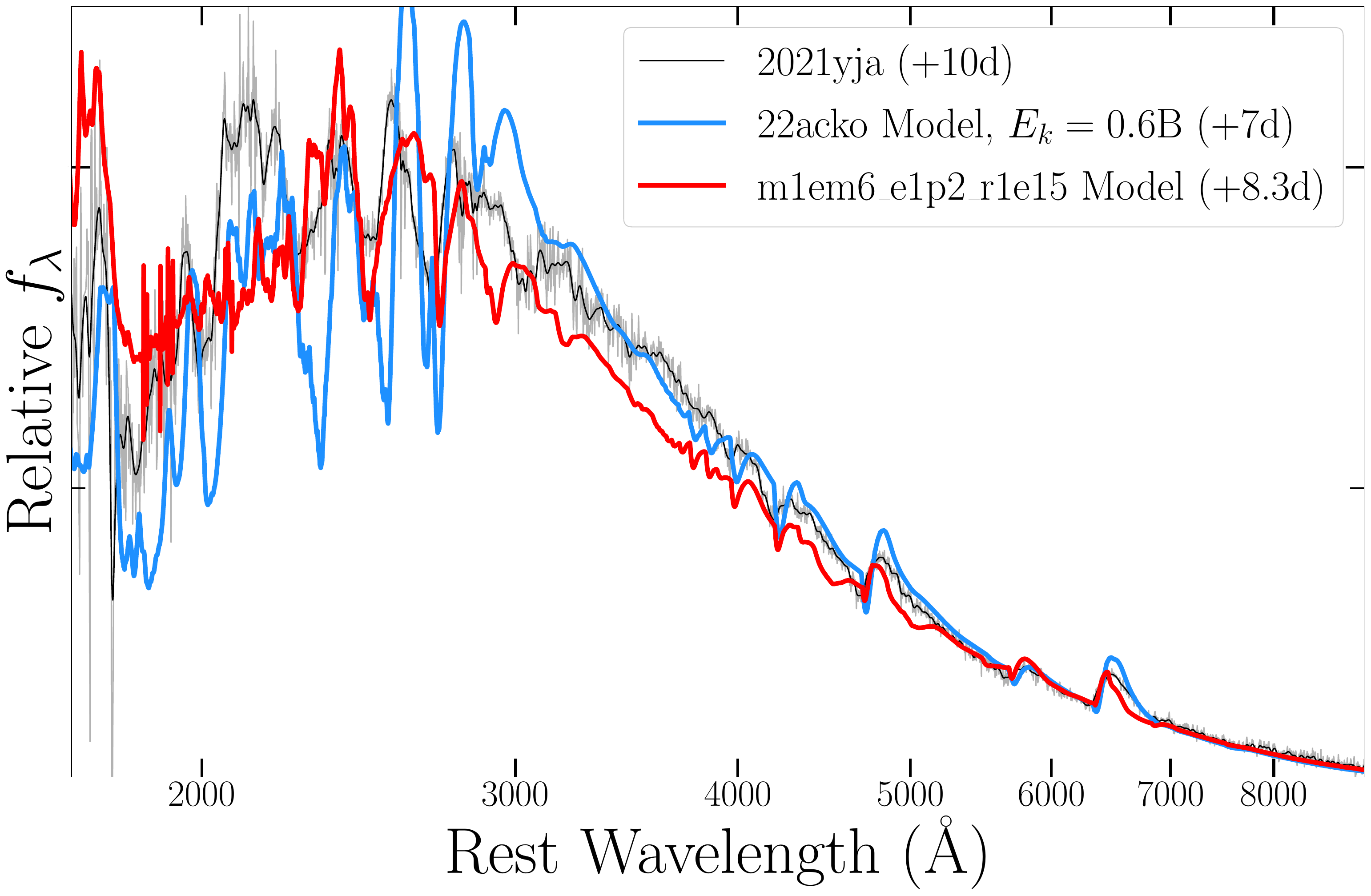}}\\
\subfigure{\includegraphics[width=0.49\textwidth]{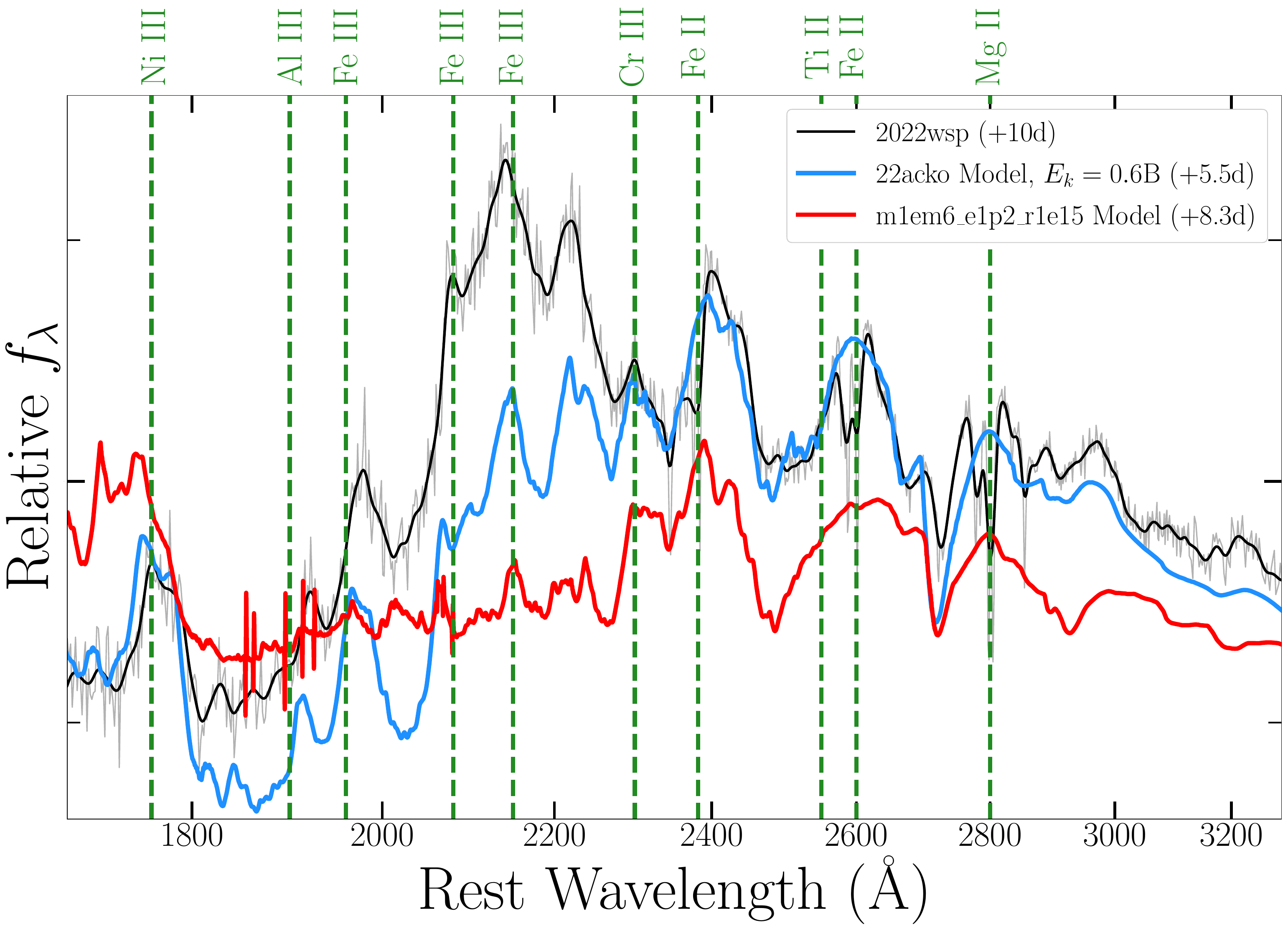}}
\subfigure{\includegraphics[width=0.49\textwidth]{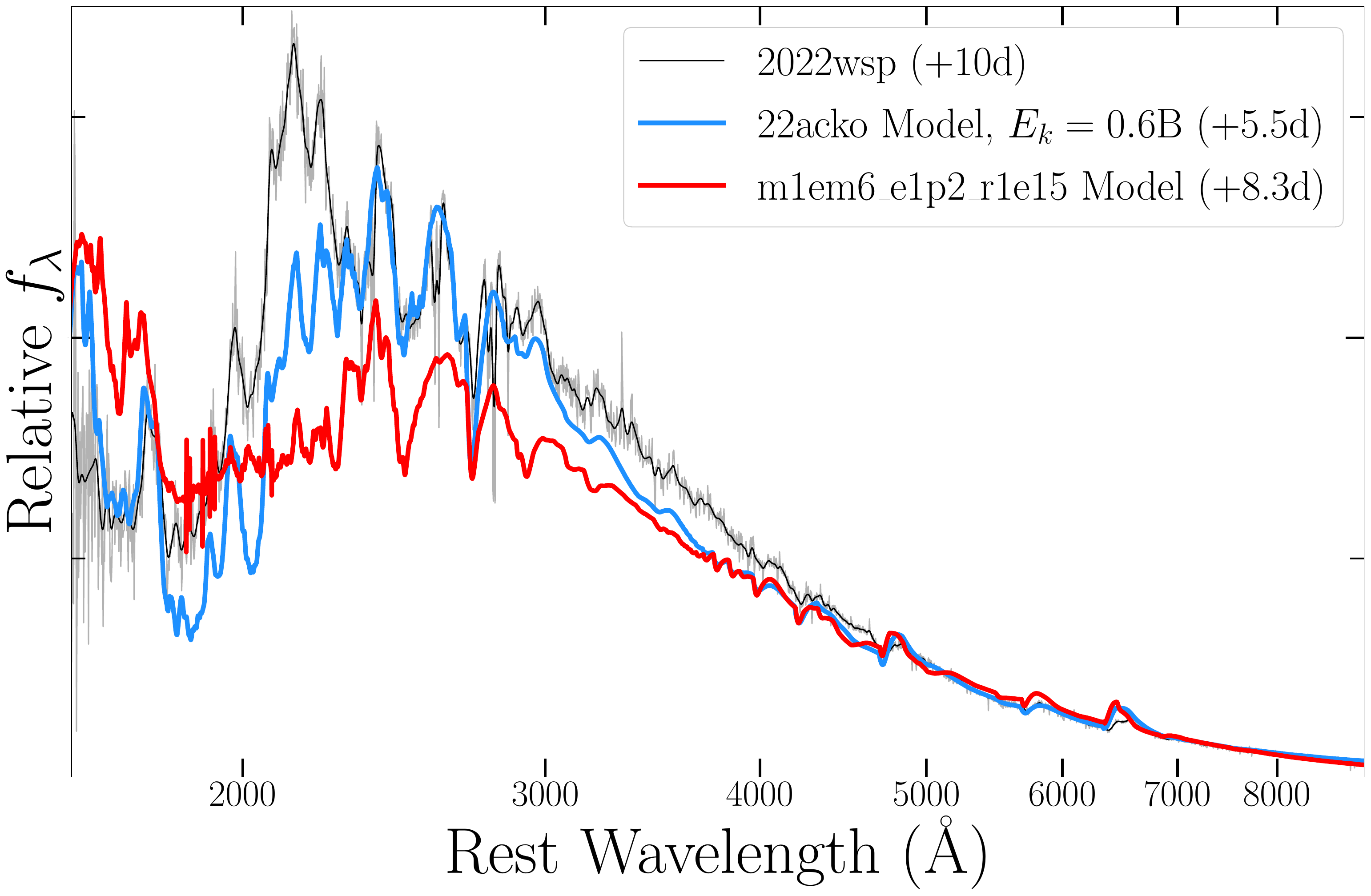}}
\caption{Comparison of sample SNe~II (black) to the best-matched \cmfgen\ model spectra at varying phases as described in \S\ref{sec:SNIImodels}. Model spectra developed for SN~II 2022acko with $E_k = 0.6$~B shown in blue and $E_k = 0.4$~B shown in yellow \citep{bostroem22acko} as well as a low mass-loss model ($10^{-6}$~\mdot) shown in red. Spectra are normalized to the optical continuum in order to determine the best-match to the UV continuum. Discrepant phases between models and data indicate that a larger progenitor model radius is required to match the observations.  \label{fig:snII_cmfgen2} }
\end{figure*}

We present model comparisons in Figures \ref{fig:snII_cmfgen1} and \ref{fig:snII_cmfgen2}, where each \cmfgen\ spectrum shown is first normalized to the optical continuum of each observed SN spectrum in order to best understand the model match to the UV continuum. This also allows us to confirm that both the optical Balmer series and UV metal profiles are well-represented by a given model spectrum. For Type II SNe 2021yja, 2022wsp, 2024ryv, 2025adj, and 2025ryz, we find that the model spectra with negligible CSM density (i.e., $\dot M = 10^{-6}$~\mdot, $v_w = 50~\kms$) can match both the optical \ion{H}{i} lines and a number of features in the near-UV that arise from blends of Fe, Cr, Ti, Al and Mg. This model grid was not tailored to specific SNe so there are some inconsistencies related to line strength and ionization states that can be addressed in future modeling studies (e.g., metallicity variations). Overall, the UV/optical line velocities of most sample SNe~II are reproduced with kinetic energies of $E_k = 0.6 - 1.2$~B, but SN~2021yja displays significantly higher velocities derived from a likely more energetic explosion (i.e., $E_k >1.2$~B). Only SN~2025ryz requires the lowest kinetic energy model with $E_k = 0.4$~B in order to match the lower UV and optical line velocities, which is similar to the low luminosity SN~II 2022acko \citep{bostroem22acko}. 

An intriguing property of this model examination is the inability of some model spectra to match the UV continuum flux when compared at the same phase as the UV/optical observations. For example, the UV flux in SN~2025adj at $\delta t = 8.0$~days cannot be reproduced (i.e., it is too faint) by the SN~2022acko model at $\delta t = 8.5$~days but it can be matched by the same model at $\delta t = 6.5$~days or the {\tt m1em6\_e1p2\_r1e15} model at $\delta t = 8.3$~days. This likely suggests that the initial progenitor radius of $500~\Rsun$ in the SN~2022acko models is too small for SN~2025adj, which requires the larger progenitor radius of $652~\Rsun$ in the {\tt m1em6\_e1p2\_r1e15} ($\dot M = 10^{-6}$~\mdot, $E_k = 1.2$~B, $R_{\rm CSM} = 10^{15}$~cm) model or the SN~2022acko 0.6~B model at an earlier phase where the increased temperature can mimic a larger radius model. A similar discrepancy can be seen in the model comparisons with SNe~2021yja and 2022wsp, which both require models at $\delta t \lesssim 7$~days to match the UV continuum despite being observed at $\delta t = 10$~days, suggesting large progenitor radii of $>650~\Rsun$. 

\begin{figure*}
\centering
\subfigure{\includegraphics[width=\textwidth]{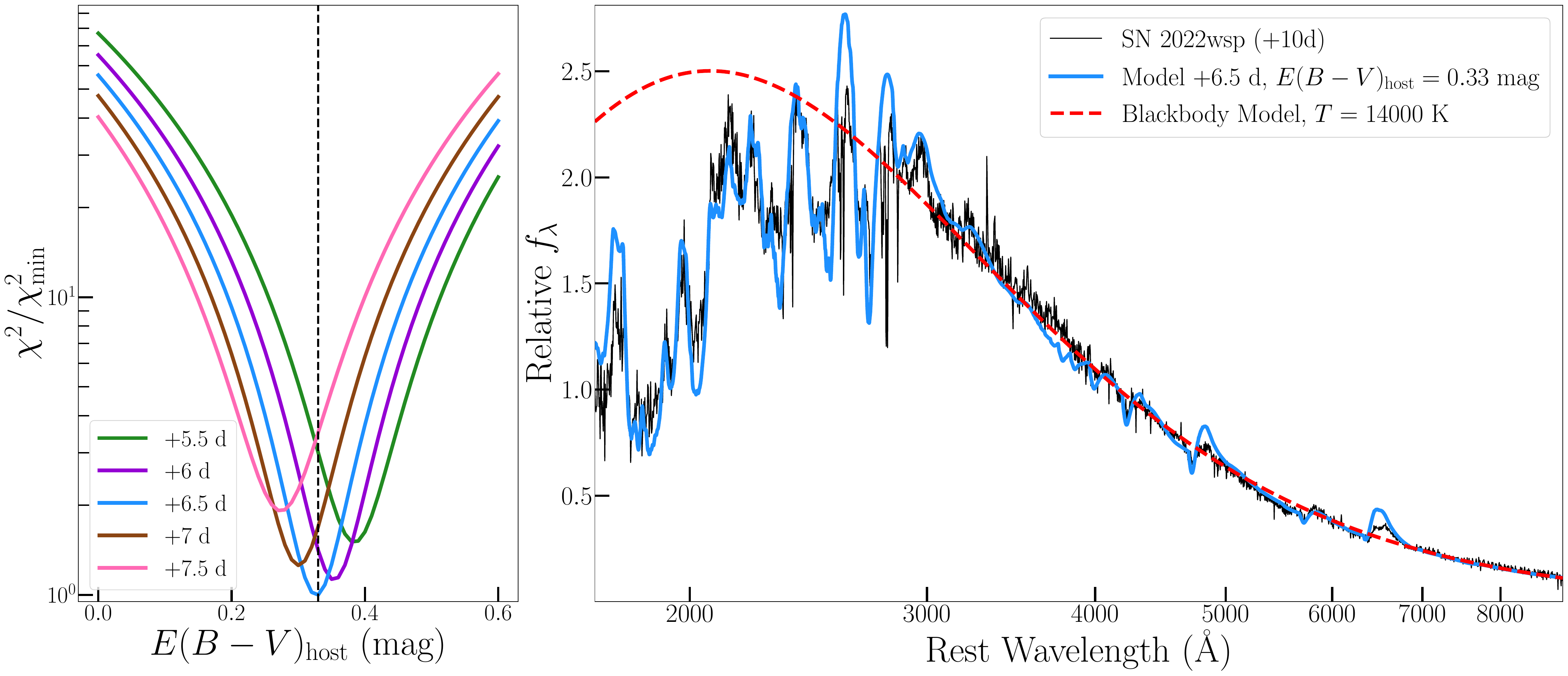}}\\
\vspace{-4mm}
\subfigure{\includegraphics[width=\textwidth]{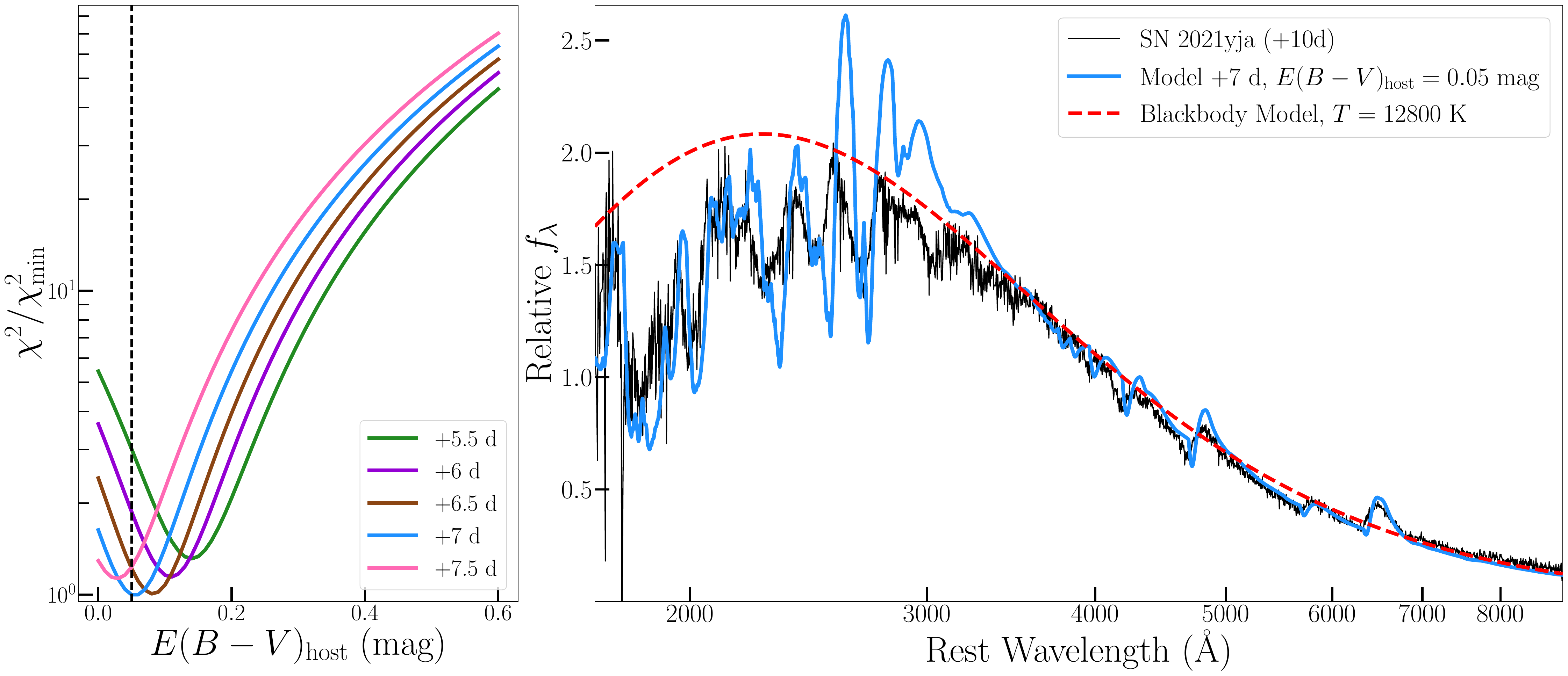}}\\
\caption{Simultaneous fits of model phase and host-galaxy reddening for SNe~II 2022wsp (top) and 2021yja (bottom), both observed at $\delta t = 10$~days, using the $E_k = 0.6$~B \cmfgen\ model time series of SN~2022acko \citep{bostroem22acko}. {\it Left:} $\chi^2$ normalized to the global minimum of all model phases and $E(B-V)_{\rm host}$ values explored in the fit as a function of $E(B-V)_{\rm host}$ for the five model phases with the lowest $\chi^2$; the dashed black line marks the best-fit $E(B-V)_{\rm host}$. Earlier, hotter model phases require larger host reddening, reflecting the degeneracy between model phase and $E(B-V)_{\rm host}$. {\it Right:} Observed UV/optical spectrum (black), corrected for MW extinction and the best-fit $E(B-V)_{\rm host}$, compared to the best-matched \cmfgen\ model spectrum (blue) and blackbody model (red) scaled by a single multiplicative factor. The best matches are the +6.5~day \cmfgen\ model with $E(B-V)_{\rm host} = 0.33$~mag for SN~2022wsp and the +7~day model with $E(B-V)_{\rm host} = 0.05$~mag for SN~2021yja, both earlier than the observed phase. \label{fig:cmfgen_ebv} }
\end{figure*}

Because earlier-phase \cmfgen\ spectra are hotter and intrinsically bluer, the preference for $\delta t \lesssim 7$~day models in SNe~2021yja and 2022wsp could instead result from an overestimate of the host-galaxy extinction, which would make the de-reddened spectra artificially blue. To test this possibility, we fit the UV/optical spectrum of each SN for both model phase and $E(B-V)_{\rm host}$, using all 13 epochs ($\delta t = 5.5 - 21.4$~days) of the SN~2022acko $E_k = 0.6$~B model series and a grid of $E(B-V)_{\rm host} = 0 - 0.6$~mag in steps of 0.01~mag. At each grid point, the observed spectrum is corrected for the fixed MW color excess and the trial $E(B-V)_{\rm host}$ using a \cite{fitzpatrick99} reddening law ($R_V = 3.1$). Both the data and models are then binned onto a common logarithmic wavelength grid with $1000~\kms$ bins spanning $1700 - 8900$~\AA. We compare the binned spectra in logarithmic flux and allow a single free multiplicative scale factor per model, solved analytically, so that the scaling cannot alter the spectral slope. As shown in Figure \ref{fig:cmfgen_ebv}, SN~2022wsp is best matched by the +6.5~day model with $E(B-V)_{\rm host} = 0.33\pm 0.01$~mag, consistent to within $\sim$1$\sigma$ of the \ion{Na}{i}~D estimate of $0.43\pm0.09$~mag, while SN~2021yja is best matched by the +7~day model with $E(B-V)_{\rm host} = 0.05^{+0.04}_{-0.01}$~mag, compared to the \ion{Na}{i}~D value of $0.085\pm0.015$~mag. In both cases the five best-matching model phases fall between $\delta t = 5.5$ and 7.5~days, and every model at $\delta t \geq 8$~days provides a poorer match at any $E(B-V)_{\rm host}$ in the grid. Earlier model phases pair with progressively larger host reddening, but lowering $E(B-V)_{\rm host}$ below the \ion{Na}{i}~D values does not allow a model at the observed phase to reproduce the UV continuum. We therefore conclude that the need for earlier, hotter model spectra is not an artifact of the adopted host extinction, which further supports progenitor radii larger than the $500~\Rsun$ assumed in the SN~2022acko models and/or additional power from confined CSM interaction.

Besides SN~2023ixf, which will be discussed in \S\ref{sec:specIIn}, no other sample SN~II shows spectroscopic evidence for ejecta-CSM interaction through the detection of narrow and IIn-like emission lines. However, as shown in Figure \ref{fig:snII_cmfgen1}, SN~II 2024faq may be the one object in our sample whose UV/optical properties are more consistent with confined CSM. Firstly, despite the UV/optical spectrum being obtained at $\delta t = 7.7$~days, SN~2024faq displays extremely weak Balmer features that are inconsistent with the strong P-Cygni profiles predicted in the SN~2022acko and {\tt m1em6\_e1p2\_r1e15} models, both of which have weak mass-loss rates of $\dot M = 10^{-6}$~\mdot. Secondly, the SN~2022acko model at a similar phase can neither match the UV continuum nor reproduce the significantly weaker absorption profiles, although the former could reflect the need for a larger progenitor radius. Furthermore, the {\tt m1em6\_e1p2\_r1e15} model can match the UV continuum of SN~2024faq but it cannot reproduce the absorption/emission profiles of \ion{Fe}{ii/iii} between $\lambda = 2000 - 2400$~\AA. Overall, these UV/optical spectral signatures can be the result of an altered outer ejecta density profile due to interaction with confined CSM \citep{dessart22}. In Figure \ref{fig:snII_cmfgen1}, we compare SN~2024faq to a model spectrum with $\dot M = 10^{-3}$~\mdot\ and $R_{\rm CSM} = 2\times 10^{14}$~cm, and find that it can reproduce the weak Balmer emission and muted UV profiles, suggesting that this SN~II may have interacted with very confined CSM of moderate density in the first $\sim$days since first light. This hypothesis is further supported by the overly luminous UV/optical light curve (e.g., see Fig. \ref{fig:LCs}) relative to other SNe~II. 

\subsection{Type IIn Supernovae}\label{sec:specIIn}

\subsubsection{Comparisons and Line Profiles}\label{sec:SNIIncomp}

\begin{figure*}
\centering
\subfigure{\includegraphics[width=0.49\textwidth]{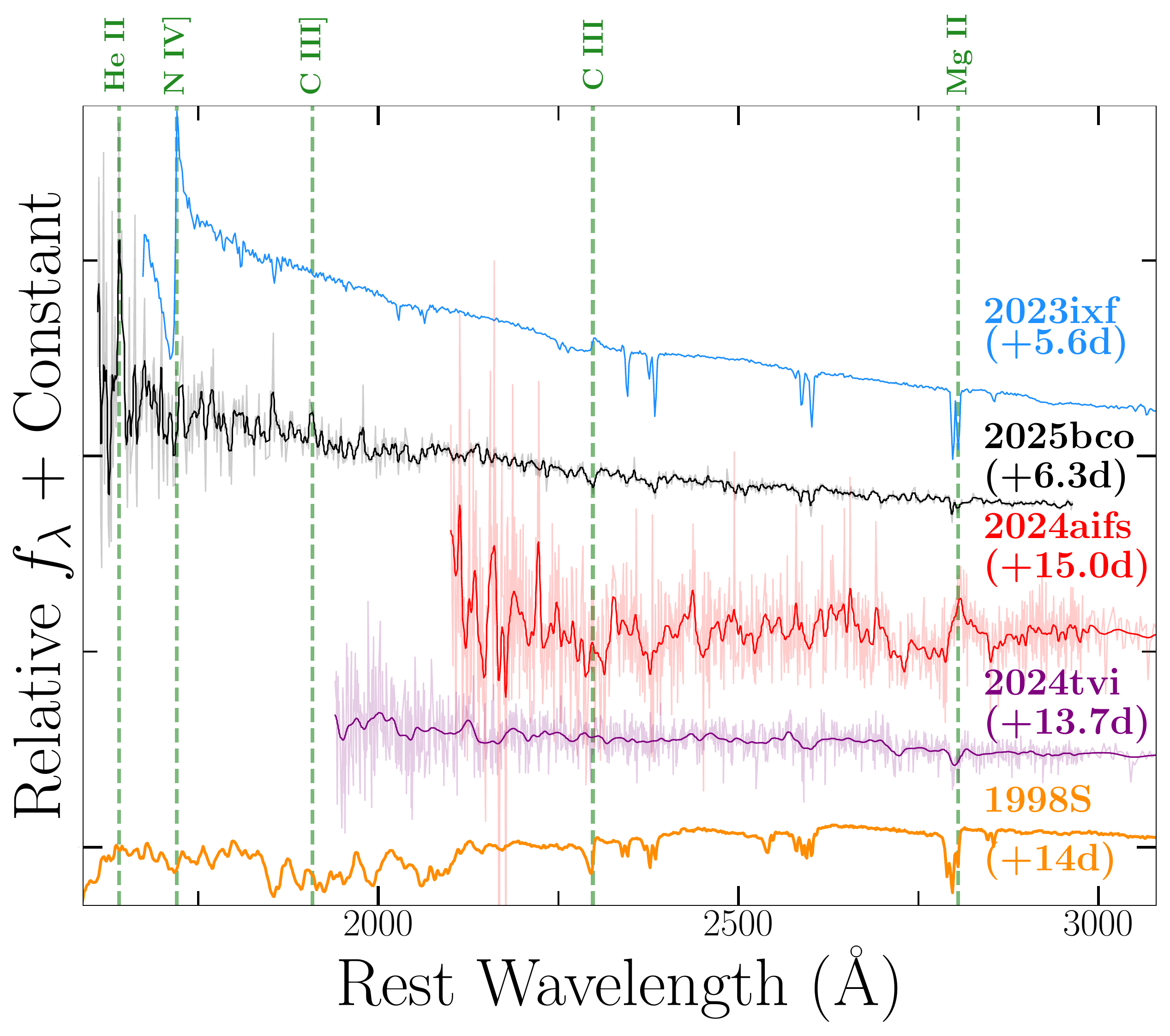}}
\subfigure{\includegraphics[width=0.49\textwidth]{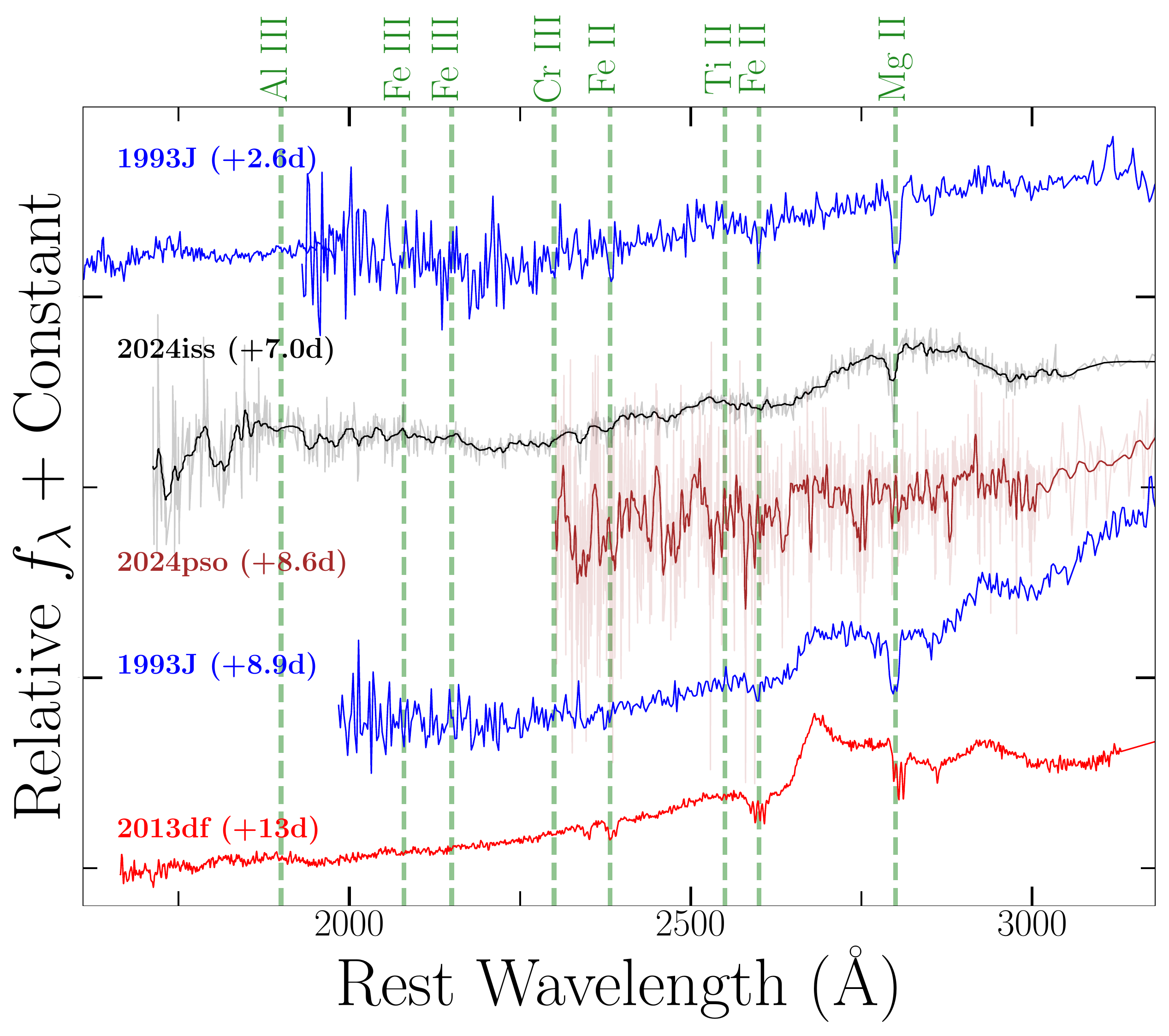}}\
\caption{{\it Left:} Early-time UV spectra of SNe IIn 2025bco (black), 2024aifs (red) and 2024tvi (green) compared to SN~II 2023ixf when narrow lines from CSM interaction are still detectable. UV spectrum of CSM-interacting SN~II 1998S at $\delta t \approx 14$~days (orange) also shown for comparison. {\it Right:} Present sample of early-time UV spectra of SNe~IIb. SNe~2024iss and 2024pso are currently the youngest SNe~IIb observed in the UV with HST. \label{fig:IInIIbcomp} }
\end{figure*}

In Figure \ref{fig:IInIIbcomp}, we compare the early-time UV spectra of sample SNe~IIn 2025bco, 2024aifs, and 2024tvi at $\delta t = 6.3 - 15.0$~days. In this comparison, we include an early-time HST UV spectrum of SN~II 2023ixf \citep{Zimmerman24} during its interaction with dense, confined CSM ($\delta t < 7$~days) and the earliest HST UV spectrum of the CSM-interacting SN~II 1998S at $\delta t \approx 14$~days \citep{fransson05}. Across all these CSM-interacting SNe, there is considerable diversity in their UV spectroscopic properties at early-time phases. For example, SN~2023ixf showed no ``IIn-like'' features in the UV despite being observed  with HST by $\delta t = 3.6$~days and despite the strong Lorentzian profiles present in the optical/NIR. A similar property is evident in the sample SNe~IIn, which show strong H/He Lorentzian emission lines in the optical/NIR for $\sim$weeks after explosion but are relatively featureless in their UV spectra. 

\begin{figure*}
\centering
\subfigure{\includegraphics[width=0.68\textwidth]{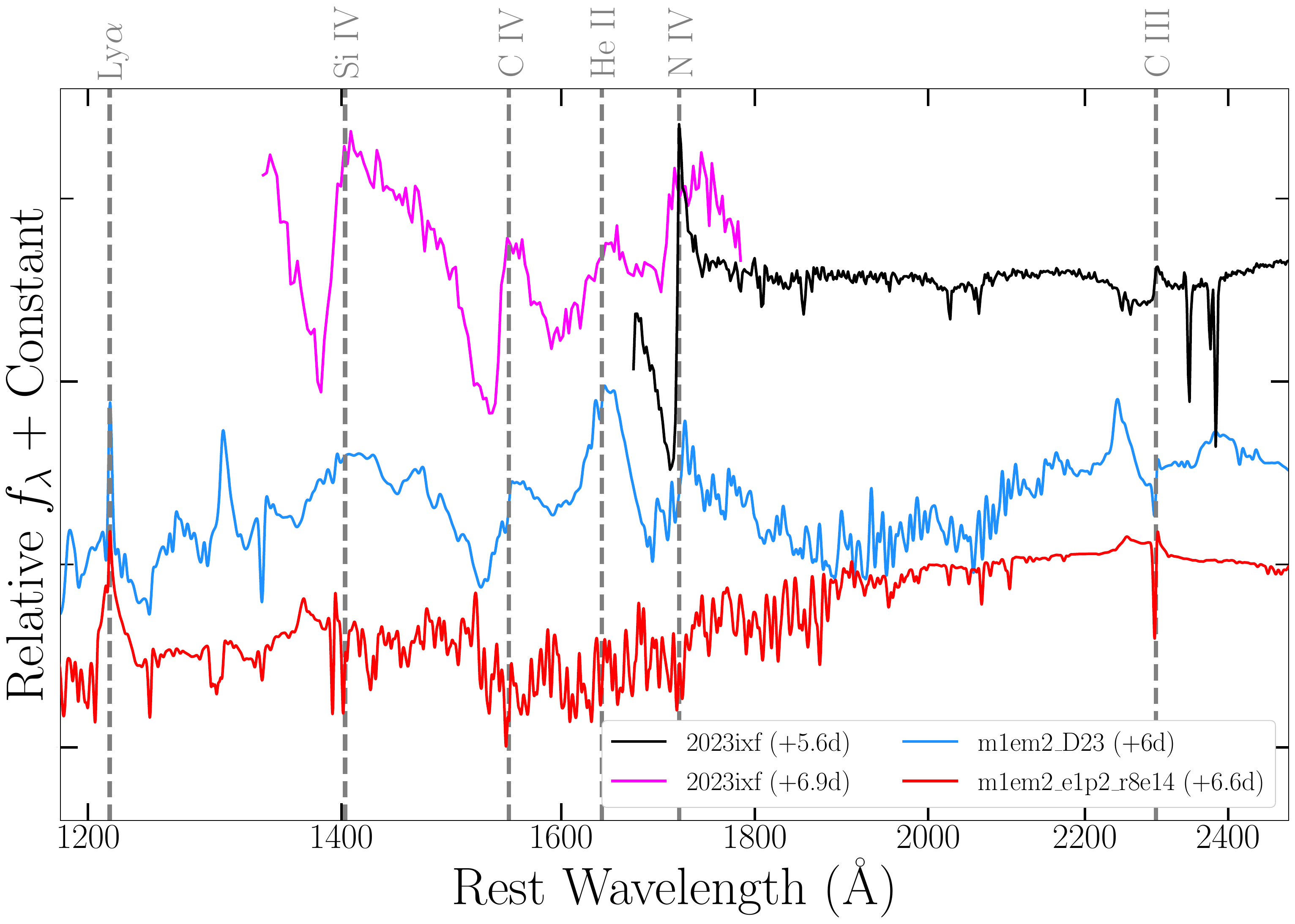}}
\subfigure{\includegraphics[width=0.31\textwidth]{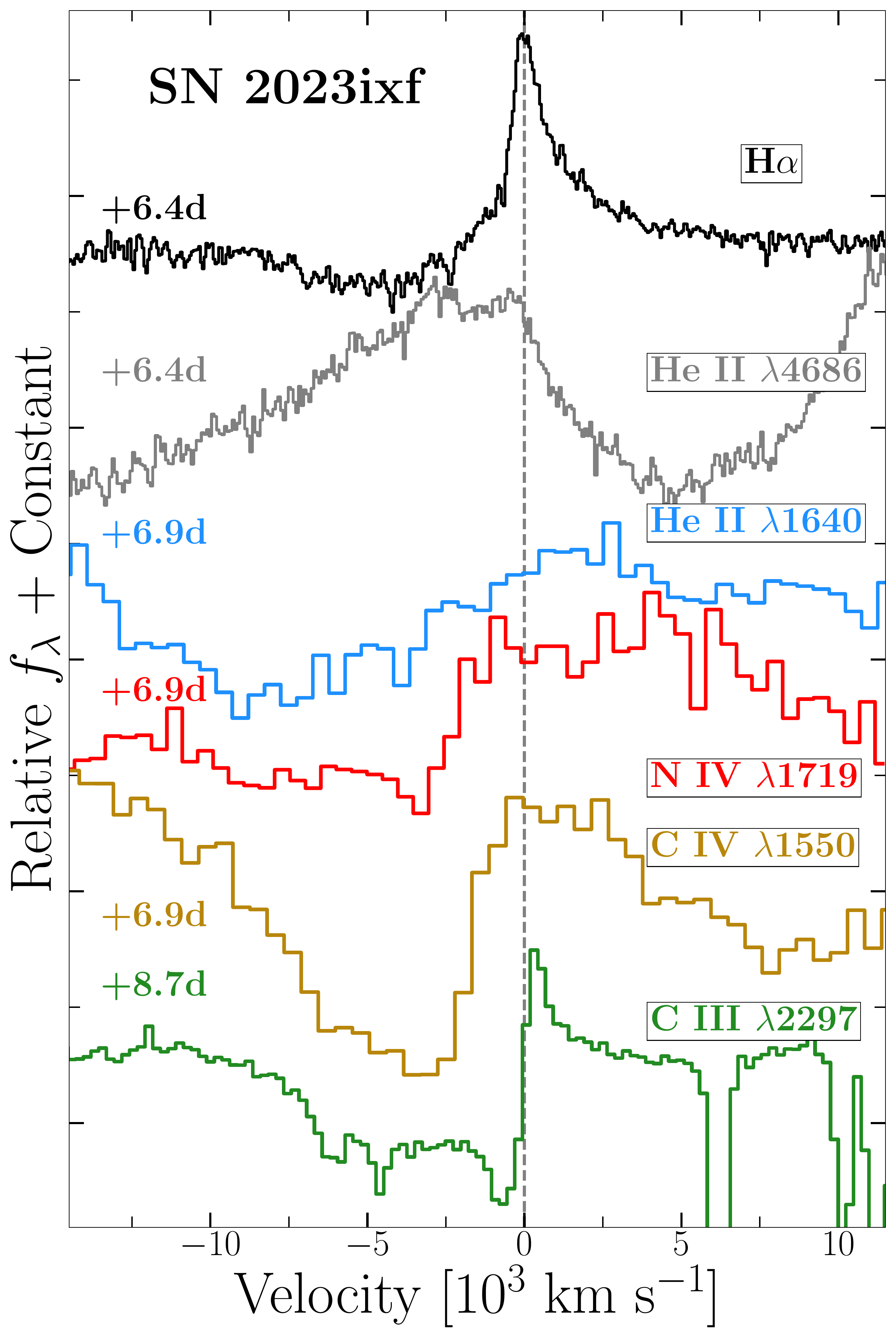}}
\caption{{\it Left:} Far-UV spectra at $\delta t = 6.9$~days (magenta) and near-UV spectra at $\delta t = 5.6$~days of SN~2023ixf \citep{Teja23, Zimmerman24} compared to $\dot M = 10^{-2}$~\mdot\ \cmfgen\ model spectra (shown in blue and red) from \cite{Dessart23} and \cite{dessart25CSMgrid}. {\it Right:} Optical and UV line velocities of SN~2023ixf at $\delta t \approx 6-9$~days. Narrow emission from pre-shock CSM is still observed in H$\alpha$ while many UV lines show the emergence of Doppler-broadened features from the fast-moving SN ejecta. \label{fig:23ixf_FUV} }
\end{figure*}

Notably, SN~2023ixf displays a prominent \ion{N}{iv}$\lambda1719$ P-Cygni profile that is visible in HST near-UV spectra obtained at $\delta t = 3.6 - 5.6$~days \citep{Zimmerman24} as well as in the Astrosat/UVIT far-UV spectrum at $\delta t = 6.9$~days \citep{Teja23} --- this feature is not visible in the next UV spectral epoch at $\delta t = 8.7$~days. In the HST spectra at $\delta t < 5.6$~days, the absorption minimum of this feature has a velocity of $\sim1000~\kms$, while the blue edge of the profile indicates material at $>5000~\kms$. As shown in Figure \ref{fig:23ixf_FUV}, the HST UV spectra of SN~2023ixf also show a \ion{C}{iii} $\lambda2297$ feature that is a combination of narrow emission at $\sim0~\kms$ with a broad absorption ``trough'' that extends to $\sim -10000~\kms$. At $\delta t = 6.9$~days, the far-UV spectrum shows broad P-Cygni profiles of \ion{He}{ii} $\lambda 1640$, \ion{N}{iv} $\lambda 1719$, \ion{C}{iv} $\lambda 1550$, and \ion{Si}{iv} $\lambda 1403$ with higher absorption velocities that also extend to $\sim -10000~\kms$. Contrary to the UV spectra, the optical spectra of SN~2023ixf show prominent Lorentzian emission profiles characteristic of electron scattering in optically thick CSM, which persist for $\sim6$~days at which point the absorption ``trough'' of the high velocity material in the cold dense shell (CDS) begins to be visible. As shown in Figure \ref{fig:23ixf_FUV}, the blueward absorption velocities of both optical and UV lines are consistent at $\delta t > 6$~days and likely both arise from fast-moving material in the SN ejecta and CDS. However, there is still pre-shocked gas ahead of the forward shock given the narrow components still visible in the Balmer series and \ion{C}{iii} $\lambda2297$ profiles at $\delta t > 7$~days. Overall, the discrepancies between early-time UV and optical profiles in SN~2023ixf can be attributed to the properties of recombination versus resonance transitions as well as CSM asymmetries, which we discuss in more detail in \S\ref{sec:CSMasymmetry}.

\begin{figure*}
\centering
\subfigure{\includegraphics[width=0.71\textwidth]{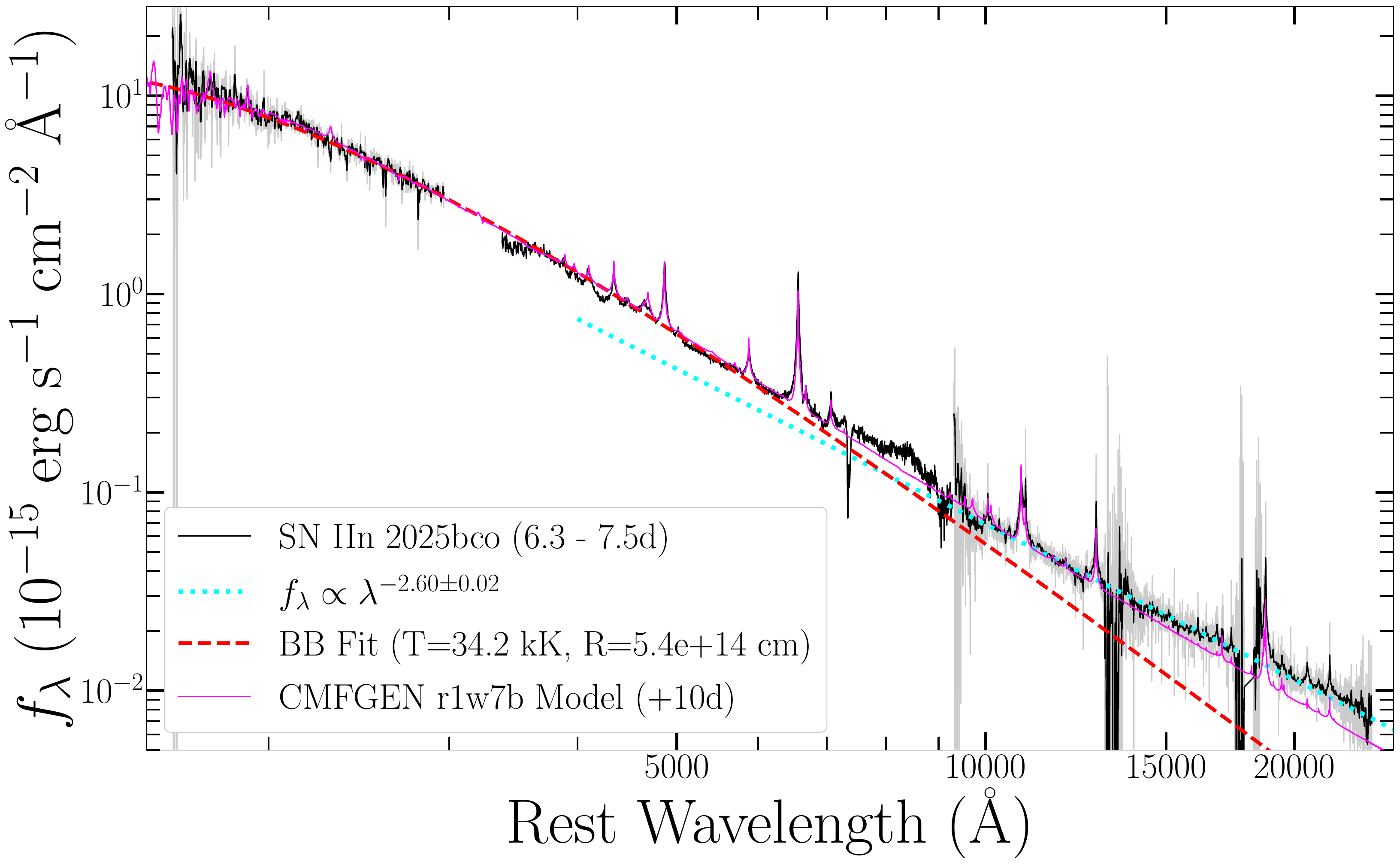}}
\subfigure{\includegraphics[width=0.28\textwidth]{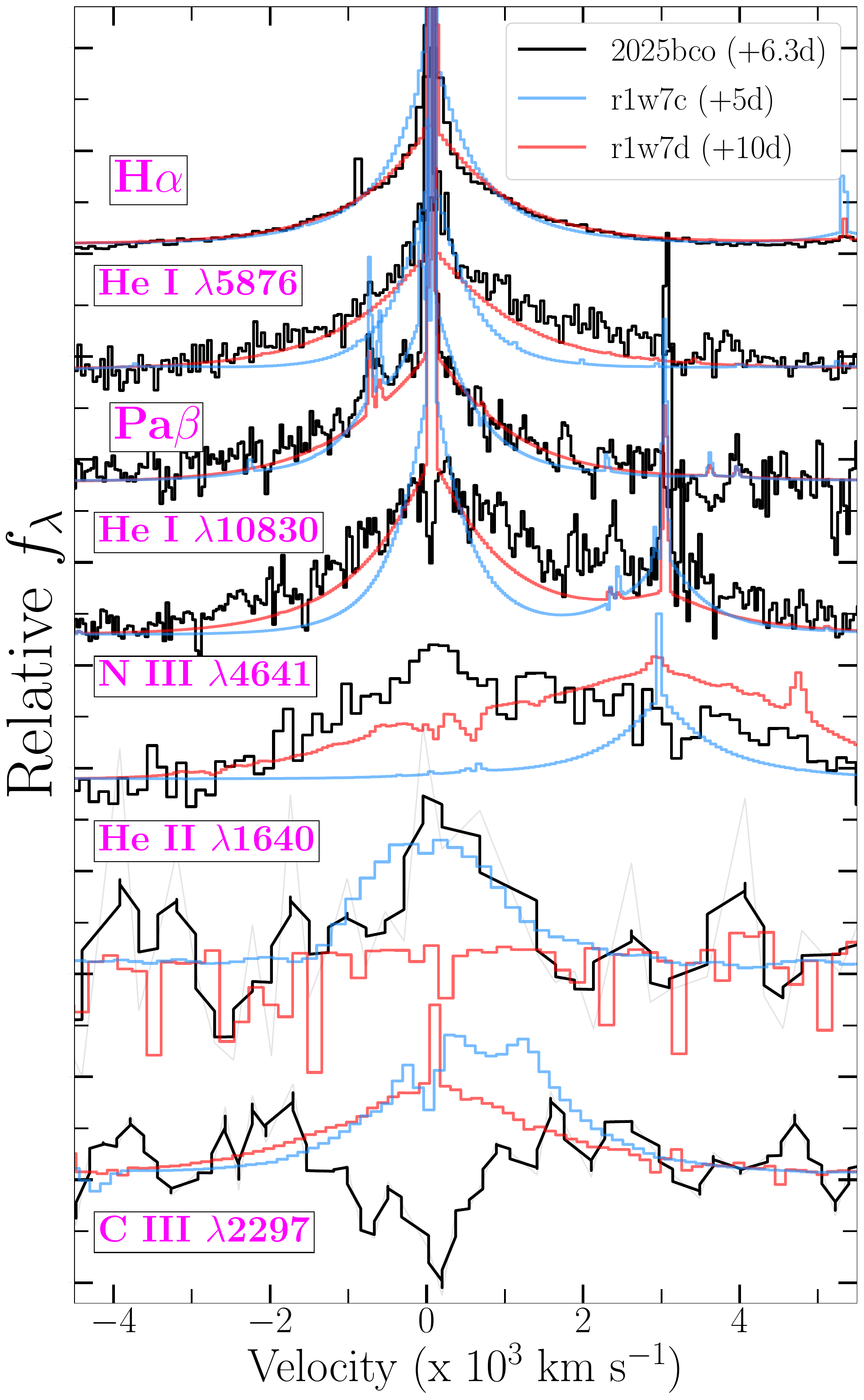}}
\caption{ Combined UV-to-NIR spectra of SN~IIn 2025bco at $\delta t = 6.3 - 7.5$~days (black) with the best-fitting blackbody model and IR excess power-law overlaid as dashed red and dotted cyan lines. {\tt CMFGEN} r1w7b model spectrum shown in magenta \citep{wjg22}, which includes mass-loss rate of $\dot M = 3\times 10^{-2}$~\mdot and CSM radius of $2\times10^{15}$~cm. {\it Right:} Early-time UV, optical and NIR line velocities of SN~2025bco (black) compared to r1w7c (blue; $R_{\rm CSM} = 4\times10^{15}$~cm) and r1w7d (red; $R_{\rm CSM} = 8\times10^{15}$~cm) models. The discrepancies between models and data at UV wavelengths despite consistent match in the optical/NIR may suggest asymmetries in the CSM interaction present in SN~2025bco. \label{fig:SED25bco} }
\end{figure*}

Similar to SN~2023ixf, the more prototypical SNe~IIn in our sample also show relatively featureless UV spectra at early times, despite persistent Lorentzian profiles being present at optical wavelengths. In the UV spectrum of SN~2025bco, we identify \ion{He}{ii} $\lambda1640$ emission as well as possible absorption from \ion{C}{iii} $\lambda2297$, the latter also being observed in SN~1998S at $\delta t = 14$~days \citep{fransson05}. In Figure~\ref{fig:SED25bco}, the UV-to-NIR spectrum of SN~2025bco at $\delta t \approx 1$~week displays prominent Balmer and Paschen series Lorentzian emission profiles as well as \ion{He}{i} lines and emission from \ion{He}{ii}~$\lambda4686$ and \ion{N}{iii}~$\lambda4641$. Notably, a blackbody model with temperature of $\sim$34~kK fails to match the SN continuum at $>7000$~\AA, which is likely the result of thermal free-free emission from the ionized CSM outside the electron-scattering photosphere -- at these wavelengths the free-free opacity rises steeply with wavelength ($\kappa_\nu^{\rm ff}\propto\nu^{-3} [1-e^{-h\nu/kT}]$; \citealt{Rybicki79}). As shown in Figure~\ref{fig:SED25bco}, the NIR excess can be fit with a power law of $f_\lambda\propto\lambda^{\beta}$, where $\beta = -2.6 \pm 0.02$. In the Rayleigh--Jeans and radio limit, the emergent spectrum goes as $F_\nu\propto\nu^{\alpha}$ with $\alpha = (4n-6.2)/(2n-1)$; $\alpha=0.6$ for a steady wind ($n=2$; \citealt{Wright75, Panagia75, Olnon75}). We find $\alpha = -(2+\beta) = 0.60\pm0.02$ and a CSM density power law index of $n = 2.0 \pm 0.02$, which is consistent with a steady-state wind density profile ($n = 2$) within $\sim$1--2\,$R_{\rm phot}$. This NIR excess from an electron-scattering atmosphere effect has been observed in other SNe~IIn (e.g., 2009ip \citealt{margutti14}) as well as fast blue optical transients (e.g., see \citealt{LeBaron26}) and tidal disruption events \citep{Roth16}.  

In the other two sample SNe~IIn, 2024aifs and 2024tvi, we do not identify any prominent emission lines associated with the ongoing CSM interaction. As shown in Figure \ref{fig:mgIIvels}, SN~2024aifs displays a \ion{Mg}{ii} profile with narrow emission at $\sim 0~\kms$ combined with a broad absorption component that extends to $\sim12000~\kms$. Comparison of this profile to H$\alpha$ at the same phase of $\delta t = 15.0$~days and at $\delta t = 32.6$~days reveals an absorption profile that extends to similar velocities despite stronger emission in H$\alpha$ than in \ion{Mg}{ii}. This  absorption ``trough'' can be attributed to the emerging CDS as the optical depth in the pre-shock CSM gas drops over time. Notably, the UV spectrum of SN~2024tvi at a similar phase does not show such a \ion{Mg}{ii} profile, but this is likely consistent with the fact that the optical emission profiles are still symmetric Lorentzians suggestive of ejecta interaction with CSM that is still completely optically thick to electron scattering.

\subsubsection{Type IIn SN Model Matching}\label{sec:SNIInmodels}

Similarly to sample SNe~II, we compare grids of \cmfgen\ model spectra for CSM-interacting SNe~II to both the UV and optical spectra of SN~2023ixf and the three SNe~IIn. We employ the same cubic grid discussed above from \cite{dessart25CSMgrid} as well as the \cmfgen\ model grids from \cite{dessart17}, \cite{wjg22}, and \cite{Dessart23} -- see Table A.2 of \cite{wjg24} for a complete summary of CSM parameters in each model set. For each of these four CSM-interacting SNe in our sample, we also attempt to find a best-matched model to the optical spectral evolution, i.e., model spectra that can effectively reproduce the IIn-like feature duration observed in the SN spectral sequence from days to weeks since first light. 

\begin{figure*}
\centering
\subfigure{\includegraphics[width=\textwidth]{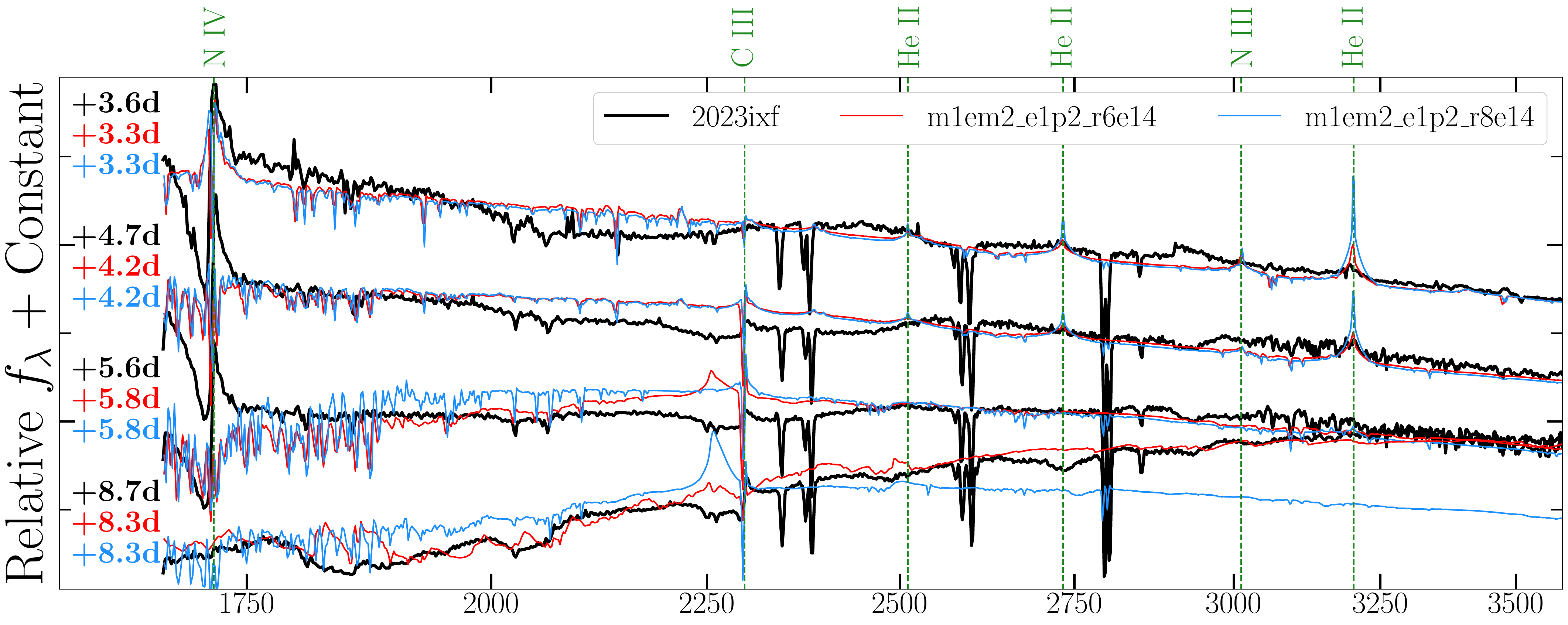}}
\subfigure{\includegraphics[width=\textwidth]{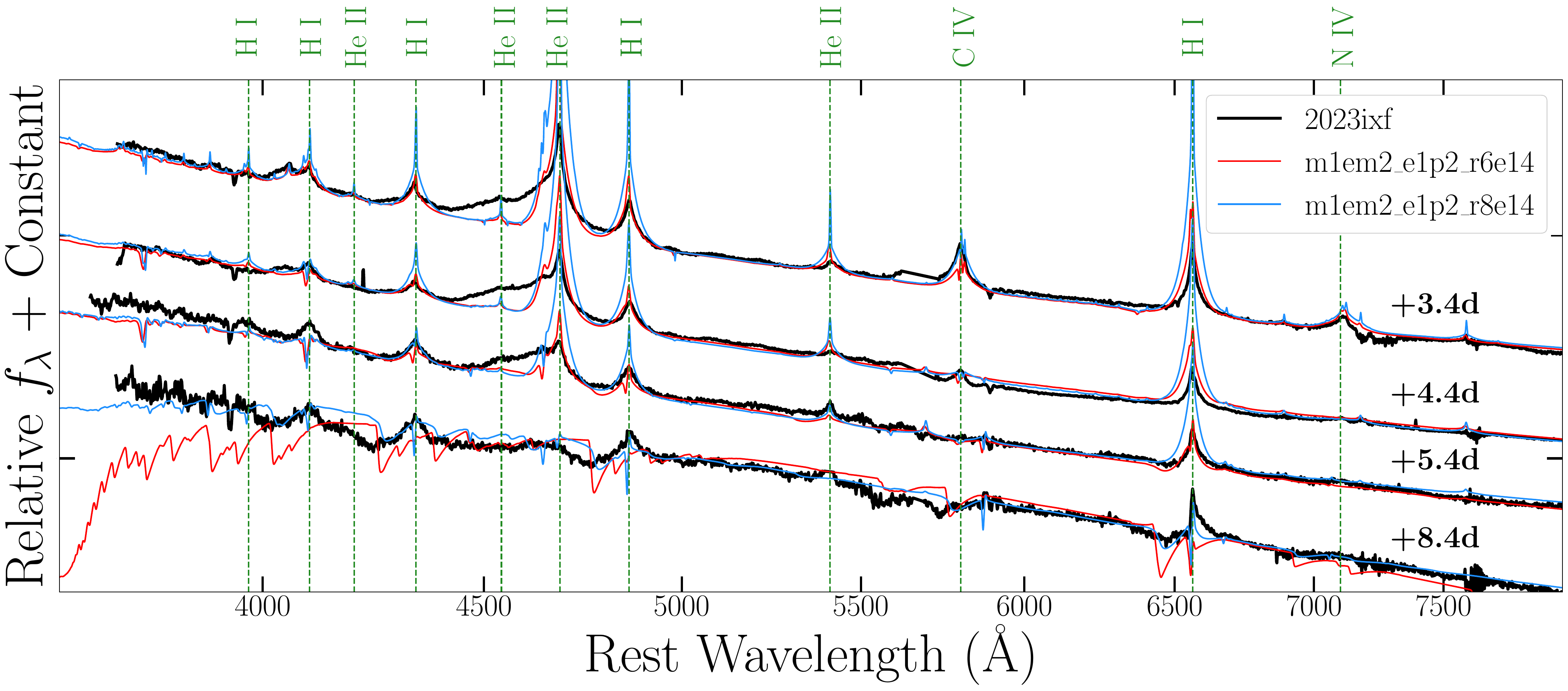}}
\caption{{\it Top:} Early-time UV and {\it Bottom:} optical spectra of CSM-interacting SN~II 2023ixf (black; \citealt{wjg23, Zimmerman24}) compared to {\tt CMFGEN} model spectra from \cite{dessart25CSMgrid}, both with $\dot M = 10^{-2}$~\mdot\ and $E_k = 1.2$~B, shown in red ($R_{\rm CSM} = 6\times 10^{14}$~cm) and blue ($R_{\rm CSM} = 8\times 10^{14}$~cm). These {\tt CMFGEN} model spectra provide a decent match the evolution of the IIn-like features in the optical/NIR but show notable differences in the near-UV, which may be caused by asymmetries in the CSM, SN ejecta, or both. Prominent, narrow UV absorption lines arise from the interstellar medium. \label{fig:23ixf_cmfgen} }
\end{figure*}

In Figure \ref{fig:23ixf_cmfgen}, we present time-series UV/optical spectra of SN~2023ixf from $\delta t = 3.6 - 8.7$~days compared to two sets of \cmfgen\ spectra at similar phases that include $\dot M = 10^{-2}$~\mdot, $R_{\rm CSM} = (6 \ \& \ 8) \times 10^{14}$~cm, and $E_k = 1.2$~B. These models are similar to the {\tt r1w6a/b} model, which provided the best match to the IIn-like feature duration and overall early-time optical spectral evolution of SN~2023ixf (e.g., see \citealt{wjg23, Bostroem23, Zhang23, Zheng25}). As shown in Figure \ref{fig:23ixf_cmfgen}, the varying CSM radius between both models causes differences in each model's ability to perfectly reproduce the SN~2023ixf spectrum. For example, the model with $R_{\rm CSM} = 6\times10^{14}$~cm does not over-predict the emission line strengths of Balmer series and optical high-ionization features (e.g., \ion{He}{ii}, \ion{C}{iv}, \ion{N}{iii/iv}) unlike the $R_{\rm CSM} = 8\times10^{14}$~cm model, but it also becomes optically-thin to electron-scattering too early compared to SN~2023ixf. 

In the near-UV, both model spectra predict strong \ion{He}{ii} and \ion{N}{iii} emission with Lorentzian profiles at $\delta t < 5$~days, neither of which is observed in SN~2023ixf at any epoch. Furthermore, there are numerous metal absorption lines from the pre-shock CSM that are prominent in both models but cannot be constrained given the resolution of the SN~2023ixf UV spectra. Compared to the first UV epoch of SN~2023ixf, both models do display strong \ion{N}{iv}~$\lambda1719$ emission comparable to the observations, but this feature does not show the broad absorption trough extending to $\sim -5000~\kms$, nor is it visible in model spectra between $\delta t = 4 - 6$~days as in SN~2023ixf. Additionally, both models do predict narrow \ion{C}{iii}~$\lambda$2297 emission consistent with the duration of the same feature observed in SN~2023ixf, confirming the continuous photo-ionization of CSM ahead of the forward shock at distances of $>8\times 10^{14}$~cm for shock and wind velocities of $10000~\kms$ and $25~\kms$, respectively. Notably, these \cmfgen\ model spectra do not show the same absorption ``trough'' observed in \ion{C}{iii}~$\lambda$2297, but rather show blue-shifted emission from the CDS, which is a product of the simulations adopting spherical symmetry \citep{D15_2n}.

\begin{figure*}
\centering
\subfigure{\includegraphics[width=0.49\textwidth]{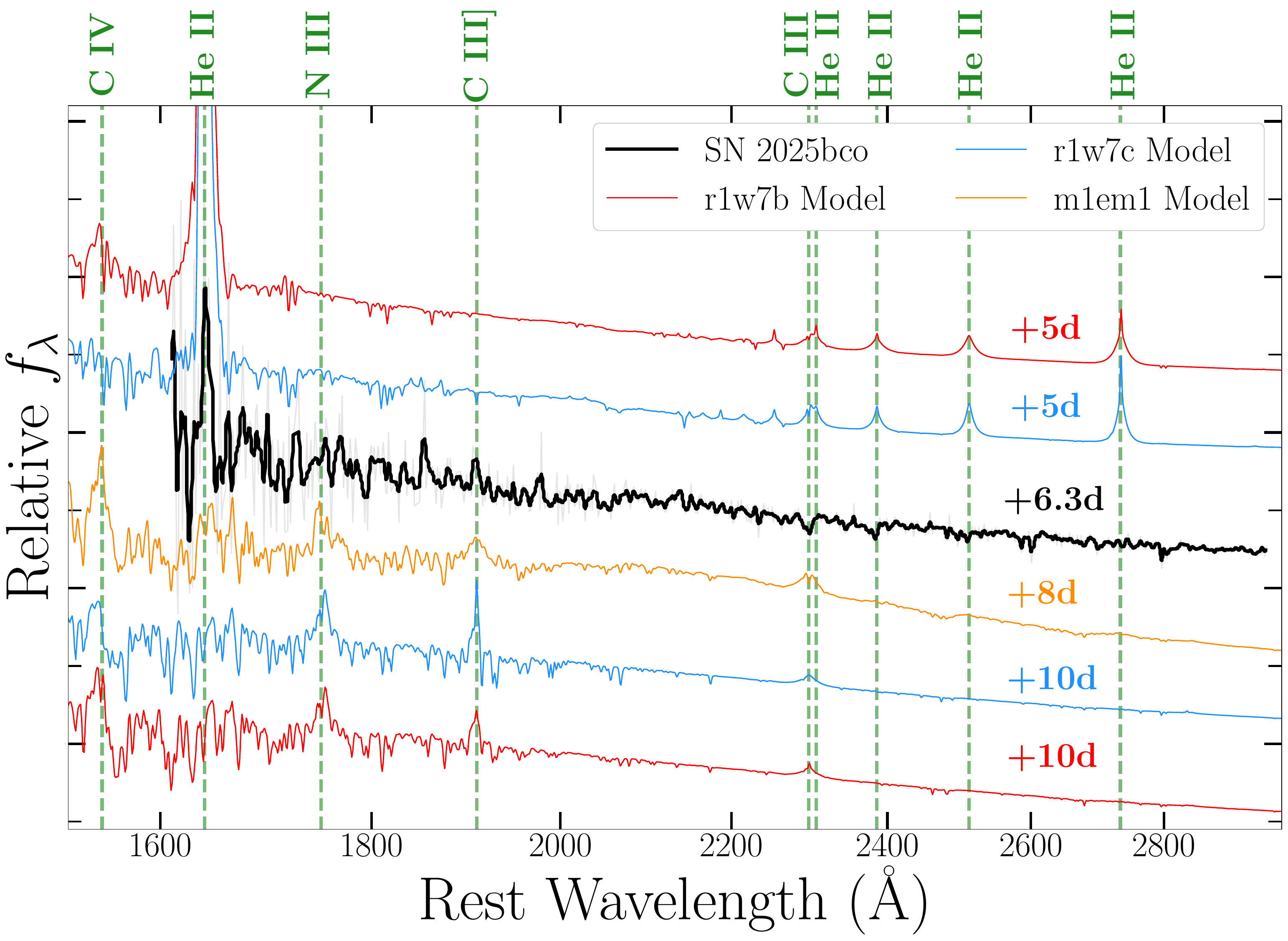}}
\subfigure{\includegraphics[width=0.49\textwidth]{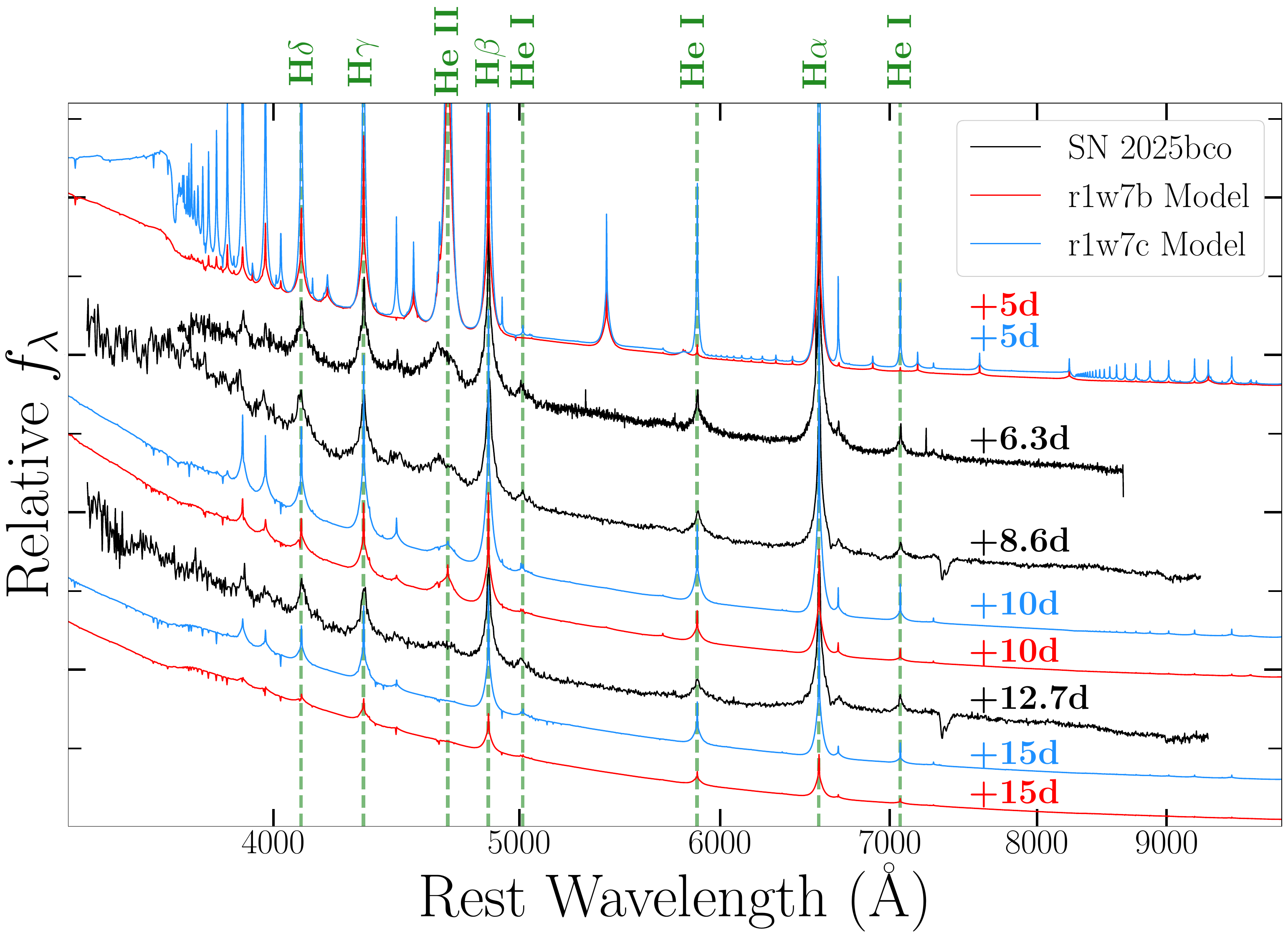}}\\
\subfigure{\includegraphics[width=0.49\textwidth]{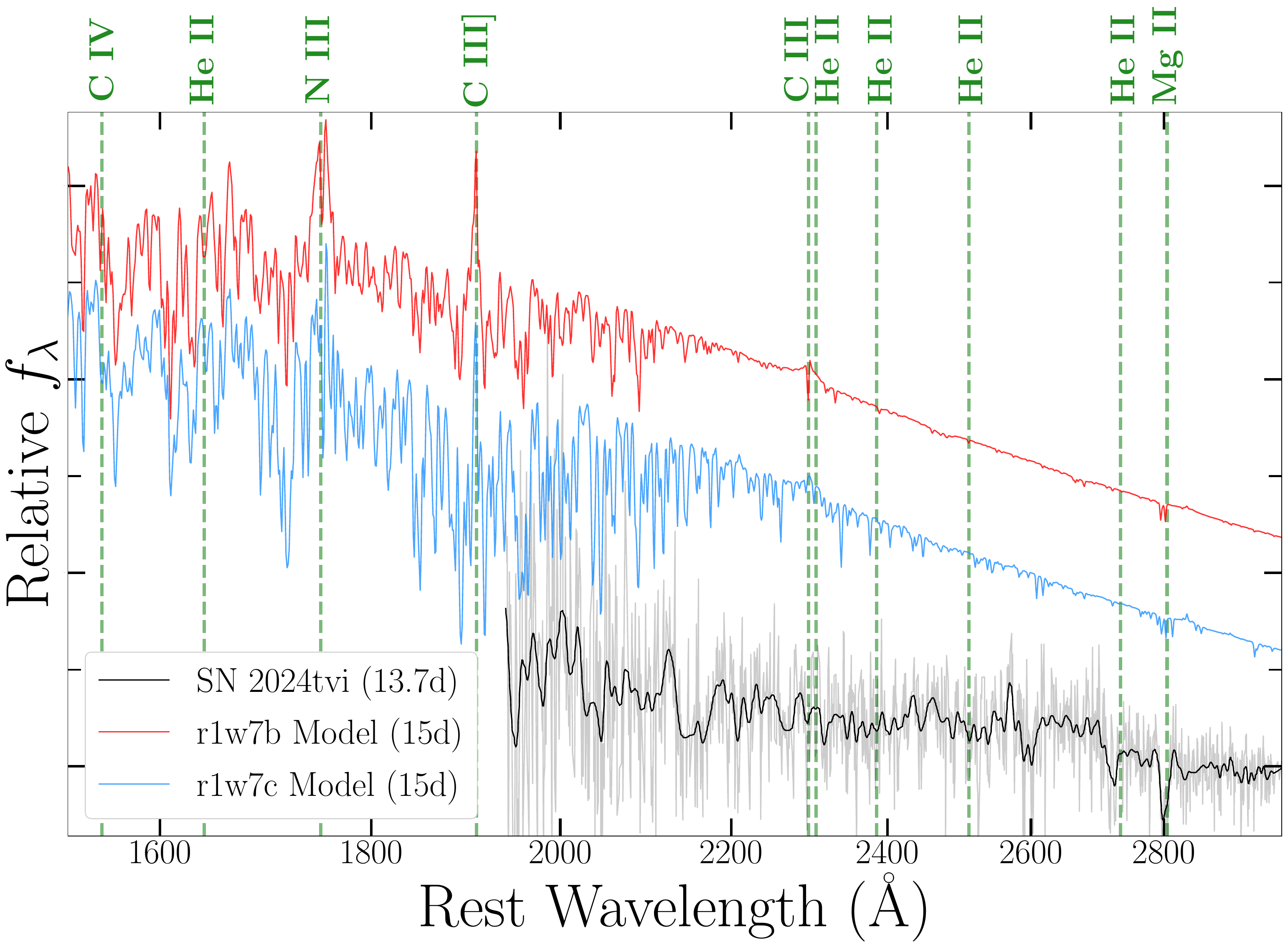}}
\subfigure{\includegraphics[width=0.49\textwidth]{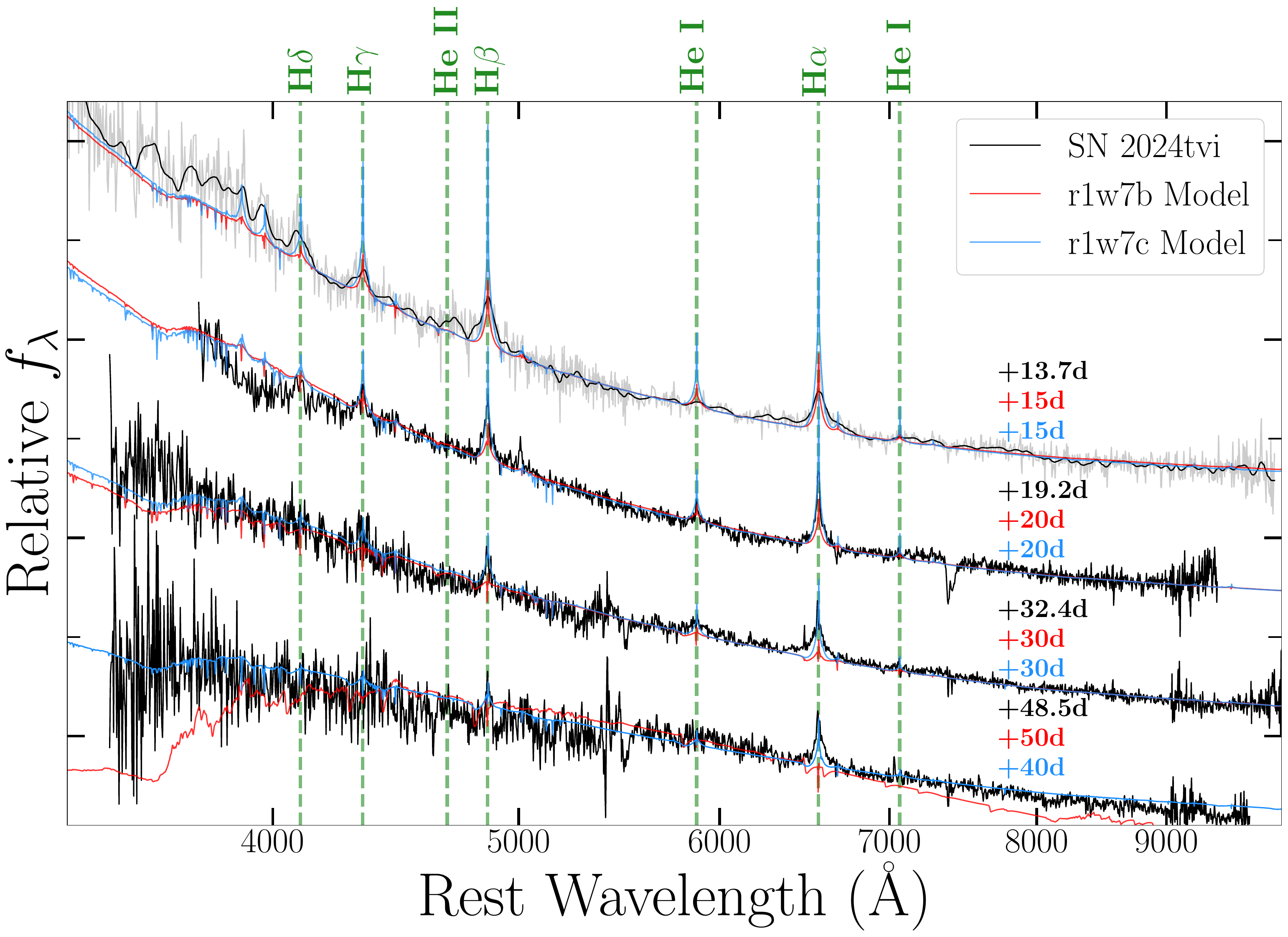}}\\
\subfigure{\includegraphics[width=0.49\textwidth]{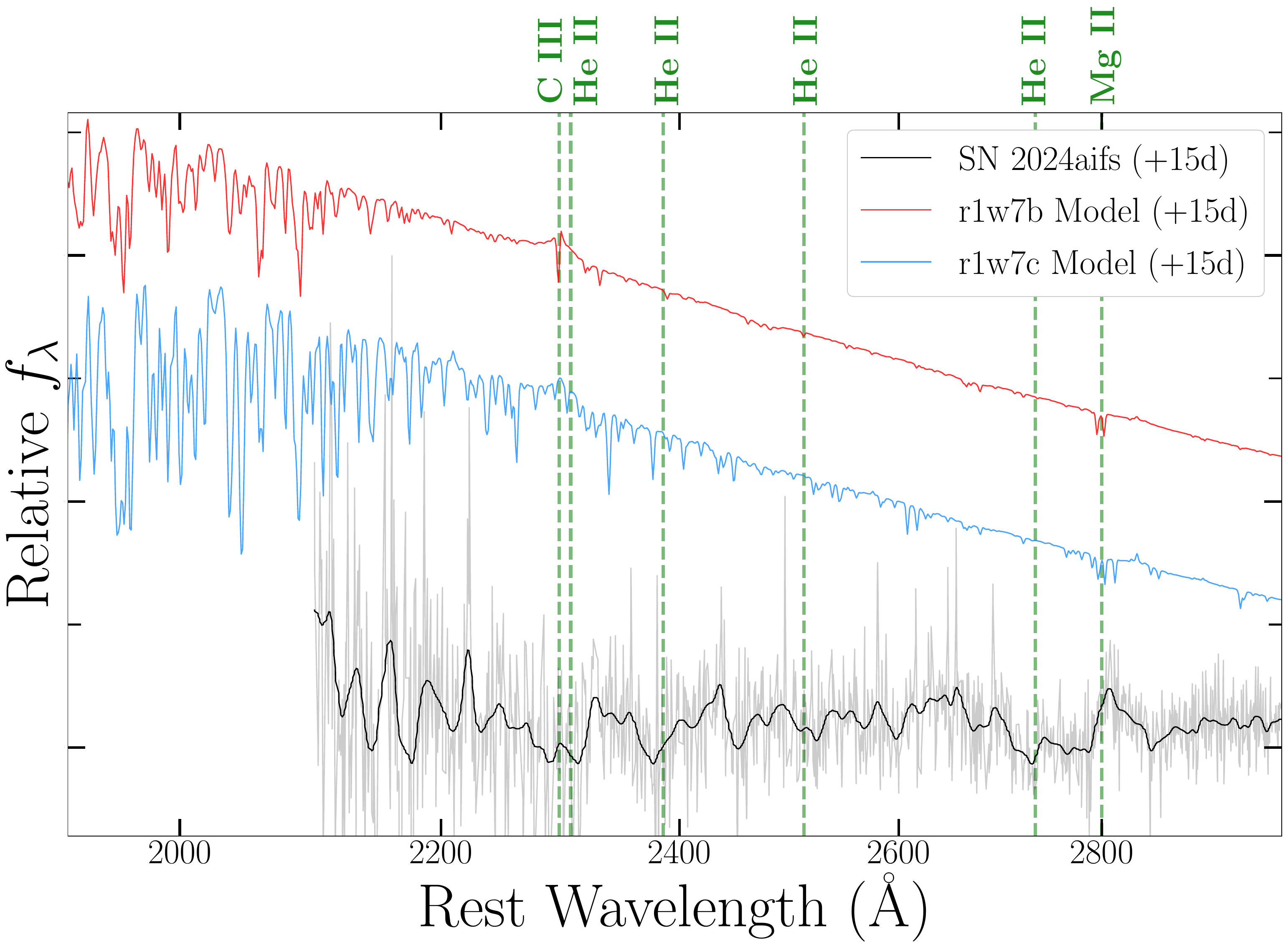}}
\subfigure{\includegraphics[width=0.49\textwidth]{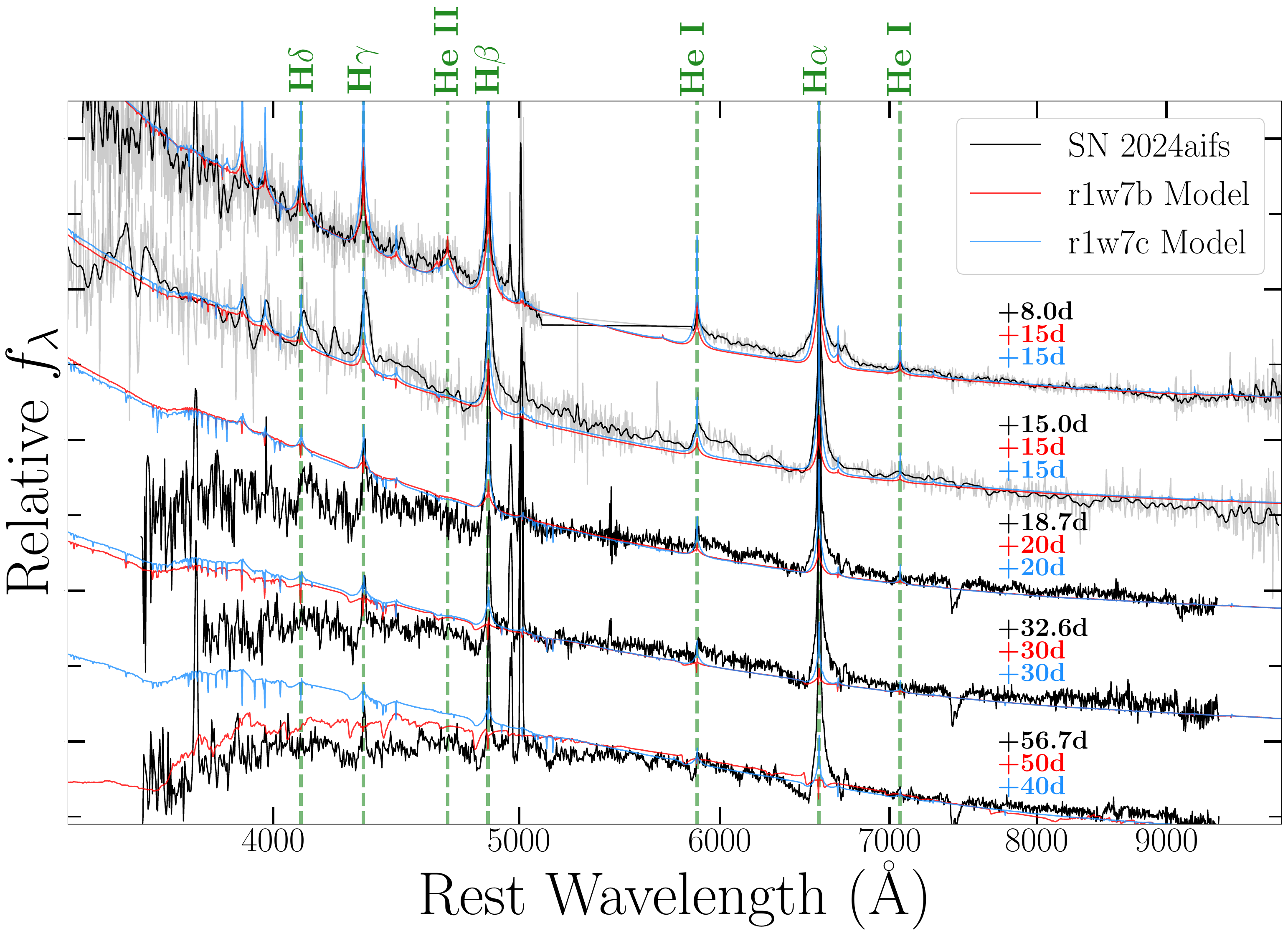}}\\
\caption{{\it Left:} Early-time UV spectra of SNe~IIn 2025bco (top panel), 2024tvi (middle panel), and 2024aifs (bottom panel) shown in black compared to the r1w7b (red; $\dot M = 3\times 10^{-2}$~\mdot, $R_{\rm CSM} = 2\times 10^{15}$~cm) and r1w7c (blue; $\dot M = 3\times 10^{-2}$~\mdot, $R_{\rm CSM} = 4\times 10^{15}$~cm) \cmfgen\ model spectra. {\it Right:} Time-series optical spectra of sample SNe~IIn (black) compared to \cmfgen\ model spectra (red/blue) at similar phases. \label{fig:IIn_cmfgen} }
\end{figure*}

In Figure \ref{fig:IIn_cmfgen}, we present the single UV and multiple optical spectral epochs of all three sample SNe~IIn compared to the {
\tt r1w7b,c} and {\tt m1em1} models \citep{wjg22, Dessart23} -- the former contains $\dot M = 3\times10^{-2}$~\mdot, $R_{\rm CSM} = (2 \ \& \ 4) \times 10^{15}$~cm and the latter contains $\dot M = 10^{-1}$~\mdot, $R_{\rm CSM} = 10^{16}$~cm. In the optical, these \cmfgen\ models with higher $\dot M$ and more extended CSM can effectively reproduce the strengths of all Balmer series and \ion{He}{i} Lorentzian emission profiles at $\delta t < 2$~weeks, but a larger CSM radius of $>4\times 10^{15}$~cm is required to match the duration of these features. For SN~2025bco, the UV spectrum at $\delta t = 6.3$~days does not contain the prominent Lorentzian-shaped \ion{He}{ii} emission lines between $\sim2200 - 2800$~\AA\ that are present in the {\tt r1w7b,c} model spectra at $\delta t = 5$~days. However, SN~2025bco does show \ion{He}{ii}~$\lambda$1640 emission like the models, but the emission strength is weaker than in the \cmfgen\ spectra at $\delta t = 5$~days -- this discrepancy is also seen in model comparison to the \ion{He}{ii}~$\lambda$4686 feature. Additionally, SN~2025bco displays \ion{C}{iii}$~\lambda$2297 absorption while the model spectra show this feature in emission out to $\delta t \approx 10$~days. Lastly, the \cmfgen\ spectra show \ion{N}{iii}$~\lambda$1750 and \ion{C}{iii}]~$\lambda$1909 emission that increases in strength by $\delta t \approx 8-10$~days, but is not detected in SN~2025bco.

The other two sample SNe~IIn, 2024aifs and 2024tvi, have later phase UV spectra at $\delta t \approx 14-15$~days, which we compare to the {\tt r1w7b,c} model spectra at $\delta t = 15$~days. As shown in Figure \ref{fig:IIn_cmfgen}, the optical evolution of both SNe out to $\delta t \approx 50$~days requires the enhanced mass-loss rate present in the models as well as a CSM radius of $R_{\rm CSM} > 4 \times 10^{15}$~cm given the persistence of the observed Lorentzian profiles. In the UV, both SNe are relatively featureless, which is consistent with the UV lines predicted in these model spectra at the observed wavelengths. Notably, SN~2024aifs shows broad \ion{Mg}{ii} absorption from the CDS (see \S\ref{sec:specIIn}), which is not present in the model spectra. Furthermore, there is weak but visible \ion{He}{ii}/\ion{C}{iii} emission near 2300~\AA\ in the model spectrum, which is not present in either SN~IIn. However, the low SNR of both HST spectra makes a robust model comparison challenging at UV wavelengths.

\subsection{Type IIb Supernovae}\label{sec:specIIb}

Our UV spectroscopy sample also includes two Type IIb SNe 2024iss and 2024pso, which were observed at $\delta t = 7.0 \ \& \ 8.6$~days, making them the youngest SNe~IIb to be observed with HST UV spectroscopy to date. In Figure \ref{fig:IInIIbcomp}, we compare the UV spectra of these objects to IUE spectra of SN~1993J at $\delta t = 2.6 \ \& \ 8.9$~days \citep{1993A&A...280L..15D, 1993AAS...182.5503S} and HST spectra of SN~2013df at $\delta t = 13$~days \citep{benami15} -- the former remains the earliest UV spectroscopy of a SN~IIb to date. A more focused, multi-wavelength analysis of SN~2024iss and its UV spectrum is presented in \cite{yete26}. Despite the small sample size, the early-time UV spectra of these SNe~IIb are consistent in their relatively featureless continuum with P-Cygni profiles from metals being significantly weaker than in SNe~II-P/L (e.g., see Fig.~\ref{fig:SNIIUV}), this likely being the result of a smaller H-rich envelope mass in SNe~IIb and/or the host enviornment metallicity (e.g., see \citealt{yete26}). However, SN~2024iss does display a prominent \ion{Mg}{ii} emission that extends to velocities of $\sim 15000~\kms$ on both the blue and red sides of the profile -- this is likely connected to the presence of confined CSM with slightly higher densities than in other SNe~IIb that do not show this feature. Blue-shifted \ion{Mg}{ii} emission is present in Type IIb SNe~1993J and 2013df at later phases and is attributed to on-going injection of shock power into a CDS that formed soon after shock breakout \citep{benami15, dessart22}. 

\subsection{Super-Luminous Supernovae}\label{sec:specSLSN}

\begin{figure*}
\centering
\subfigure{\includegraphics[width=0.49\textwidth]{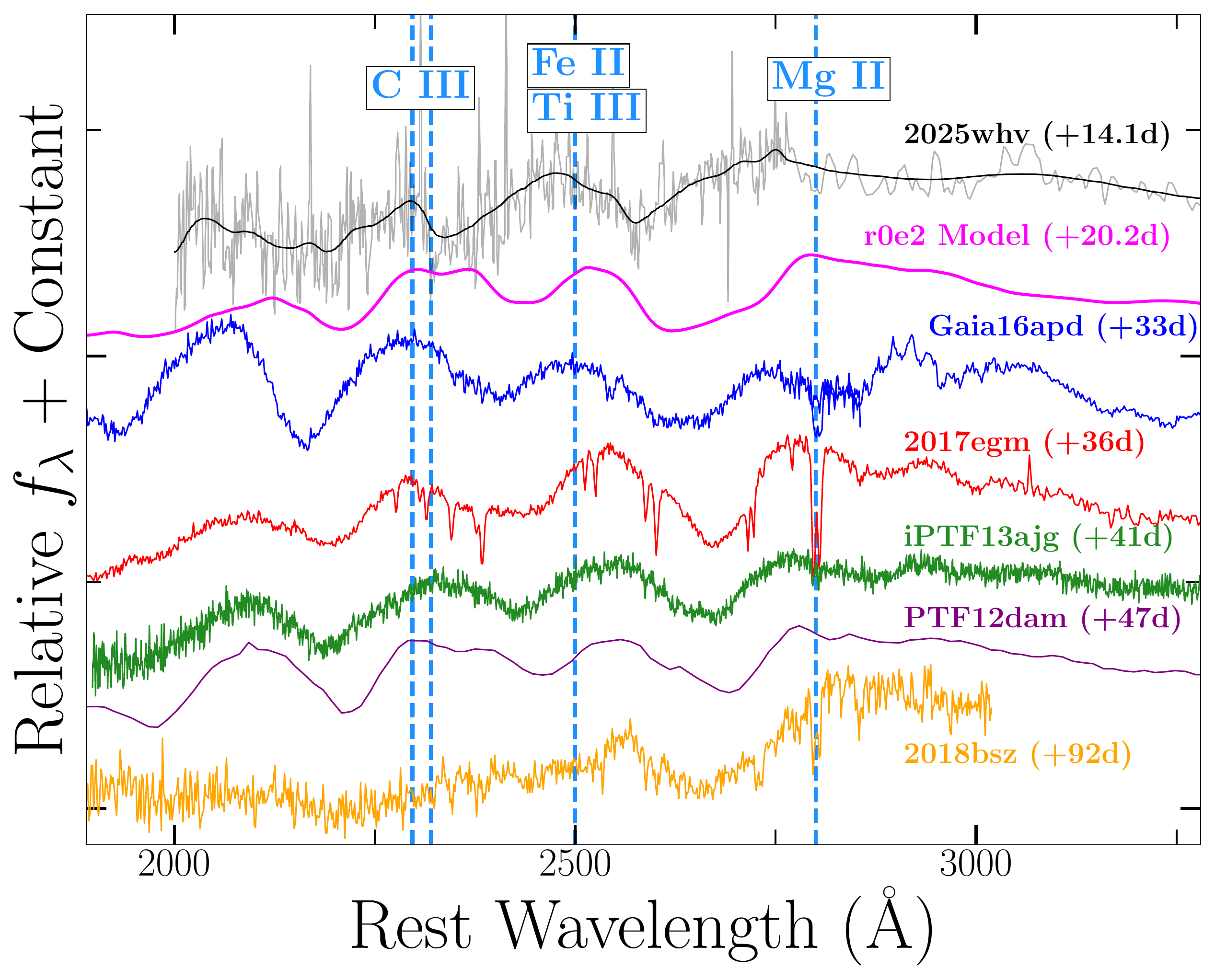}}
\subfigure{\includegraphics[width=0.49\textwidth]{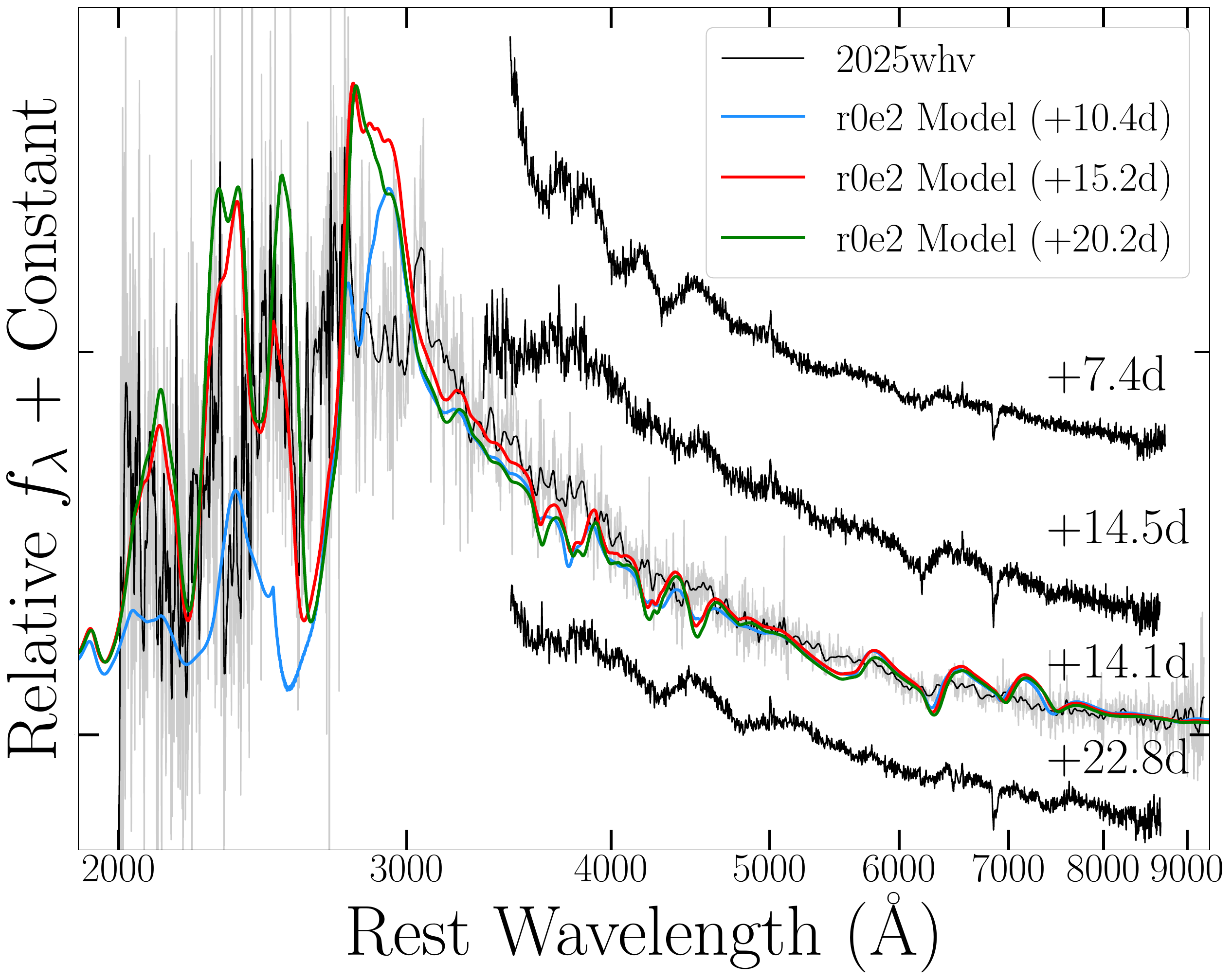}}\\
\caption{{\it Left:} Compilation of all previously published SLSNe-I with UV spectroscopy compared to the newly obtained HST UV spectrum of SN~2025whv. Superluminous Type Ic SN model spectrum from \cite{Dessart19} shown in magenta. {\it Right:} Early-time UV and optical spectra of SN~2025whv (black) compared to {\tt r0e2} SLSN-I model (magnetar power injection with $M_{\rm ej} = 9.7~\Msun$, $E_{\rm pm} = 4\times 10^{50}$~erg, $P_{\rm pm} = 7$~ms, and $B_{\rm pm} = 3.5\times 10^{14}$~G) at different phases shown as blue, red and green lines. \label{fig:SLSNe} }
\end{figure*}

Through our ``Flex Day'' HST program, we obtained one UV spectrum of Type Ic SLSN 2025whv at $\delta t = 15.7$~days, which represents one of the earliest UV spectra of such a SN type to date. In Figure \ref{fig:SLSNe}, we compare the UV spectrum of SN~2025whv to UV spectra of Type I SLSNe Gaia16apd ($\delta t = 33$~days; \citealt{yan17}), SN~2017egm ($\delta t = 36$~days; \citealt{yan18}), PTF12dam ($\delta t = 47$~days; \citealt{Nicholl13}), and SN~2018bsz ($\delta t = 92$~days; \citealt{Pursiainen22}). We also compare the UV/optical spectra of SN~2025whv to Type Ic SLSN \cmfgen\ {\tt r0e2} model spectra from \cite{Dessart19}, which include power injection from a magnetar with rotational energy of $4\times 10^{50}$~erg, period of 7~ms and magnetic field of $3.5\times 10^{14}$~G within $9.7~\Msun$ of SN ejecta. Overall, SN~2025whv is consistent with the UV spectra of other Type I SLSNe and with the model spectra as it displays broad P-Cygni profiles of \ion{Mg}{ii}, \ion{Fe}{ii}, \ion{Ti}{iii} and \ion{C}{iii}. In particular, the velocities of these features in SN~2025whv appear to be higher than in both the model and comparison object spectra, which could be due to the early phase of the UV spectrum for SN~2025whv. 

\section{Discussion}\label{sec:discussion}

\subsection{The Heterogeneity of SN~II UV Spectra and RSG Progenitor Constraints}\label{sec:SNIIsame}

Examination of the full early-time sample of HST UV core-collapse SN spectra enables further constraints on the diversity of young SNe~II. Despite being far from a homogeneous and/or complete sampling of early SN II UV spectra, we can identify observational differences in the UV line profiles that can be mapped to differences in the explosion and progenitor star. For example, examination of UV line velocities in Figures \ref{fig:vels} and \ref{fig:mgIIvels} reveals variance of $\sim 8000~\kms$ in the measured velocities of absorption profile minima at $\delta t \lesssim 10$~days across the SN~II sample objects. As confirmed by model comparisons in \S\ref{sec:SNIImodels}, this indicates that variations in the kinetic energy are necessary to capture the UV and optical properties of young SNe~II e.g., a lower kinetic energy was already shown to be needed for accurate \cmfgen\ modeling of the full UV spectrum of SN~2022acko \citep{bostroem22acko}. Furthermore, our attempts to match both the optical and UV spectra of sample objects (e.g., Figs. \ref{fig:snII_cmfgen1} and \ref{fig:snII_cmfgen2}) with \cmfgen\ spectra demonstrate that some SNe~II require ``younger'' models to reproduce the UV brightness than the observed UV spectral phase. This phenomenon is likely connected to the initial radius of the progenitor star wherein a larger radius is needed to match the increased UV brightness -- this property has been explored extensively in the context of early-time UV/optical light curve modeling of shock cooling emission for SNe~II (e.g., see \citealt{dessart10, sapir17, piro21, morag_shock_2023, Irani24}). 

\begin{figure*}
\centering
\subfigure{\includegraphics[width=0.49\textwidth]{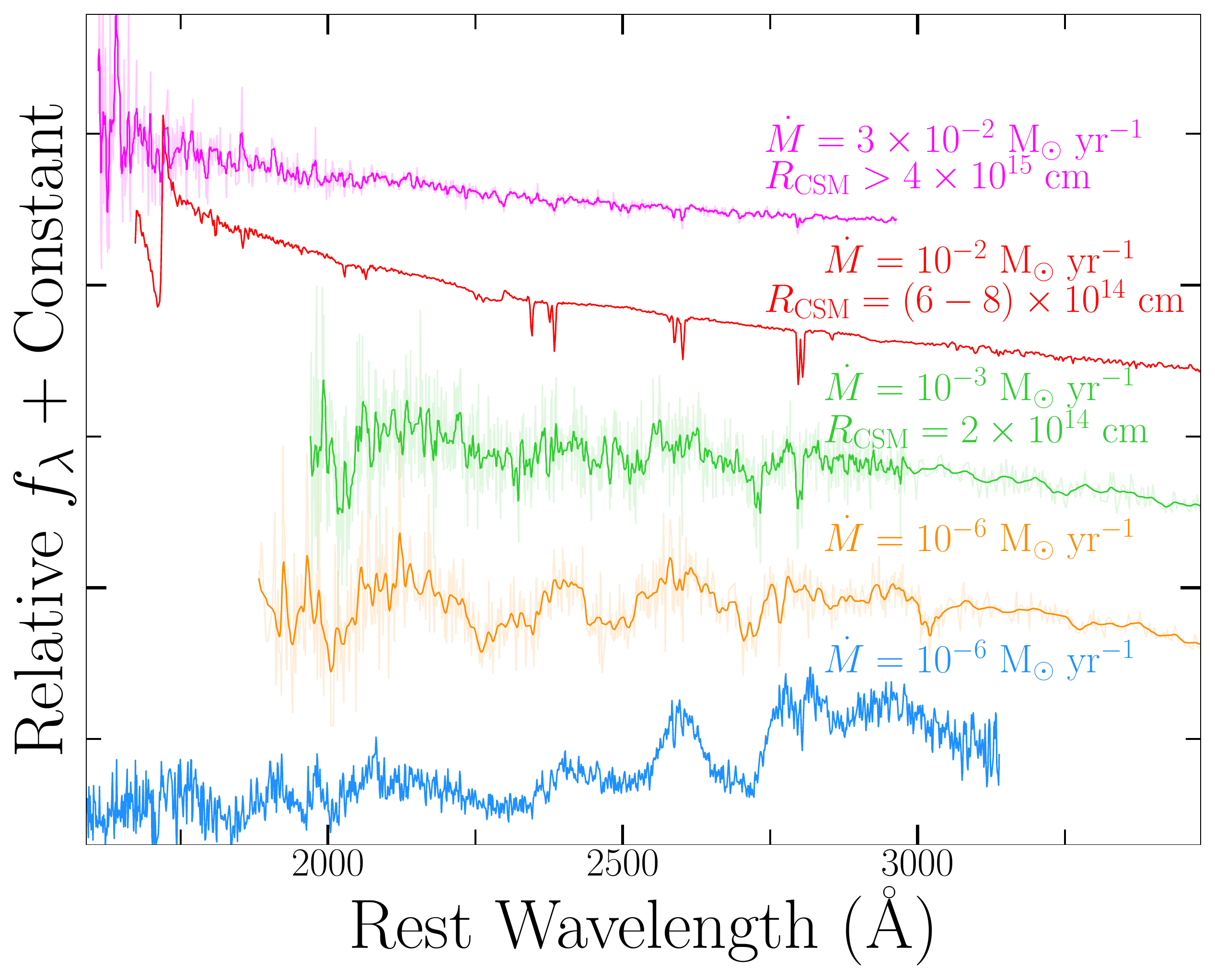}}
\subfigure{\includegraphics[width=0.49\textwidth]{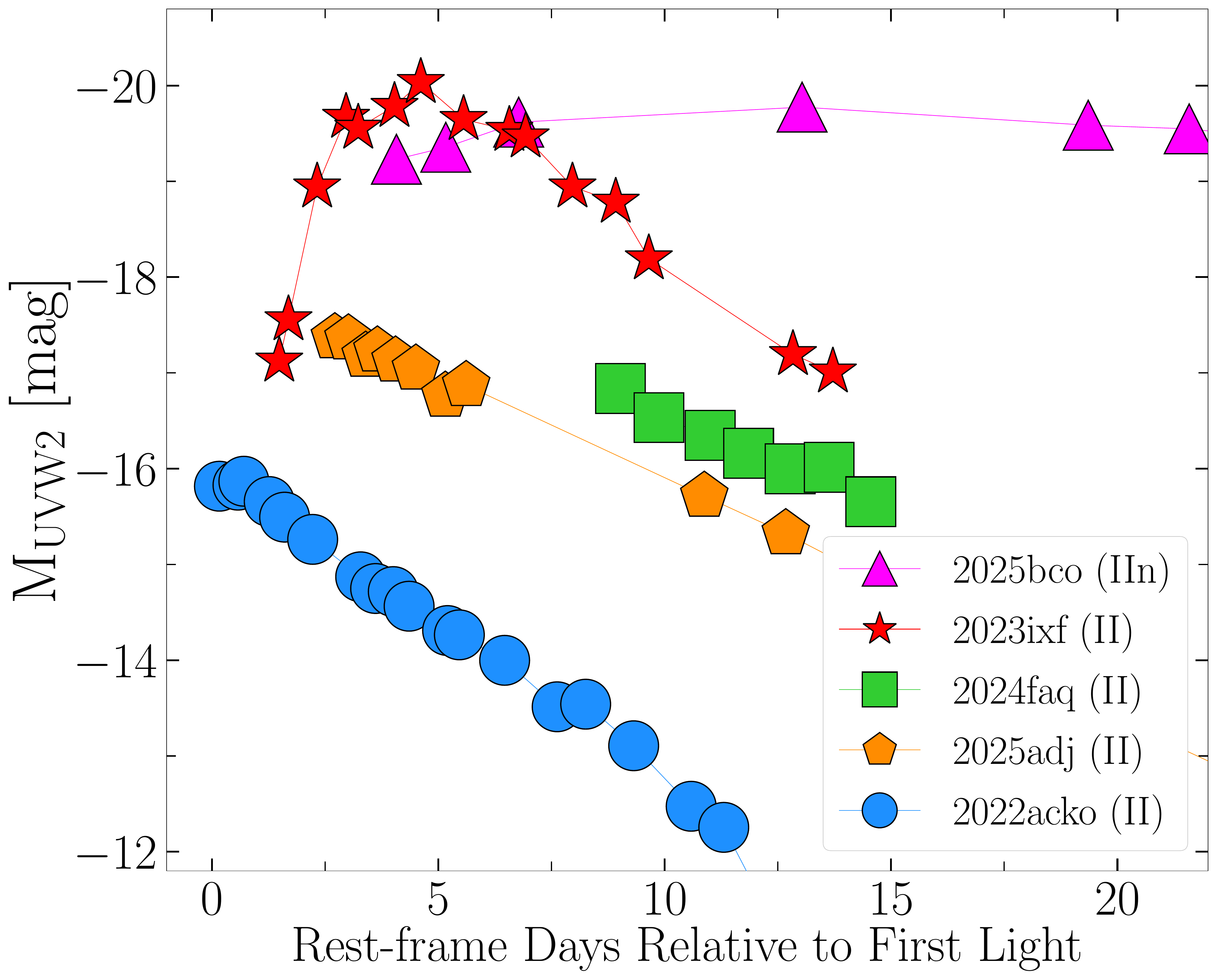}}\\
\caption{{\it Left:} Comparison of sample SNe~II/IIn UV spectra ordered by decreasing mass-loss rate as determined from best-match \cmfgen\ model spectrum matching in \S\ref{sec:SNIImodels} and \S\ref{sec:SNIInmodels}. SN~2025bco at $\delta t = 6.3$~days shown in magenta, SN~2023ixf at $\delta t = 5.6$~days in red, SN~2024faq at $\delta t = 7.7$~days shown in green, SN~2025adj at $\delta t = 8.0$~days in orange, and SN~2022acko at $\delta t = 7.3$~days in blue.  {\it Right:} Early-time UVW2 light curves of the same sample SNe~II/IIn. \label{fig:mdot} }
\end{figure*}

Additionally, the UV heterogeneity within our sample can also be linked to differences in the amount of confined CSM in each SN progenitor system. For example, the UV/optical spectra of most of the sample SNe~II can be matched with models containing weak mass-loss rates of $10^{-6}$~\mdot, but some objects do require higher mass-loss rates and more extended CSM radii e.g., the weaker optical and UV spectral features combined with the more luminous UV light curve in SN~2024faq are more consistent with ejecta interaction with confined CSM ($\dot M = 10^{-3}$~\mdot, $R_{\rm CSM} = 2\times 10^{14}$~cm) that then altered the outer ejecta density profile. In SNe~II such as SN~2023ixf, the CSM is even higher density and extended ($\dot M = 10^{-2}$~\mdot, $R_{\rm CSM} > 6\times 10^{14}$~cm) such that narrow emission features are present in the UV/optical from the CSM interaction for $\sim$1~week and there is spectroscopic evidence for swept up CSM in a CDS in the broad absorption troughs present in both UV and optical transitions (Fig. \ref{fig:23ixf_FUV}). In Figure \ref{fig:mdot}, we present the UV spectra and light curves of selected SNe~II/IIn that range from high to low CSM densities, as inferred from model matching. While the sample size remains too limited to robustly confirm a continuum of UV spectral properties linked to CSM interaction, this motivates the creation of larger, homogeneous samples of early-time SN~II UV spectra as well as the construction of large \cmfgen\ model spectral grids that explore how varying explosion and CSM properties affect UV spectral features (e.g., such as \citealt{dessart25CSMgrid}). 

\subsection{Evidence of Asymmetry in CSM-interacting SNe~II}\label{sec:CSMasymmetry}

As shown in the \cmfgen\ models, the far- and near-UV spectra of CSM-interacting SNe can be rich in narrow and Lorentzian-shaped emission lines from photo-ionized gas ahead of the forward shock as well as broadened absorption troughs from fast-moving gas in the CDS \citep{dessart2016, dessart17}. However, unlike optical recombination lines (e.g., H$\alpha$), whose photons are emitted after recombination and are then broadened mainly by electron scattering in the CSM, many UV lines are resonance transitions from the ground state, whose much larger line optical depths cause their photons to be repeatedly absorbed and re-emitted (i.e., line-scattered). Consequently, UV emission is directional and geometry-sensitive, i.e., UV photons can probe out to larger radii and lower CSM densities than optical photons \citep{Fransson84, Bowen00, dessart22}. Therefore, discrepancies between optical and UV emission lines directly reflect the fact that UV emission paints a unique picture of the dynamical evolution of the CSM interaction, as well as of gas asymmetries and ionization/temperature stratification, all of which the optical struggles to fully reveal \citep{Fransson89, Fransson84b, dessart22}. This phenomenon is clearly demonstrated for sample SNe~2023ixf (SN~II with confined CSM) and 2025bco (SN~IIn with extended CSM) -- both objects show strong emission lines in the optical that are not present at UV wavelengths. UV lines in these SNe arise as both narrow emission from optically-thin, pre-shock gas as well as higher-velocity absorption from the post-shocked gas in the CDS \citep{Dessart23b}. While this property allows for unique probes of asymmetries in the CSM and SN ejecta, 1-D \cmfgen\ model spectra cannot capture such 2- or 3-D effects and therefore the over-predicted UV line fluxes shown in Figures \ref{fig:23ixf_cmfgen} and \ref{fig:IIn_cmfgen} are a result of this model geometric limitation. 

\begin{figure*}
\centering
\subfigure{\includegraphics[width=0.49\textwidth]{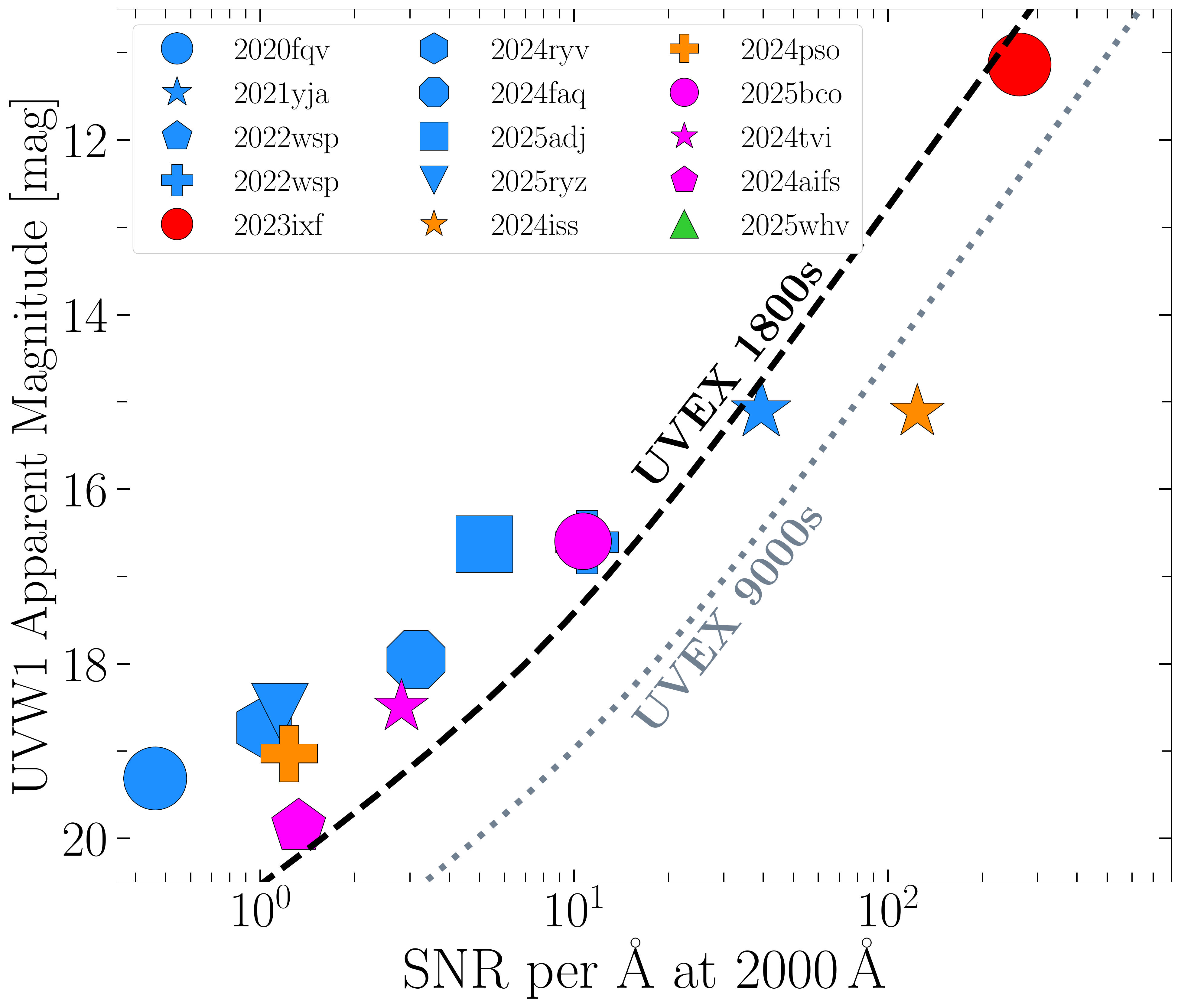}}
\subfigure{\includegraphics[width=0.49\textwidth]{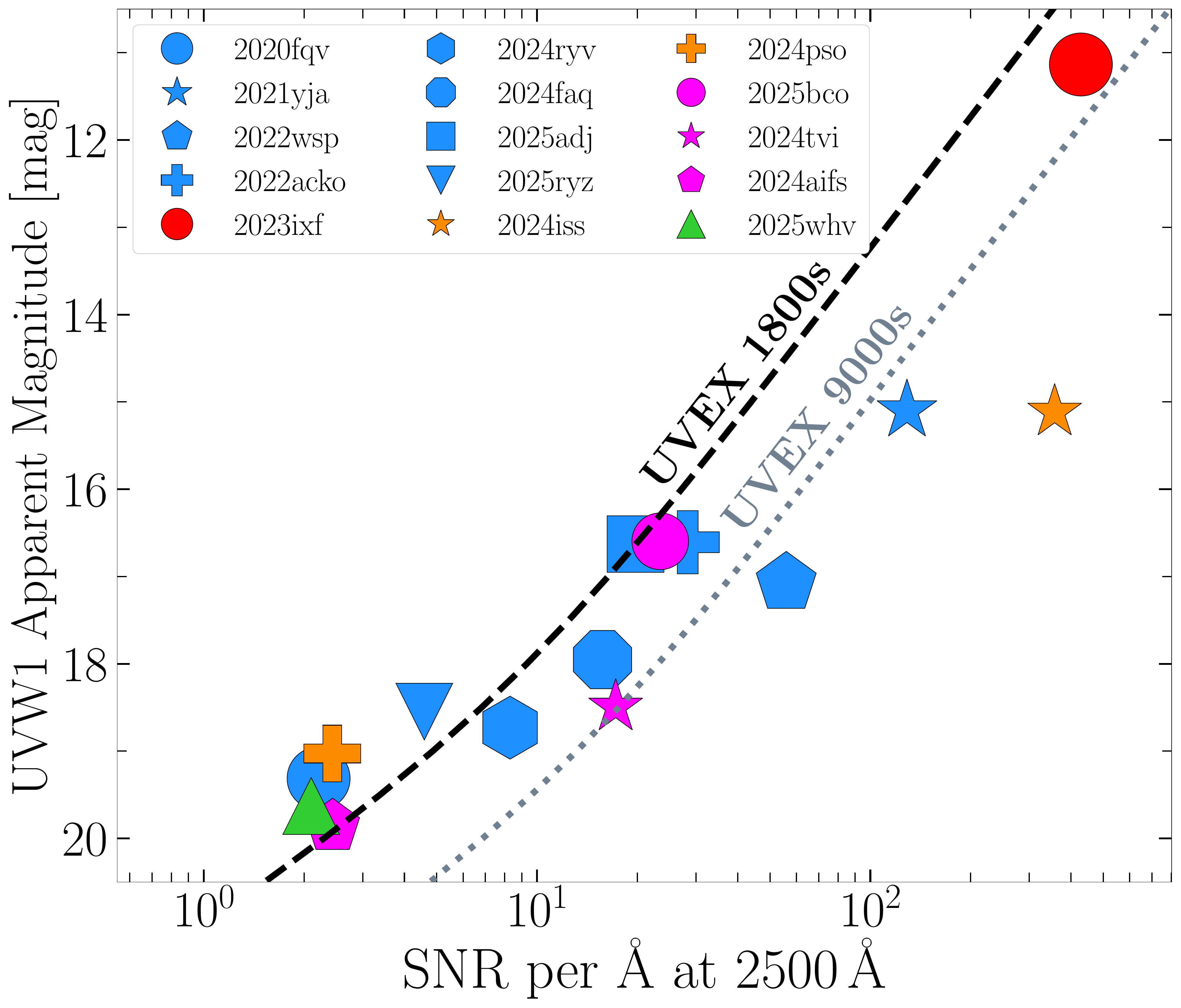}}\\
\caption{Observed Swift-UVOT $UVW1$ magnitude (y-axis) at the same phase as the signal-to-noise ratio per \AA\ in each early-time HST UV spectrum computed at 2000 (left) and 2500 (right)~\AA\ (x-axis). UVEX sensitivity simulation calculated for 1800s (dashed black line) and 9000s (dotted gray line) exposures using the SN~2022acko \cmfgen\ model spectrum at $\delta t = 7$~days. SNe~II shown in blue, SN~2023ixf in red, SNe~IIn in magenta, SNe~IIb in orange and SLSNe-I in green. \label{fig:snr} }
\end{figure*}

\subsection{Future Prospects for Early-time UV SN Spectroscopy}\label{sec:future}

Based on this work, expanding the current sample of early-time UV spectra of core-collapse SNe requires both decreased telescope response time and increased UV sensitivity. We visualize the latter in Figure \ref{fig:snr} by plotting the observed Swift-UVOT $UVW1$ magnitude at the same phase as the HST/STIS spectrum SNR at 2000 and 2500~\AA. While dependent on the SN phase and extinction, it is clear that the HST/STIS sensitivity prohibits near-UV spectroscopy of young SNe~II with decent SNR for $UVW1$ magnitudes of $\gtrsim 19$~mag. Furthermore, this trend towards low SNR gets worse at shorter wavelengths and the quality of some sample object spectra at $<2000$~\AA\ is sub-optimal and/or unusable for spectral analysis. Issues with low UV sensitivity also arise from the fact that most of the UV spectra obtained on young core-collapse SNe are from STIS/CCD which has both lower resolution and sensitivity than STIS/MAMA. Beyond SNR, the ToO response time of HST is a significant limiting factor in obtaining UV spectra of core-collapse SNe at $\delta t < 1$~week. Even with the ``FlexDay'' initiative, the number of SNe with UV spectra at $\delta t \lesssim 1$~week only increased by a factor of 2.7 and the number of SNe observed at $\delta t \lesssim 2$~weeks increased by a factor of 2.4. Given the limited range of HST visibility from the reduced gyro mode and the limited number of disruptive ToO triggers available for UV SN spectroscopy, future missions like the UltraViolet EXplorer (UVEX) will be essential for creating high-quality, early-time UV spectral time-series of core-collapse SNe \citep{Kulkarni21}.

\section{Conclusions} \label{sec:conclusion}

In this paper, we have presented UV/optical/NIR observations and modeling of the largest sample to date of core-collapse SNe with early-time HST UV spectroscopy. Below we summarize the primary observational findings from our sample analysis.  

\begin{itemize}

\item Our sample includes 19 objects in total (12 SNe II, 3 SNe IIb, 3 SNe IIn, 1 SLSN-I), including 11 new/unpublished UV spectra from the ``Flexible Thursdays'' ToO program obtained during HST Cycles 31–32. These programs increased the number of core-collapse SNe with early-time HST UV spectra at $\delta t \lesssim 2$~weeks by a factor of $\sim 2.4$ and highlight the value of a designated ToO program paired with a coordinated infant SN discovery effort using time-domain surveys such as ZTF. Some notable sample objects include SN 2024ryv (now the third-youngest SN II with HST UV spectra), SNe~2024iss and 2024pso (youngest SNe IIb to date with HST UV spectra), and SN~2025whv (youngest SLSN-I to date with HST UV spectra).

\item The early-time near-UV spectra of most sample SNe~II show similar Doppler-broadened P-Cygni profiles of metals such as Mg, Ti, Cr, Fe. However, there is notable diversity in the velocities of these features and their absorption strength relative to the UV continuum. 

\item Matching of early-time UV/optical spectra to \cmfgen\ model grids reveals that most sample SNe~II are consistent with low CSM densities ($\dot M = 10^{-6}$~\mdot) and standard kinetic energies ($E_k = 0.6 - 1.2$~B), but the UV line velocities of SN~II 2021yja are more consistent with a higher energy explosion ($E_k > 1.2$~B) and those in SN~II 2025ryz with a low energy explosion ($E_k = 0.4$~B). Several SNe~II require \cmfgen\ models at earlier phases than the SN observations in order to match the SN continuum from optical to UV, which may be connected to structural differences (e.g., $R_{\star}$) or shock power not accounted for in some simulations (e.g., \citealt{bostroem22acko}), or the way the shock power is treated.

\item The UV/optical spectra of two sample SNe~II, 2024faq and 2023ixf, are consistent with SN ejecta interaction with confined, high-density CSM. SN~2024faq shows weak Balmer/UV features and a luminous UV light curve that are matched by \cmfgen\ models with $\dot M = 10^{-3}$~\mdot\ and $R_{\rm CSM} = 2\times10^{14}$~cm. The UV/optical spectral series of SN~2023ixf requires more extended, higher density CSM of $\dot M = 10^{-2}$~\mdot\ and $R_{\rm CSM} = (6-8) \times10^{14}$~cm. 

\item SN~II 2023ixf and sample SNe~IIn 2024aifs, 2024tvi and 2025bco all show relatively featureless early-time near-UV spectra despite prominent optical Lorentzian emission lines of \ion{H}{i}, \ion{He}{i/ii}, \ion{N}{iii/iv}, and \ion{C}{iii/iv} from electron scattering in optically thick CSM. Nonetheless, notable UV features include narrow \ion{He}{ii}$\lambda$1640 emission in SN~2025bco, narrow/intermediate-width absorption/emission of \ion{N}{iv}~$\lambda 1719$ and \ion{C}{iii}~$\lambda 2297$ in SN~2023ixf, and broad \ion{Mg}{ii}~$\lambda 2803.5$ absorption in SN~2024aifs. Variations between UV and optical during the phase of CSM interaction may be attributed to properties of UV resonance versus optical recombination transitions, as well as ionization stratification and intrinsic asymmetries in the CSM. 

\item UV/optical spectra of the sample SNe~IIn are best matched to \cmfgen\ model spectra with $\dot M = (0.3 - 1)\times 10^{-1}$~\mdot\ and $R_{\rm CSM} \gtrsim 4 \times10^{15}$~cm. However, despite reproducing optical emission line profiles and evolution, the current \cmfgen\ models systematically over-predict the line strengths of key near-UV transitions (e.g., \ion{He}{ii}, \ion{N}{iii}, \ion{C}{iii}) during the phase of ejecta interaction with dense CSM -- this may be the result of the assumption of spherical symmetry in present simulations and the real multi-dimensional asymmetries present in the observed CSM-interacting SNe.

\end{itemize}

The study highlights the potential for using early-time UV spectroscopy to uncover both progenitor and explosion properties of core-collapse SNe. However, the limited number of SNe with such observations prohibits a more statistical analysis of early-time UV spectroscopic phenomena across SN classes. Future missions such as UVEX will enable systematic construction of high-fidelity, early-time UV spectral time-series for SNe II, which, paired with larger grids of non-LTE radiative transfer simulations, will provide an unprecedented picture of the outermost ejecta layers of young SNe II and the mass-loss histories of their progenitor stars.


\facilities{Neil Gehrels Swift Observatory (UVOT/XRT), Shane Telescope (Kast), Keck Observatory (LRIS/NIRES), Hale Telescope (DBSP/NGPS), Zwicky Transient Facility (ZTF), Hubble Space Telescope (STIS)}

\software{IRAF (Tody 1986, Tody 1993), YSE-PZ \citep{Coulter22, Coulter23}, \cmfgen\ \citep{hillier12, dessart22}, Lpipe \citep{Perley19}, HEAsoft \citep{HEAsoft} }

\section{Acknowledgments} \label{Sec:ack}

We thank Patricia Royle, Mercedes Lopez-Morales, and Tala Monroe for all their assistance with our Flex Day programs. 

W.J.-G.\ is supported by NASA through Hubble Fellowship grant HSTHF2-51558.001-A awarded by the Space Telescope Science Institute, which is operated for NASA by the Association of Universities for Research in Astronomy, Inc., under contract NAS5-26555. R.J.Y. acknowledges support from the Caltech Summer Undergraduate Research Fellowship (SURF) program. This work was granted access to the HPC resources of TGCC under the allocation 2024 -- A0170410554 made by GENCI, France. R.M.\ acknowledges support via program HST-GO-17507  provided by NASA through a grant from the Space Telescope Science Institute, which is operated by the Association of Universities for Research in Astronomy, Inc., under NASA contract NAS5-26555." R.M. and R.C. acknowledge the hospitality of the Kavli Institute for Theoretical Physics at UC Santa Barbara, where portions of this work were completed. This research was supported in part by the National Science Foundation under PHY-2309135. C.D.K.\ gratefully acknowledges support from the NSF through AST-2432037, the HST Guest Observer Program through HST-SNAP-17070 and HST-GO-17706, and from JWST Archival Research through JWST-AR-6241 and JWST-AR-5441. H.S. acknowledges partial salary support from a Moore Foundation Postdoctoral Fellowship Grant to Rutgers University. K.C.P is funded by the Moore Foundation. N.R. is supported by a Northwestern University Presidential Fellowship Award. Zwicky Transient Facility access was supported by Northwestern University and the Center for Interdisciplinary Exploration and Research in Astrophysics (CIERA). D.A.C. acknowledges funding through JWST program grants JWST-GO-06541, JWSTGO-06585, and JWST-GO-05324. The UCSC transients team is supported in part by a fellowship from the David and Lucile Packard Foundation to R.J.F. K.W.D.\ is supported in part by a University of California President's Pre-Professoriate Fellowship (PPPF). A.G-Y’s research is supported by ISF, IMOS and BSF grants, as well as the André Deloro Institute for Space and Optics Research, the Center for Experimental Physics, a WIS-MIT Sagol grant, the Norman E Alexander Family M Foundation ULTRASAT Data Center Fund, and Magnus Konow in honor of his mother Olga Konow Rappaport;  A.G-Y is the incumbent of the Arlyn Imberman Professorial Chair. A.S. acknowledges support from the Knut and Alice Wallenberg Foundation through the ``Gravity Meets Light'' project. J.H.T. acknowledges Horizon Europe ERC grant no. 101125877.

Based on observations made with the Nordic Optical Telescope. We acknowledge support from the Swedish Research Council (VR) through the Oskar Klein Centre.

Some of the data presented in this paper were obtained from the Mikulski Archive for Space Telescopes (MAST) at the Space Telescope Science Institute. STScI is operated by the Association of Universities for Research in Astronomy, Inc., under NASA contract NAS5–26555. Support to MAST for these data is provided by the NASA Office of Space Science via grant NAG5–7584 and by other grants and contracts. The specific observations analyzed can be accessed via \dataset[https://doi.org/10.17909/5ktt-v982]{https://doi.org/10.17909/5ktt-v982}.

This research is based on observations made with the NASA/ESA Hubble Space Telescope obtained from the Space Telescope Science Institute, which is operated by the Association of Universities for Research in Astronomy, Inc., under NASA contract NAS 5–26555. These observations are associated with programs HST-GO-17507 (PI Jacobson-Gal\'an), HST-GO-17807 (PI Jacobson-Gal\'an) and HST-GO-17788 (PI Howell).

We acknowledge the use of public data from the Swift data archive. We are grateful for all that the Swift telescope and the mission team has provided to the study of supernova science.

Some of the data presented herein were obtained at Keck Observatory, which is a private 501(c)3 non-profit organization operated as a scientific partnership among the California Institute of Technology, the University of California, and the National Aeronautics and Space Administration. The Observatory was made possible by the generous financial support of the W.\ M.\ Keck Foundation.
The authors wish to recognize and acknowledge the very significant cultural role and reverence that the summit of Maunakea has always had within the indigenous Hawaiian community.  We are most fortunate to have the opportunity to conduct observations from this mountain.

A major upgrade of the Kast spectrograph on the Shane 3\,m telescope at Lick Observatory, led by Brad Holden, was made possible through gifts from the Heising-Simons Foundation, William and Marina Kast, and the University of California Observatories. Research at Lick Observatory is partially supported by a generous gift from Google.


Based on observations obtained with the Samuel Oschin Telescope 48-inch and the 60-inch Telescope at the Palomar Observatory as part of the Zwicky Transient Facility project. ZTF is supported by the National Science Foundation under Grants No. AST-1440341, AST-2034437, and currently Award \#2407588. ZTF receives additional funding from the ZTF partnership. Current members include Caltech, USA; Caltech/IPAC, USA; University of Maryland, USA; University of California, Berkeley, USA; Cornell University, USA; Drexel University, USA; University of North Carolina at Chapel Hill, USA; Institute of Science and Technology, Austria; National Central University, Taiwan, German Center for Astrophysics, Germany, and OKC, University of Stockholm, Sweden. Operations are conducted by Caltech's Optical Observatory (COO), Caltech/IPAC, and the University of Washington at Seattle, USA.

SED Machine is based upon work supported by the National Science Foundation under Grant No.\ 1106171. The ZTF forced-photometry service was funded under the Heising-Simons Foundation grant \#12540303 (PI: Graham). The Gordon and Betty Moore Foundation, through both the Data-Driven Investigator Program and a dedicated grant, provided critical funding for SkyPortal.

\bibliographystyle{aasjournal} 
\bibliography{references} 


\clearpage
\appendix
\counterwithin{figure}{section}

\section{Tables and Supplementary Figures} 
\begin{figure*}
\centering
\subfigure{\includegraphics[width=0.48\textwidth]{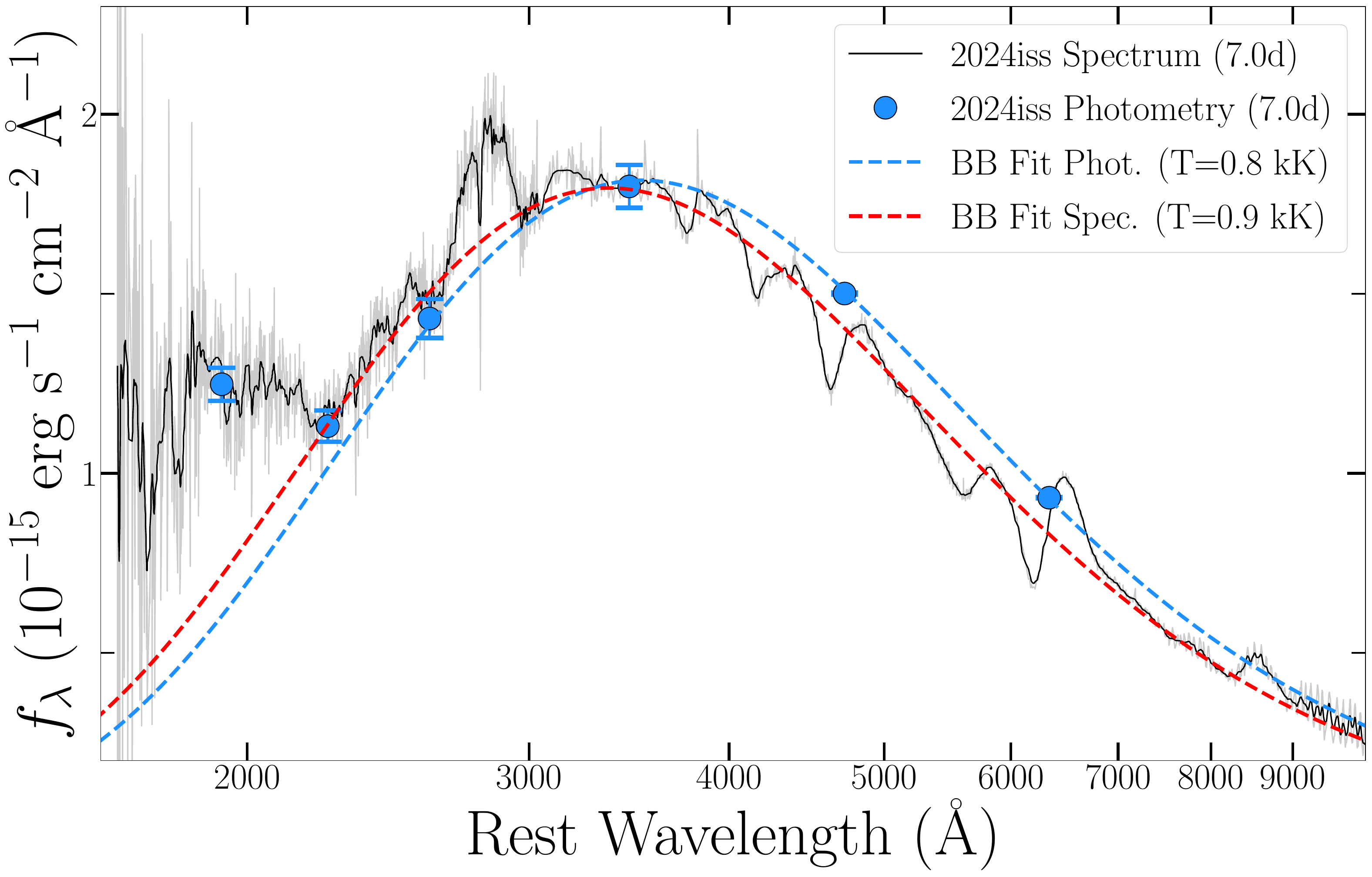}}
\subfigure{\includegraphics[width=0.49\textwidth]{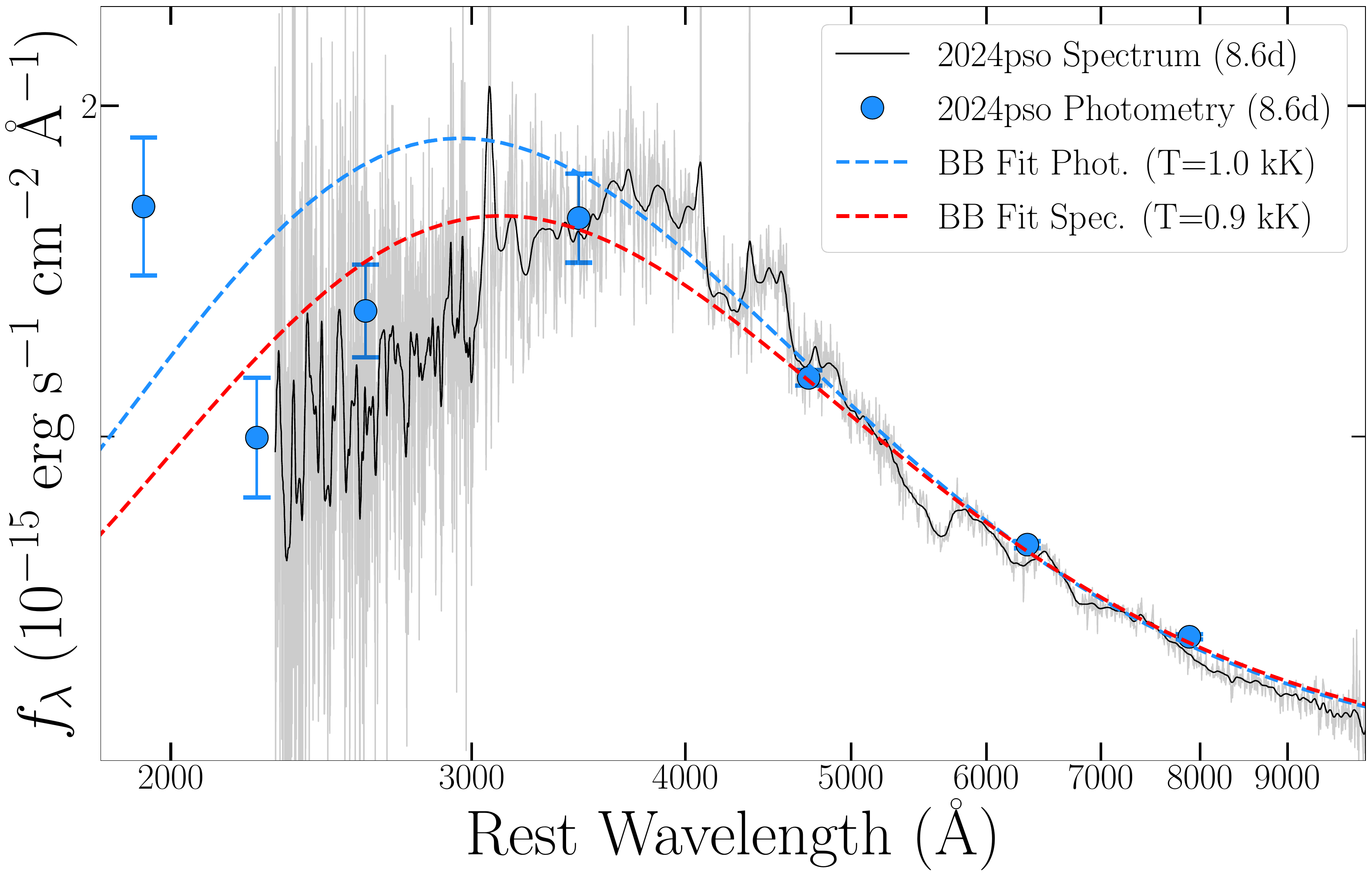}}\\
\subfigure{\includegraphics[width=0.49\textwidth]{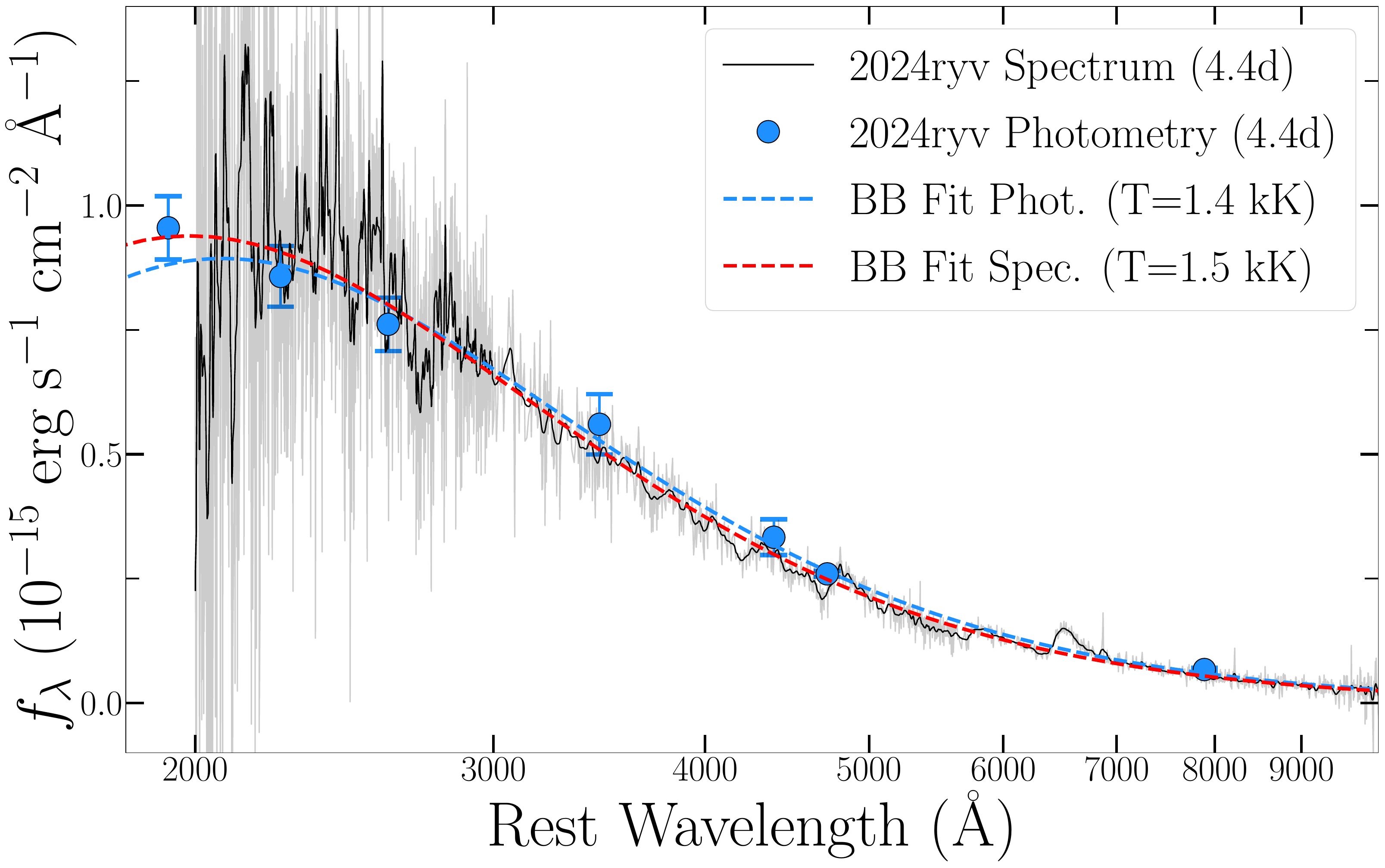}}
\subfigure{\includegraphics[width=0.49\textwidth]{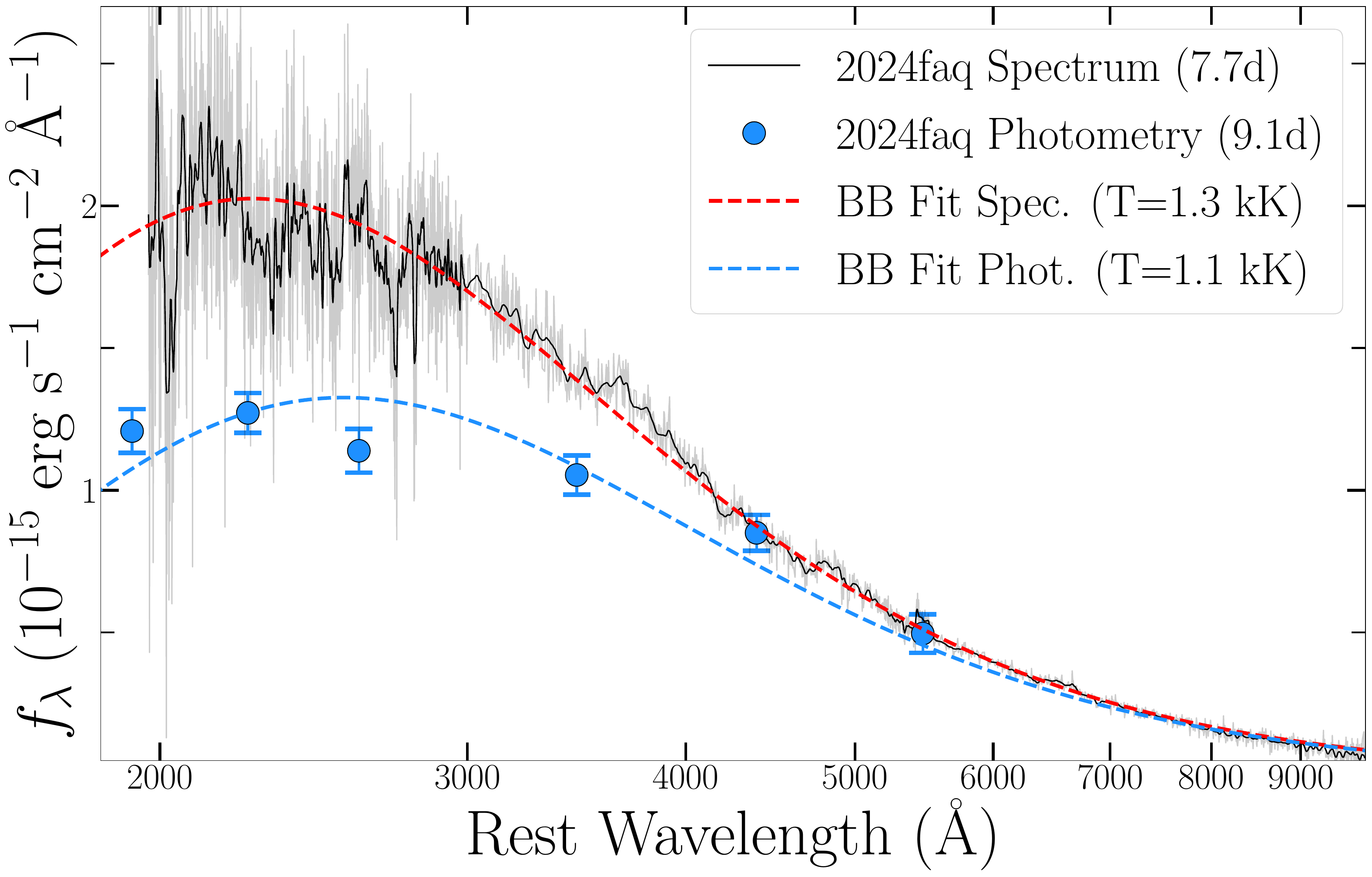}}\\
\subfigure{\includegraphics[width=0.49\textwidth]{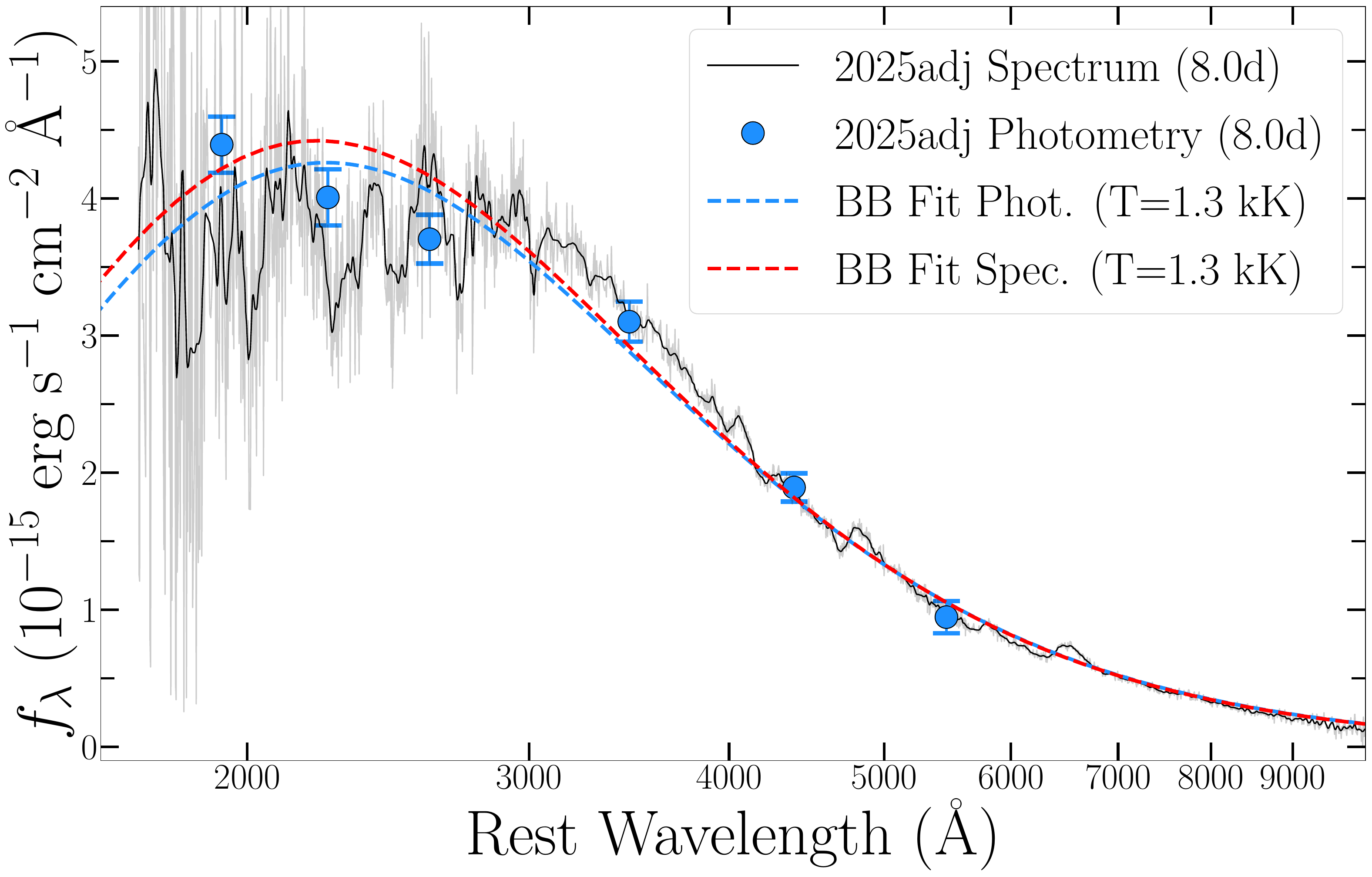}}
\subfigure{\includegraphics[width=0.49\textwidth]{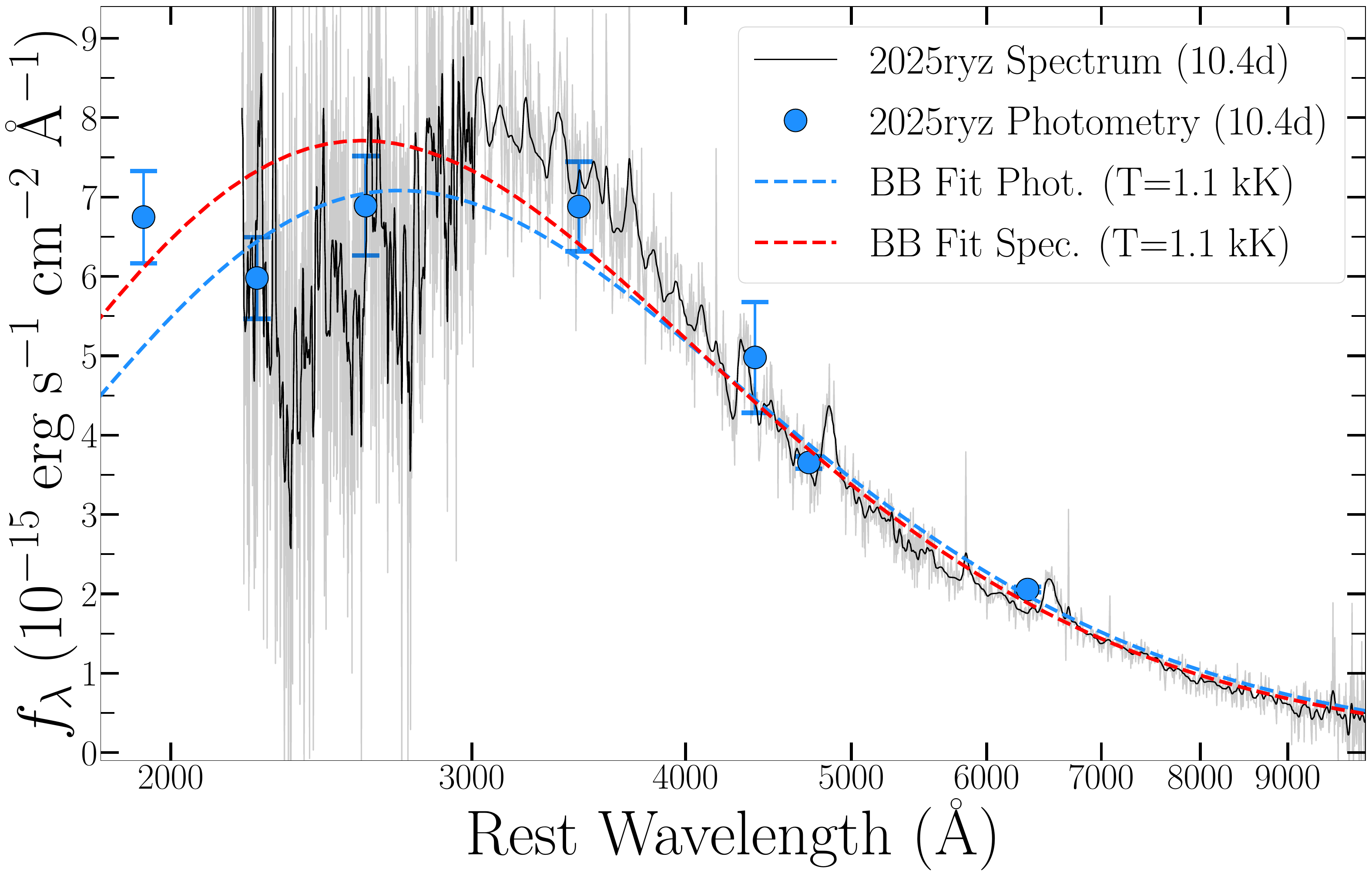}}\\
\caption{{\it Top:} UV/optical spectra (black) and photometry (blue circles) of sample SNe~IIb with best-fit blackbody models shown as dashed blue (photometric fit) and red (spectral fit) lines. {\it Middle/Bottom:} UV/optical spectra and photometry of sample SNe~IIP/L compared to best-fit blackbody models.  \label{fig:BB1} }
\end{figure*}

\begin{figure*}
\centering
\subfigure{\includegraphics[width=0.49\textwidth]{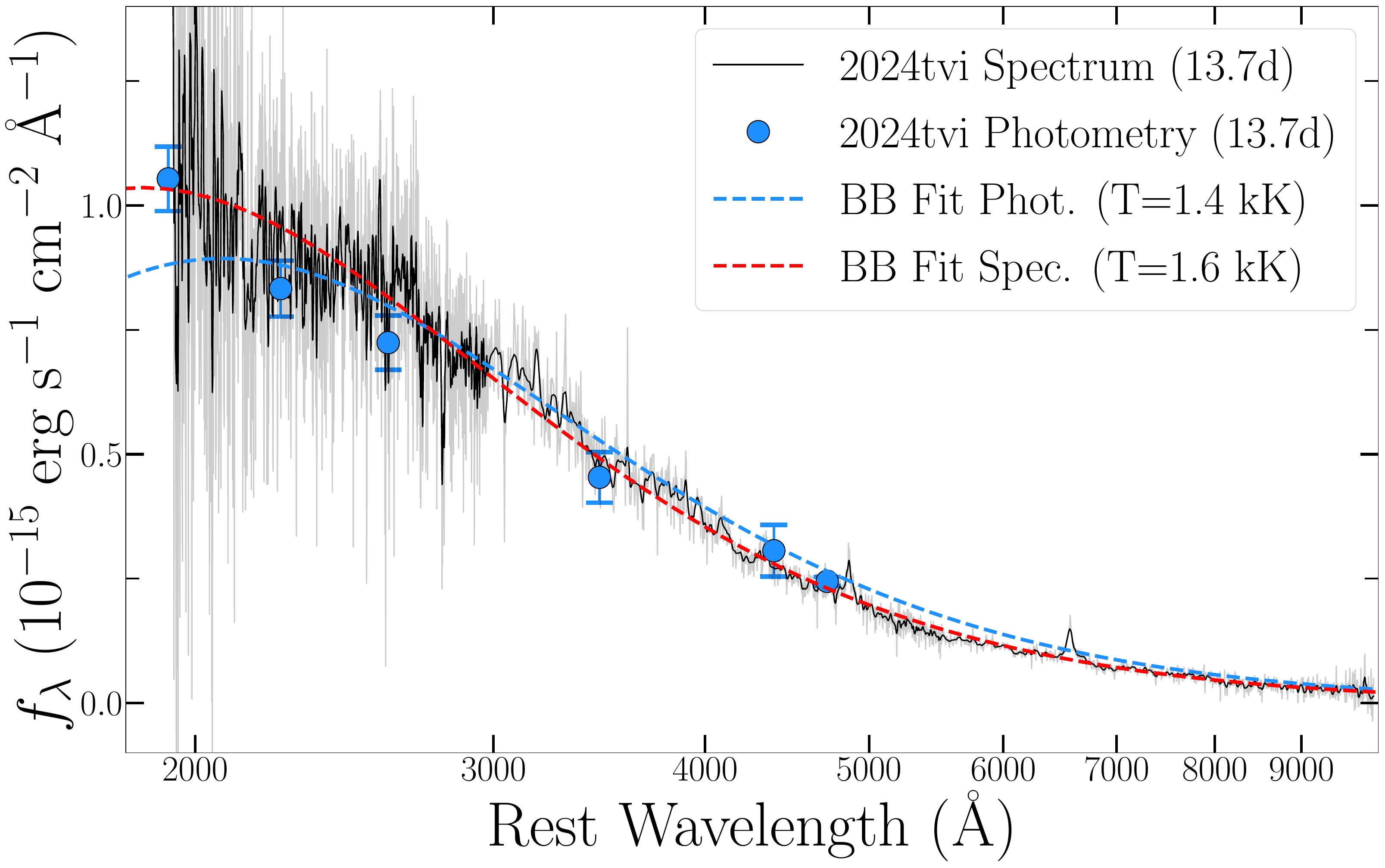}}
\subfigure{\includegraphics[width=0.49\textwidth]{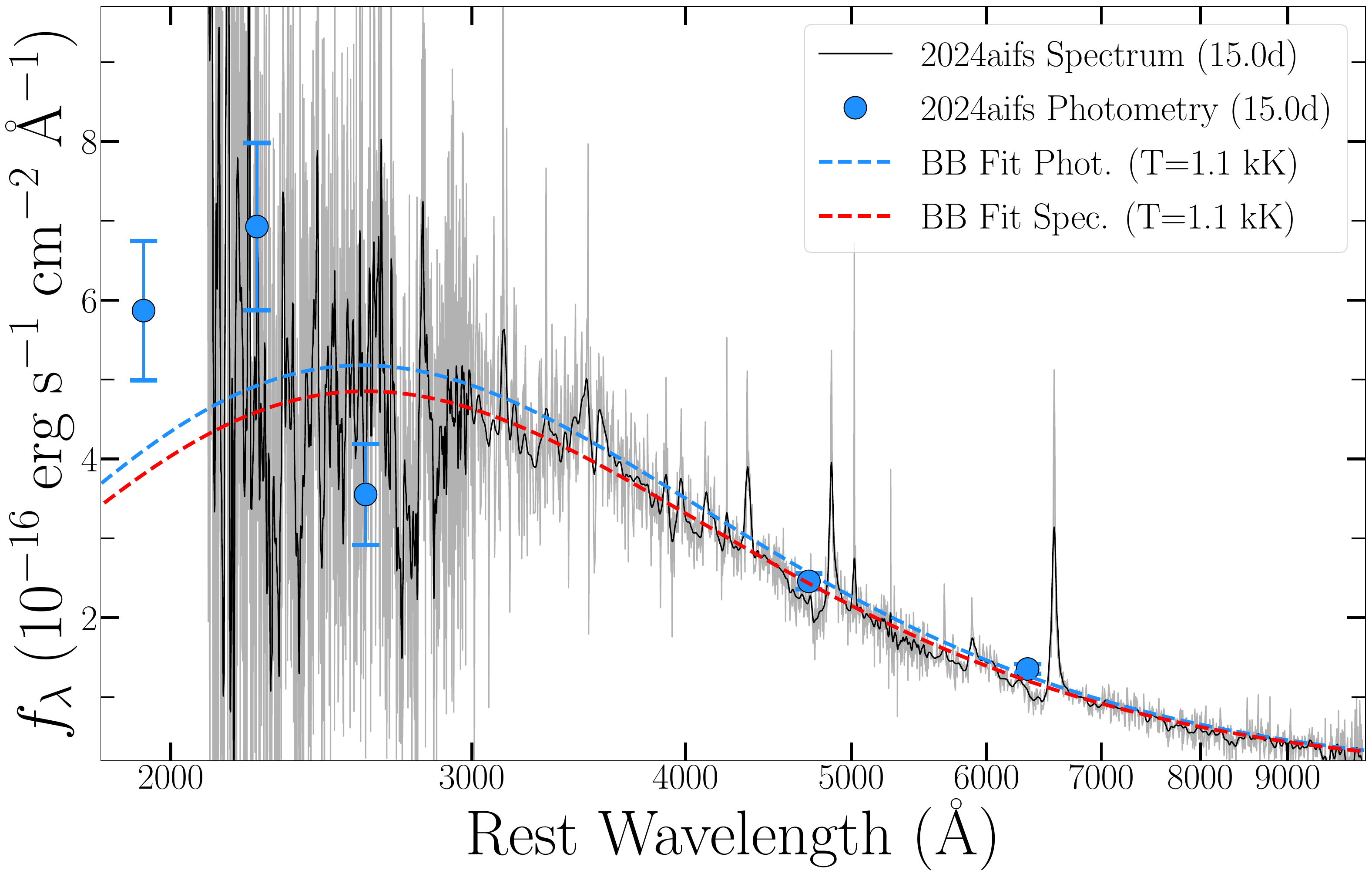}}\\
\caption{ UV/optical spectra (black) and photometry (blue circles) of SNe~IIn 2024tvi and 2024aifs with best-fit blackbody models shown as dashed blue (photometric fit) and red (spectral fit) lines.  \label{fig:BB2} }
\end{figure*}

\begin{figure*}
\centering
\subfigure{\includegraphics[width=0.49\textwidth]{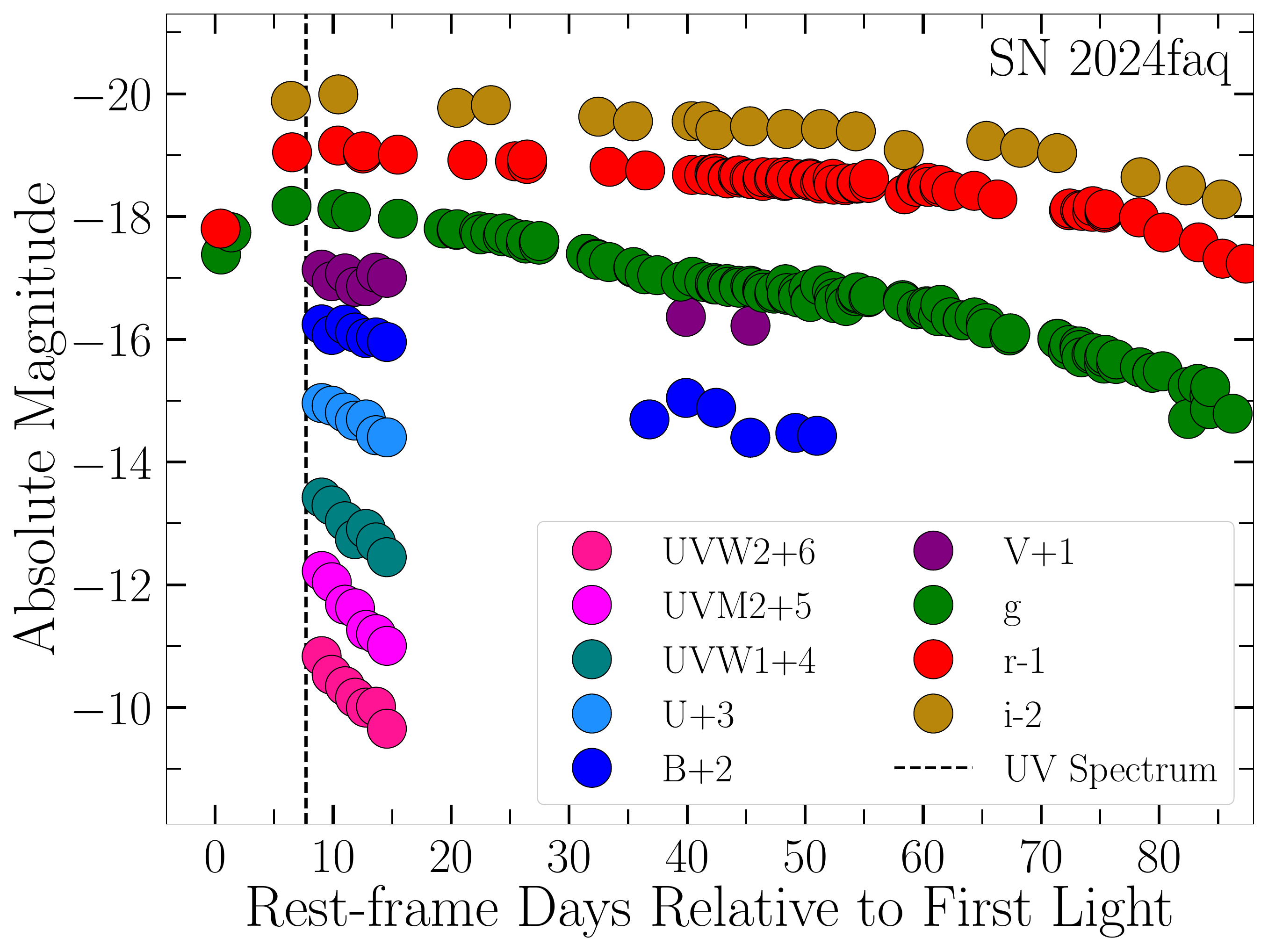}}
\subfigure{\includegraphics[width=0.47\textwidth]{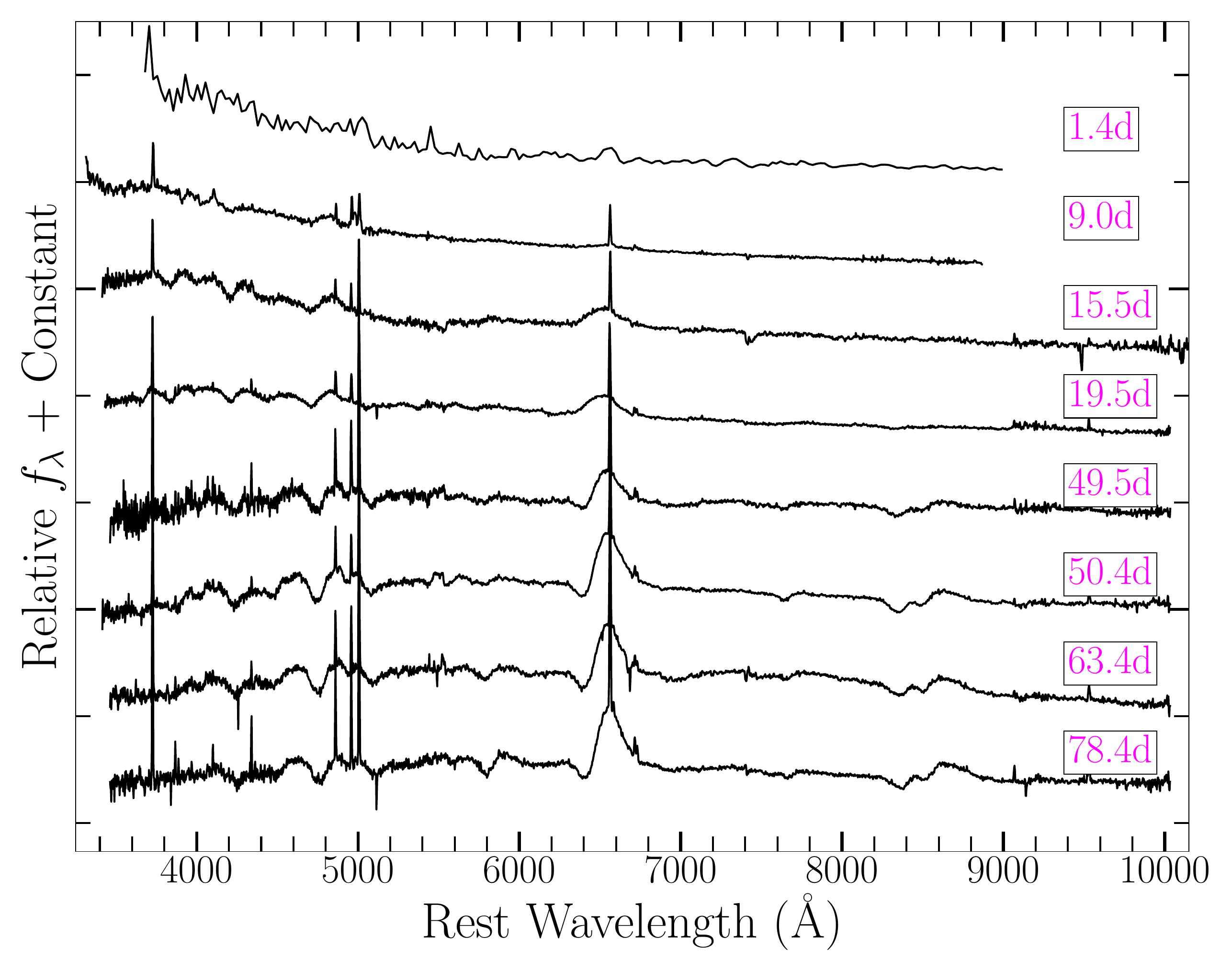}}\
\caption{Multi-color light curves (left) and optical spectral series (right) of SN~II 2024faq. Phase of UV spectrum marked as vertical dashed black line in light curve plot.    \label{fig:24faq} }
\end{figure*}

\begin{figure*}
\centering
\subfigure{\includegraphics[width=0.49\textwidth]{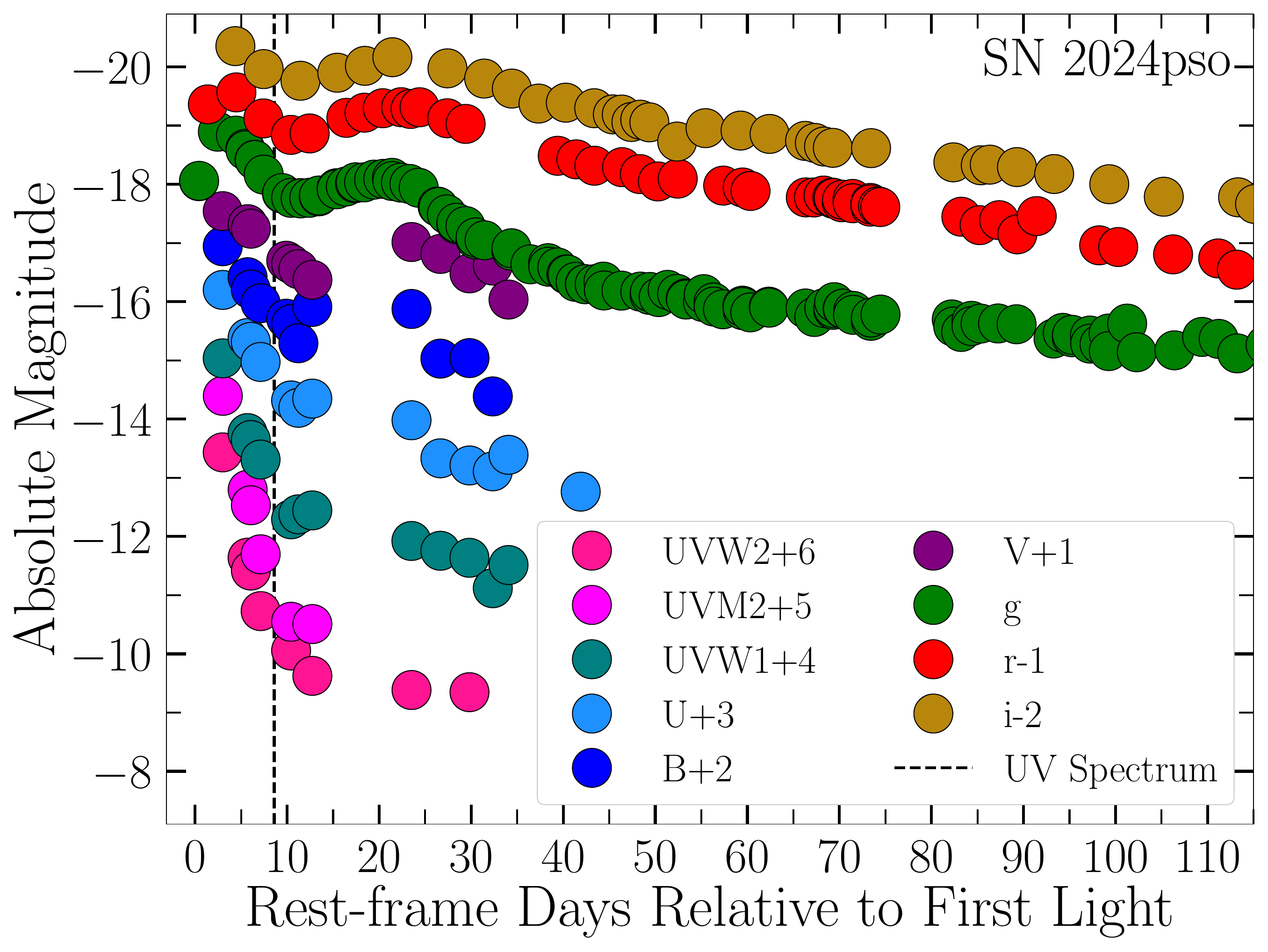}}
\subfigure{\includegraphics[width=0.47\textwidth]{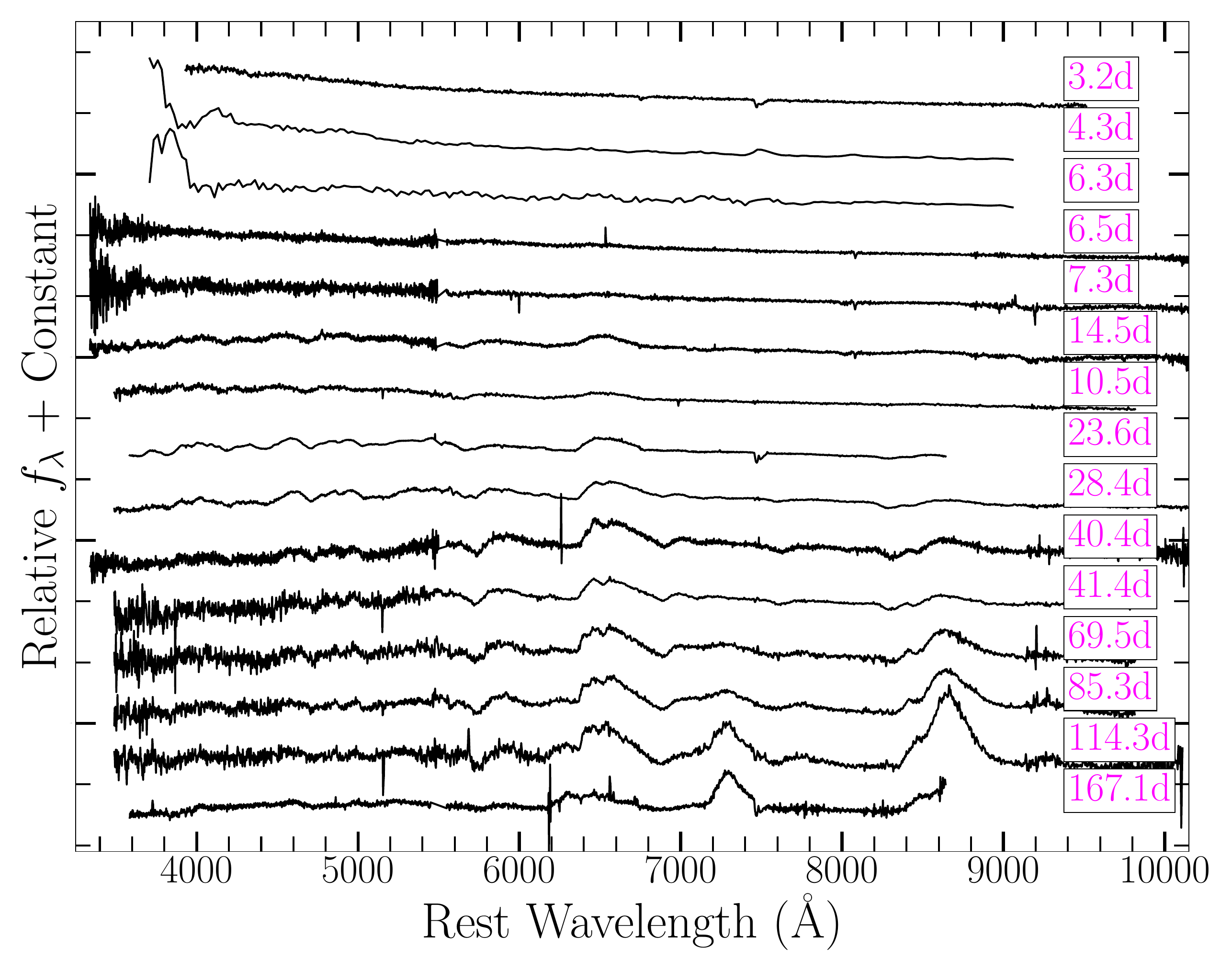}}\
\caption{Multi-color light curves (left) and optical spectral series (right) of SN~IIb 2024pso. Phase of UV spectrum marked as vertical dashed black line in light curve plot.    \label{fig:24pso} }
\end{figure*}

\begin{figure*}
\centering
\subfigure{\includegraphics[width=0.49\textwidth]{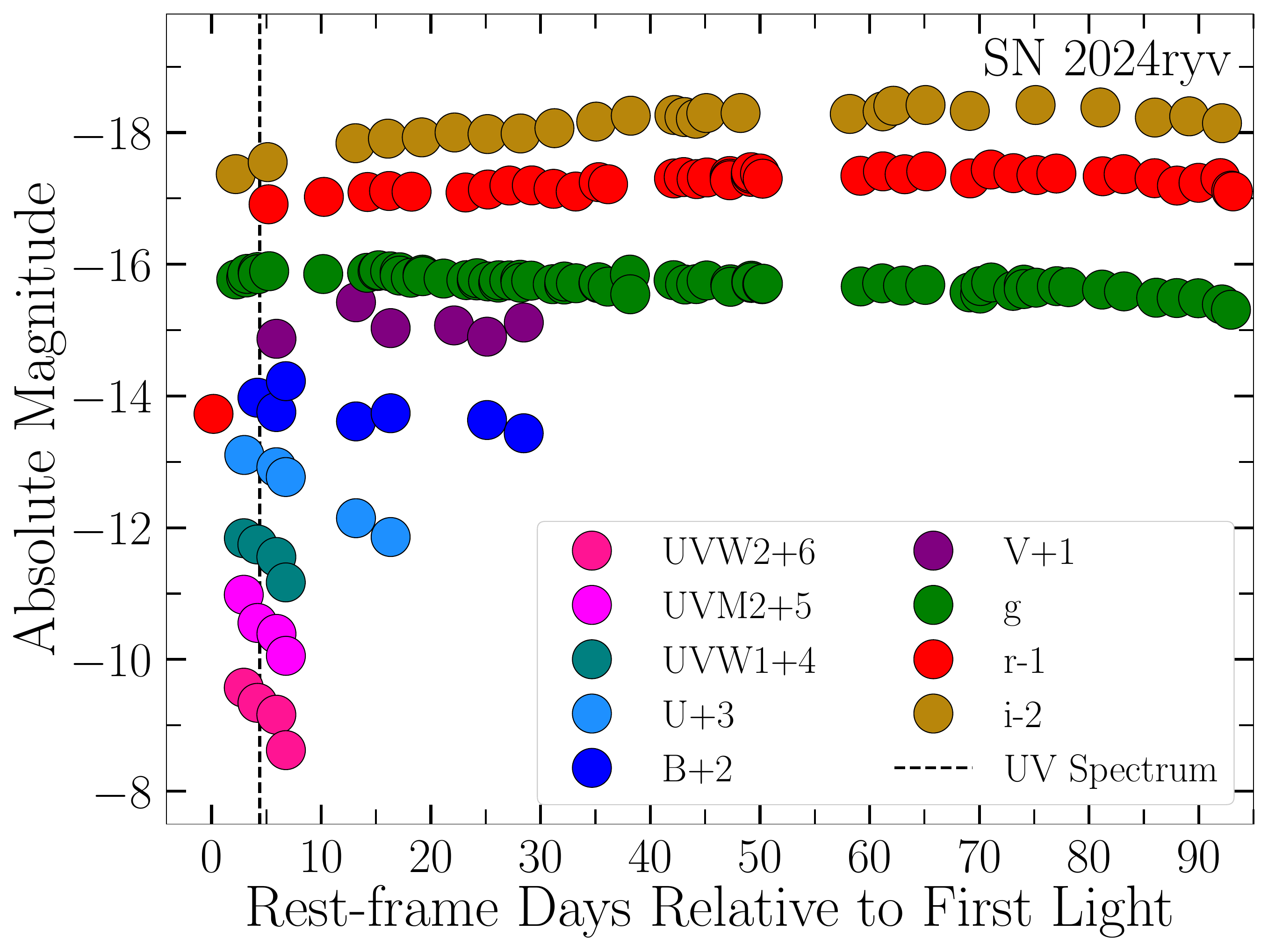}}
\subfigure{\includegraphics[width=0.47\textwidth]{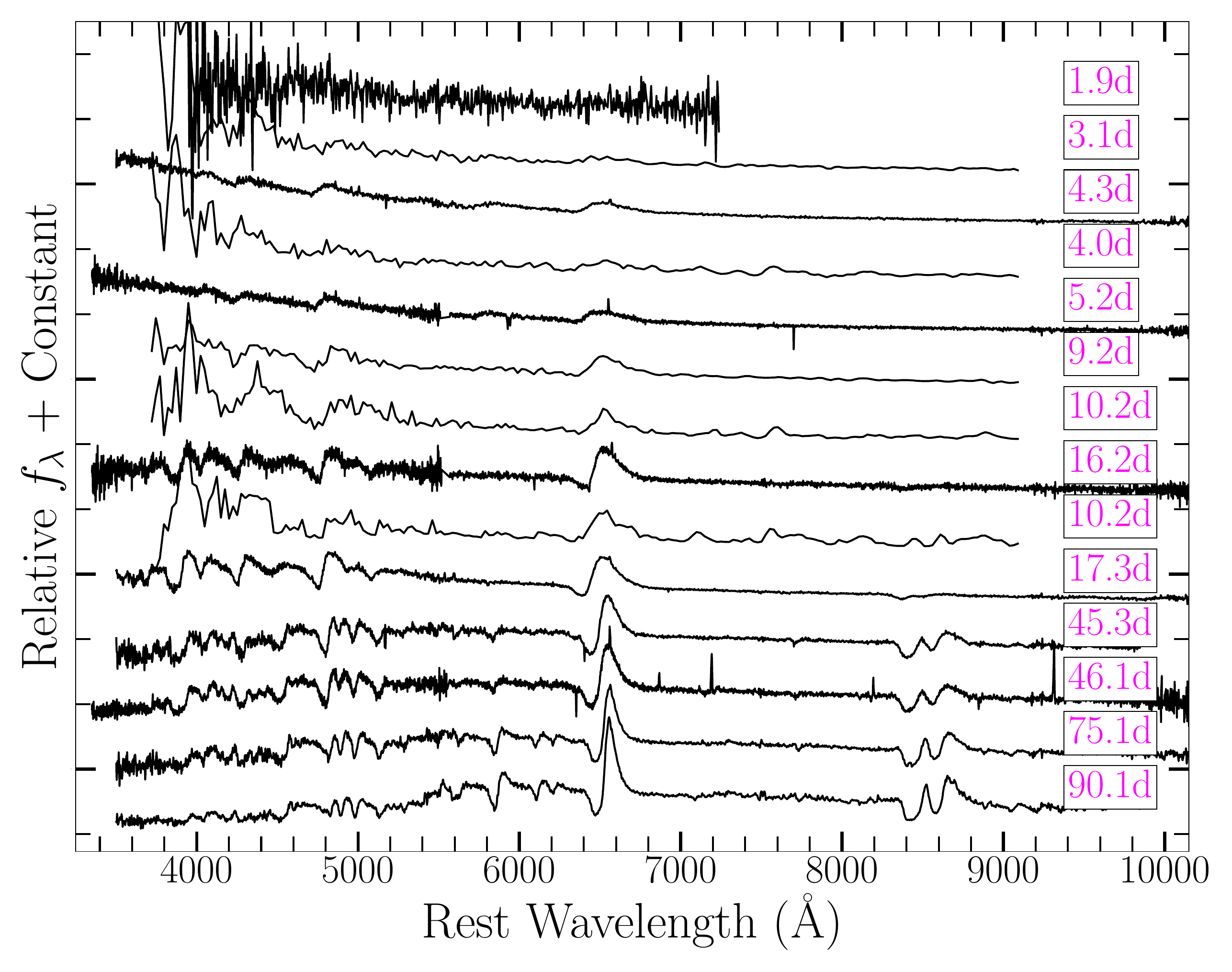}}\
\caption{Multi-color light curves (left) and optical spectral series (right) of SN~II 2024ryv. Phase of UV spectrum marked as vertical dashed black line in light curve plot.    \label{fig:24ryv} }
\end{figure*}

\begin{figure*}
\centering
\subfigure{\includegraphics[width=0.49\textwidth]{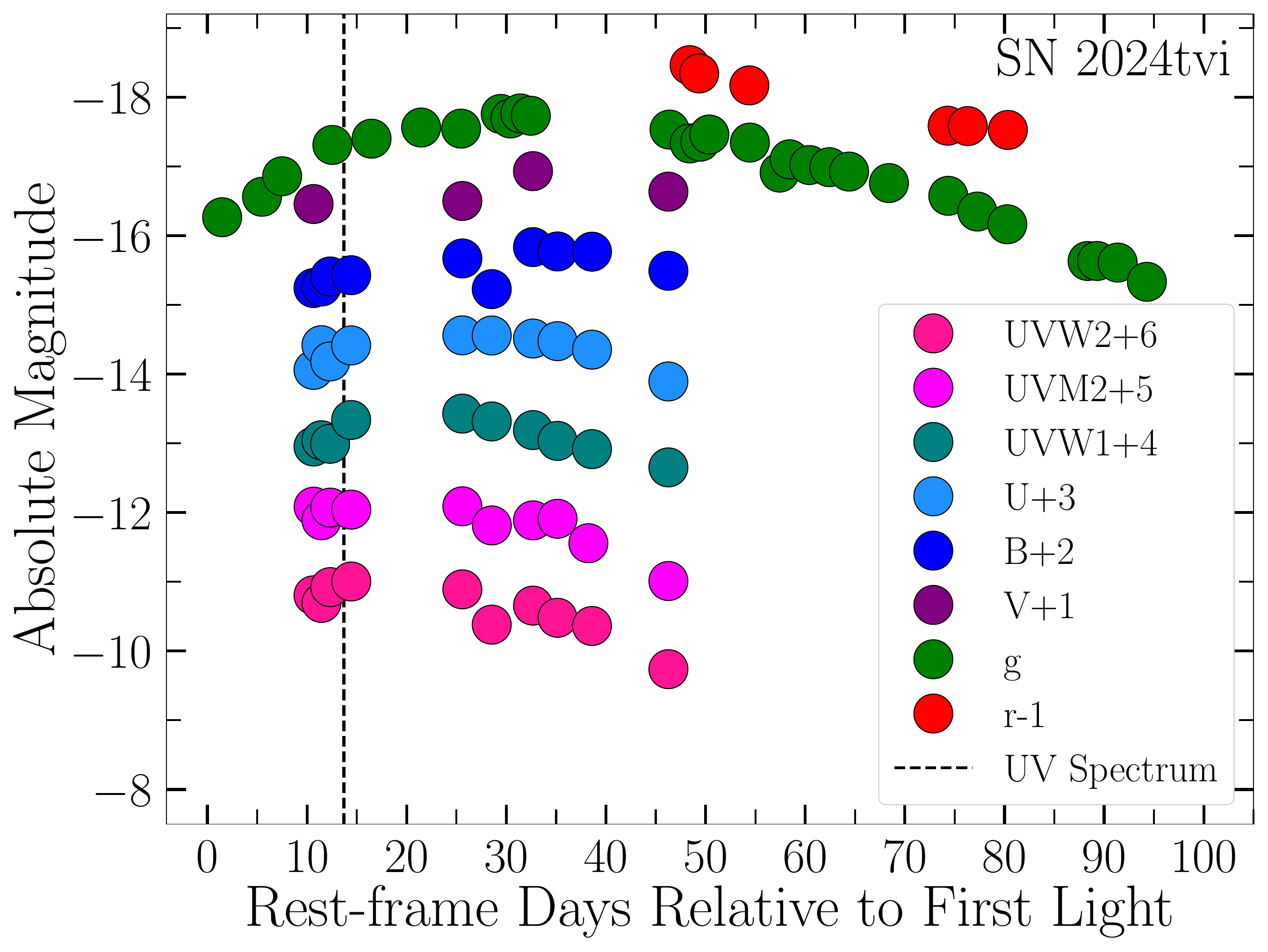}}
\subfigure{\includegraphics[width=0.47\textwidth]{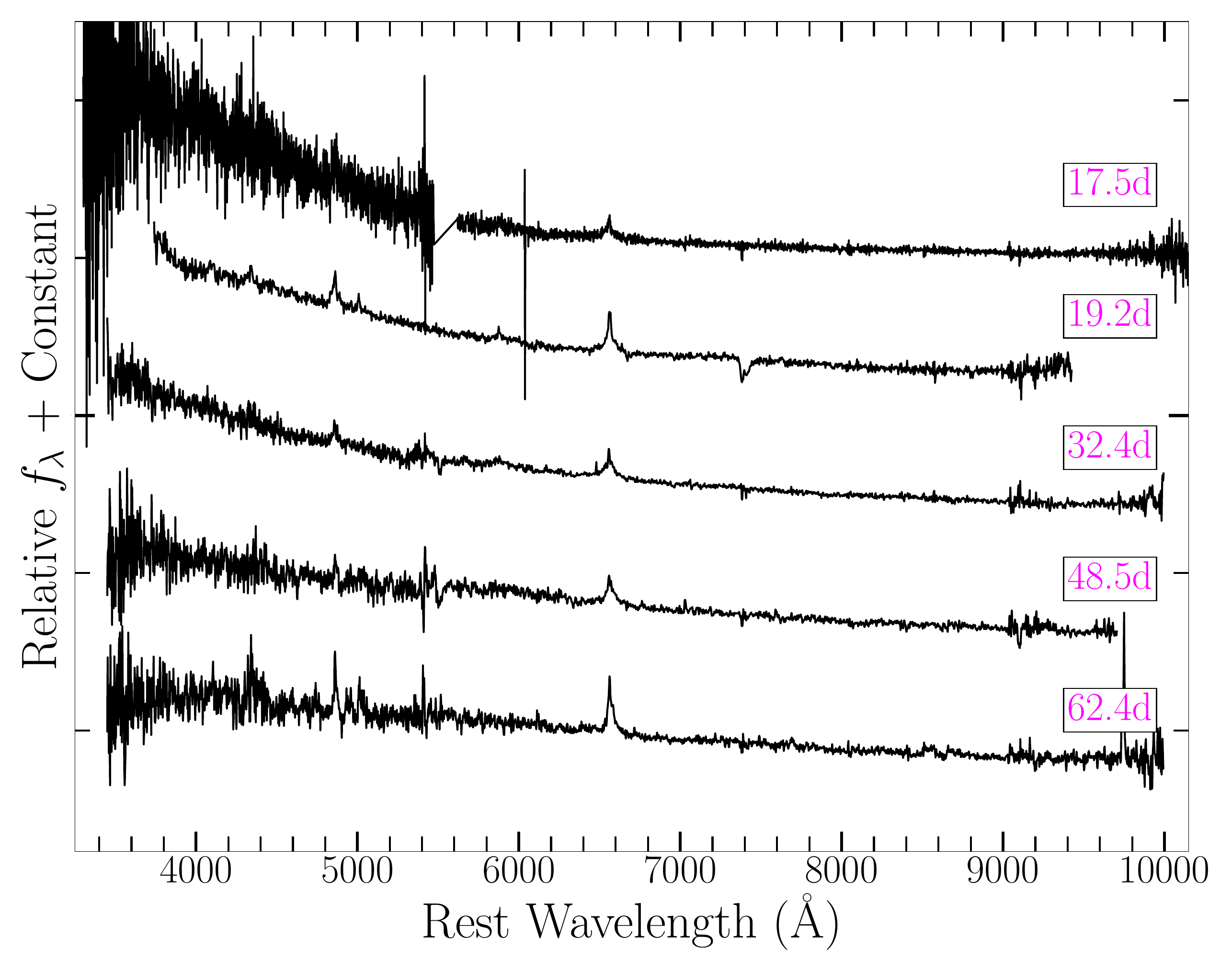}}\
\caption{Multi-color light curves (left) and optical spectral series (right) of SN~IIn 2024tvi. Phase of UV spectrum marked as vertical dashed black line in light curve plot.    \label{fig:24tvi} }
\end{figure*}

\begin{figure*}
\centering
\subfigure{\includegraphics[width=0.49\textwidth]{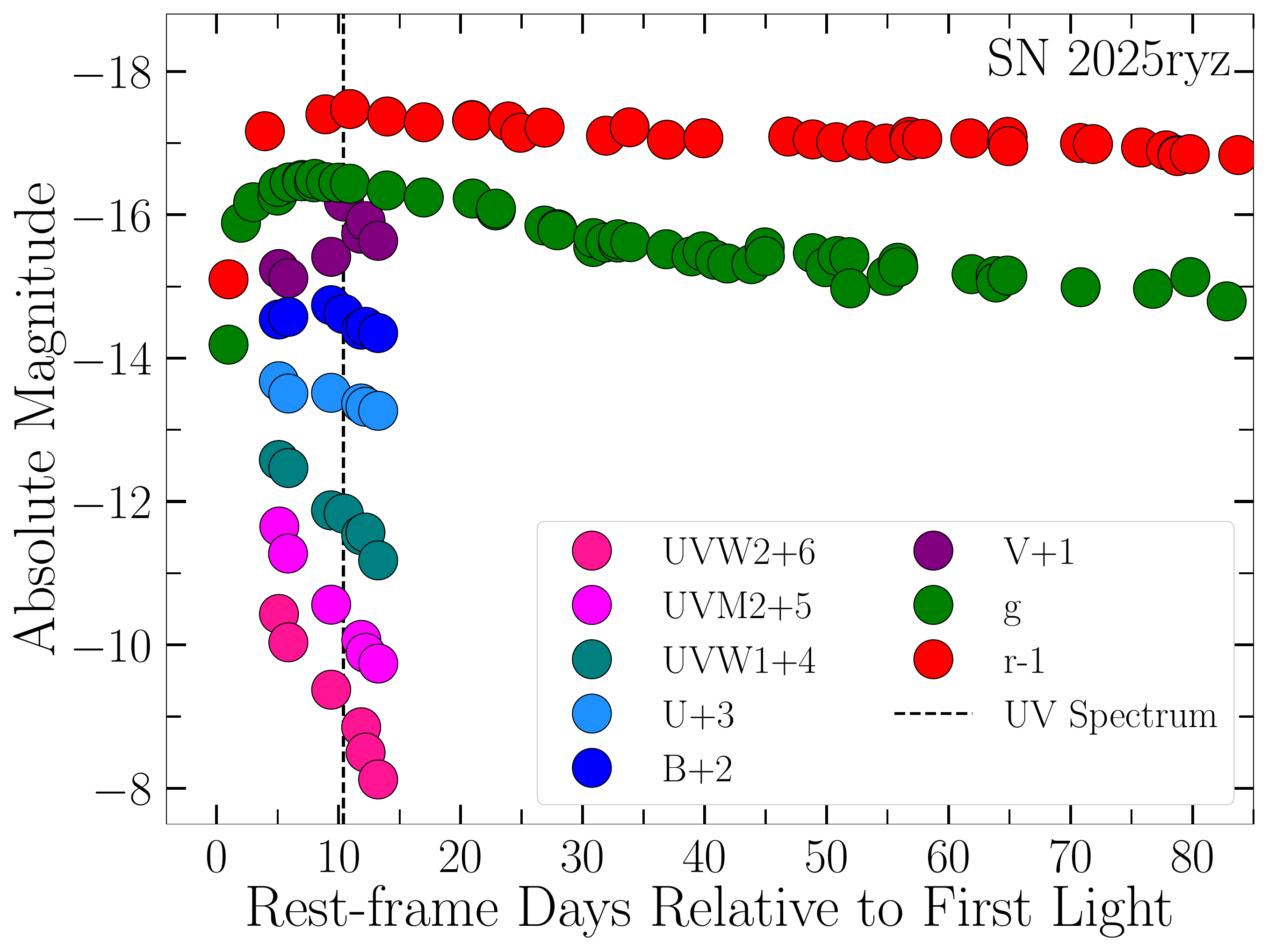}}
\subfigure{\includegraphics[width=0.47\textwidth]{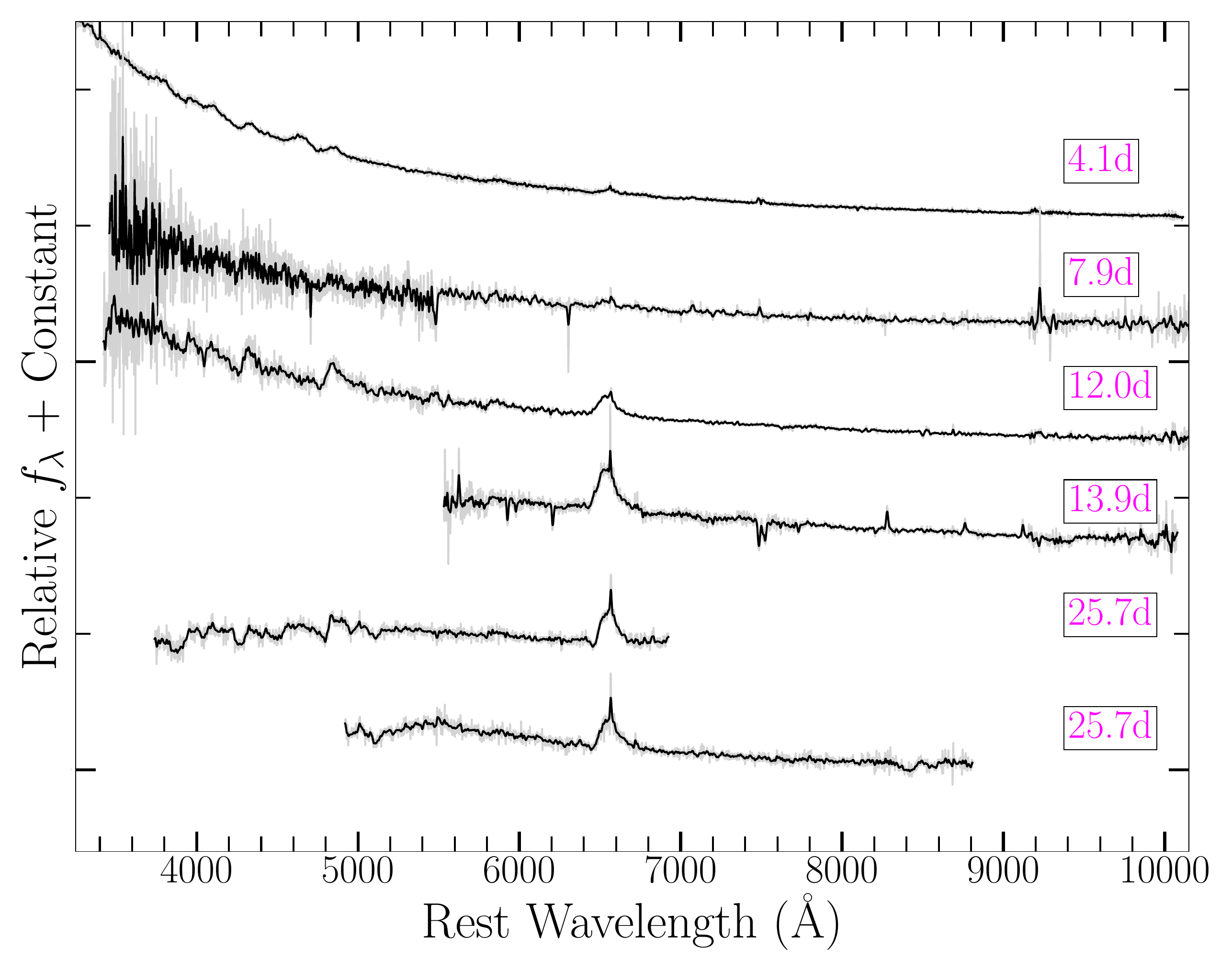}}\
\caption{Multi-color light curves (left) and optical spectral series (right) of SN~II 2025ryz. Phase of UV spectrum marked as vertical dashed black line in light curve plot.    \label{fig:25ryz} }
\end{figure*}

\begin{figure*}
\centering
\subfigure{\includegraphics[width=0.49\textwidth]{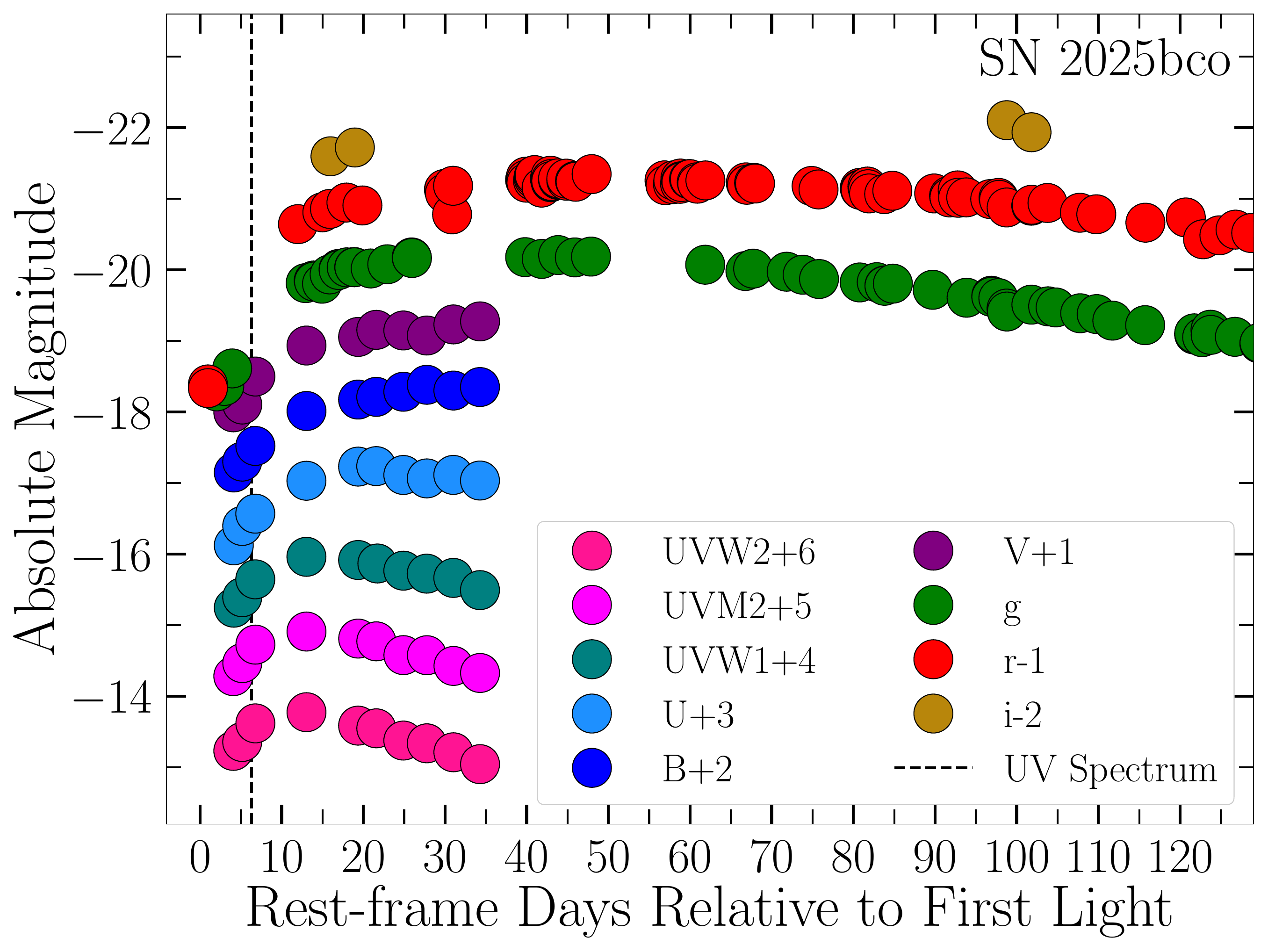}}
\subfigure{\includegraphics[width=0.47\textwidth]{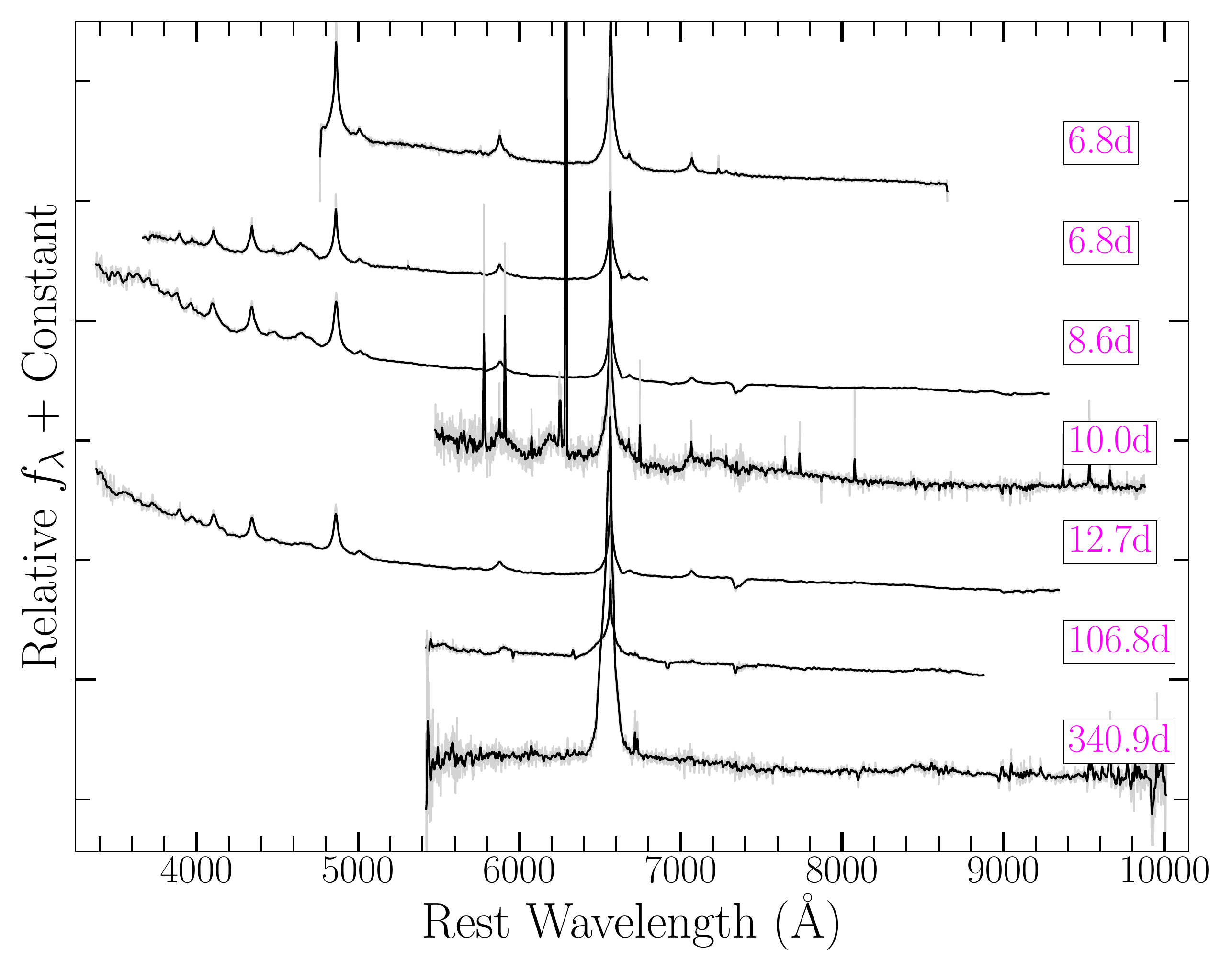}}\
\caption{Multi-color light curves (left) and optical spectral series (right) of SN~IIn 2025bco. Phase of UV spectrum marked as vertical dashed black line in light curve plot.    \label{fig:25bco} }
\end{figure*}

\begin{figure*}
\centering
\subfigure{\includegraphics[width=0.49\textwidth]{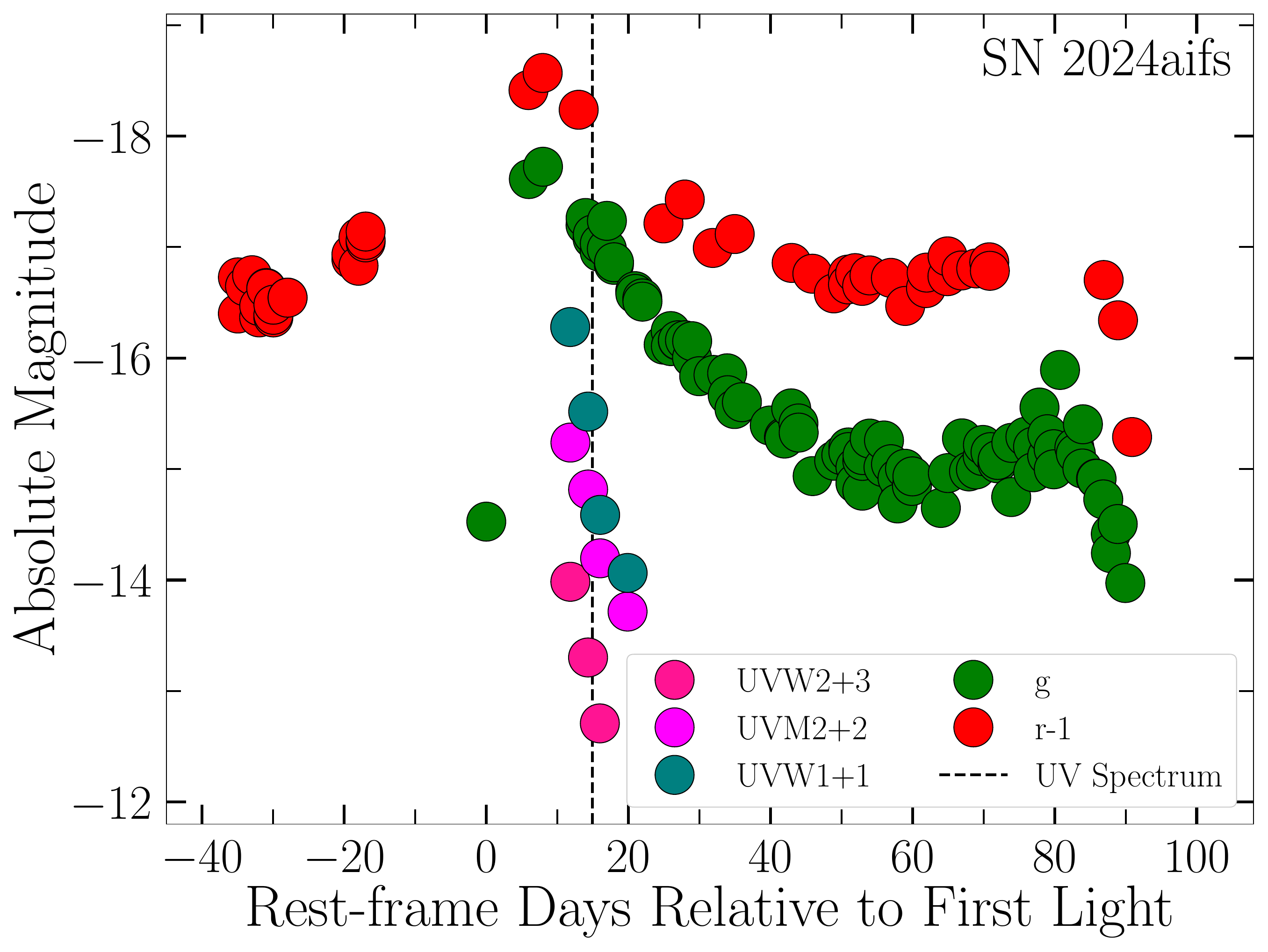}}
\subfigure{\includegraphics[width=0.47\textwidth]{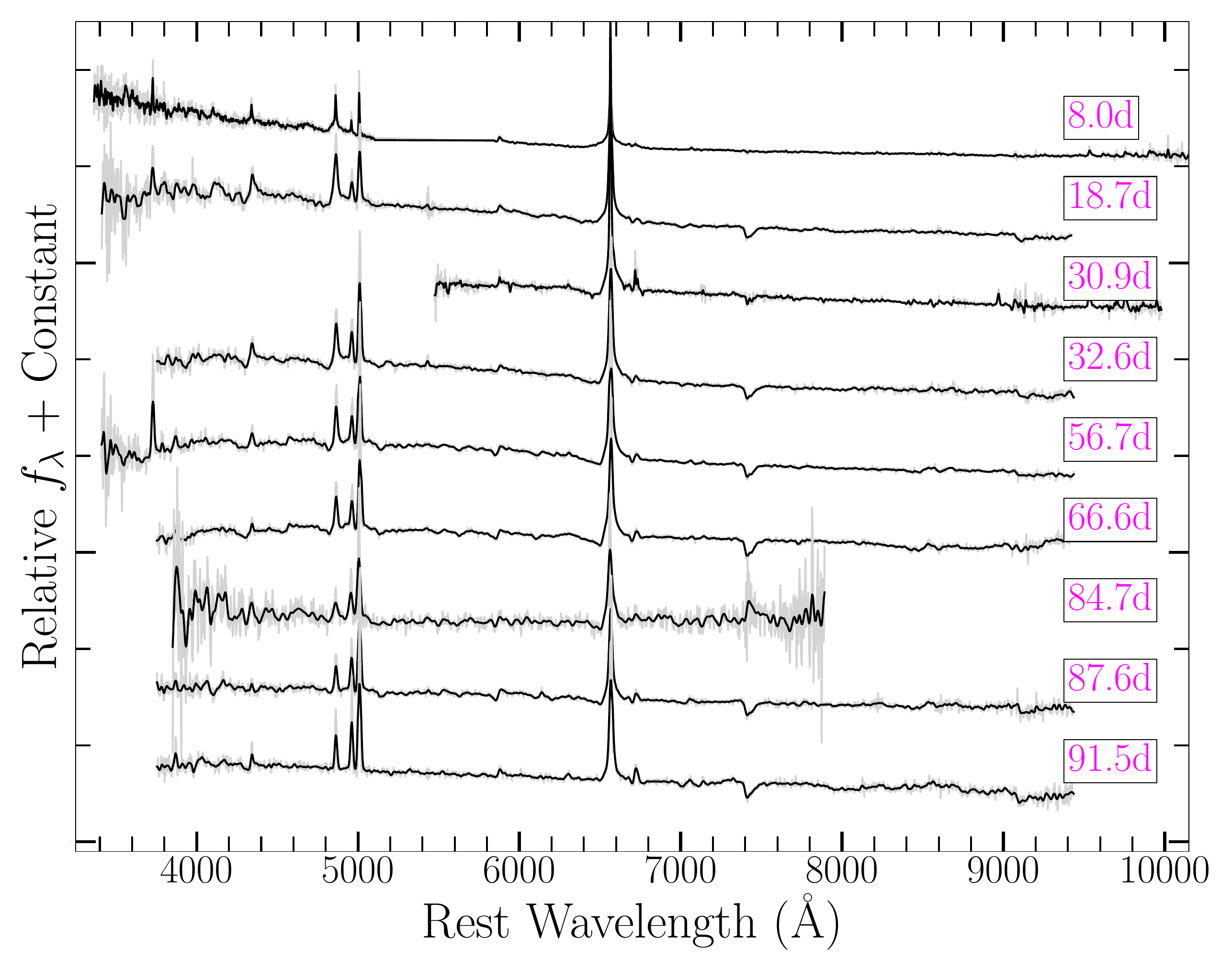}}\
\caption{Multi-color light curves (left) and optical spectral series (right) of SN~IIn 2024aifs. Phase of UV spectrum marked as vertical dashed black line in light curve plot.    \label{fig:24aifs} }
\end{figure*}

\begin{figure*}
\centering
\subfigure{\includegraphics[width=0.49\textwidth]{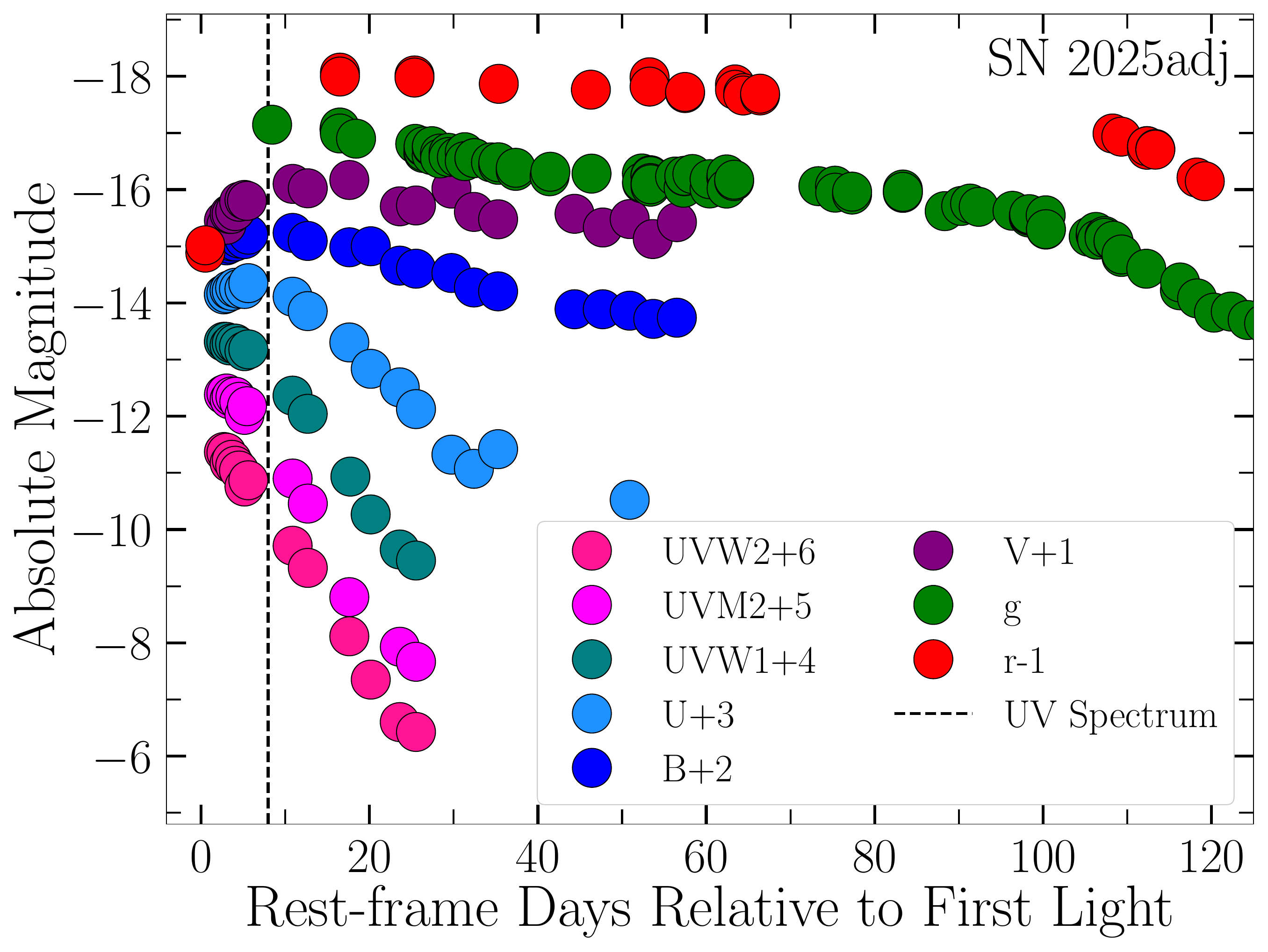}}
\subfigure{\includegraphics[width=0.47\textwidth]{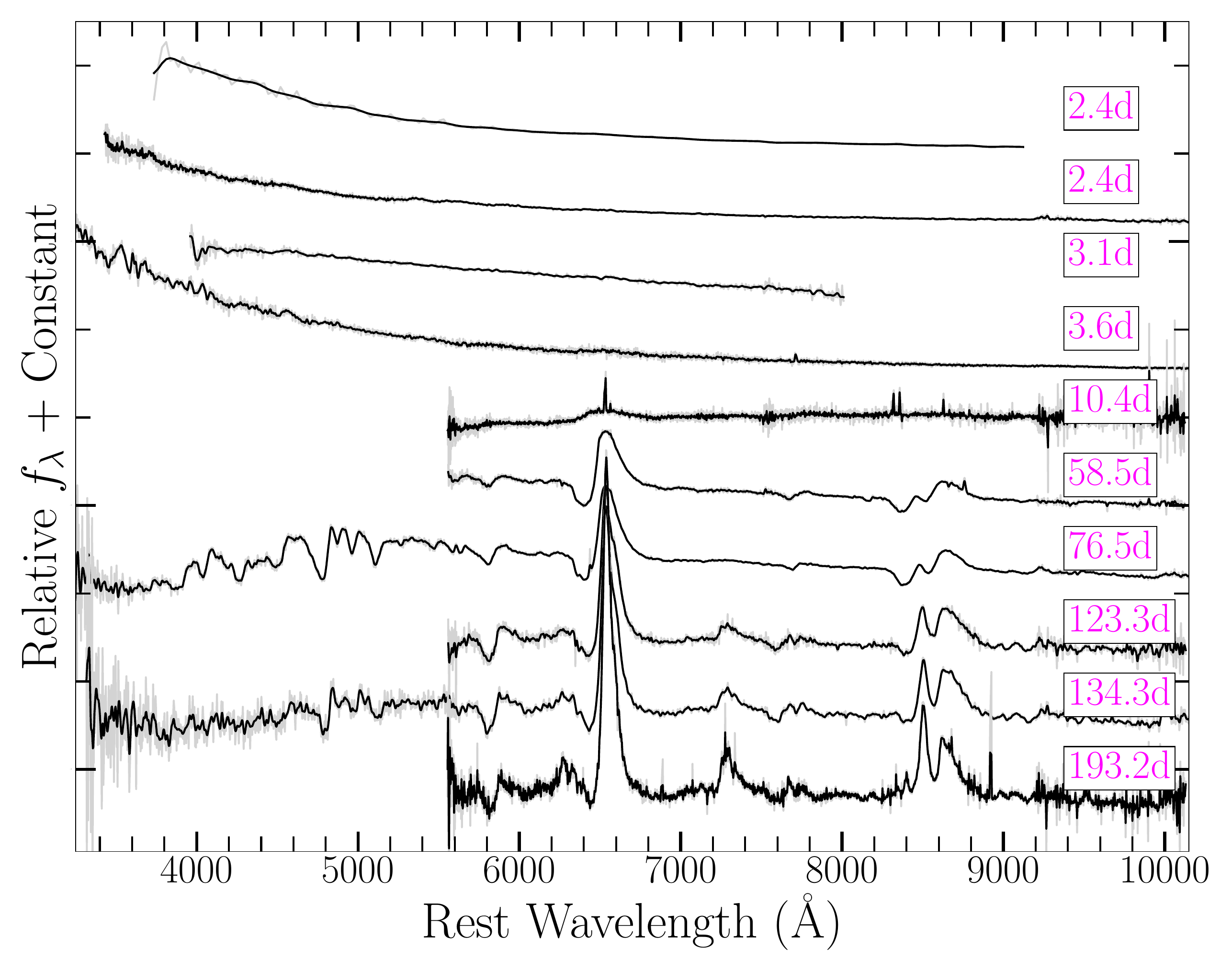}}\
\caption{Multi-color light curves (left) and optical spectral series (right) of SN~II 2025adj. Phase of UV spectrum marked as vertical dashed black line in light curve plot.    \label{fig:25adj} }
\end{figure*}

\begin{figure*}
\centering
\subfigure{\includegraphics[width=0.49\textwidth]{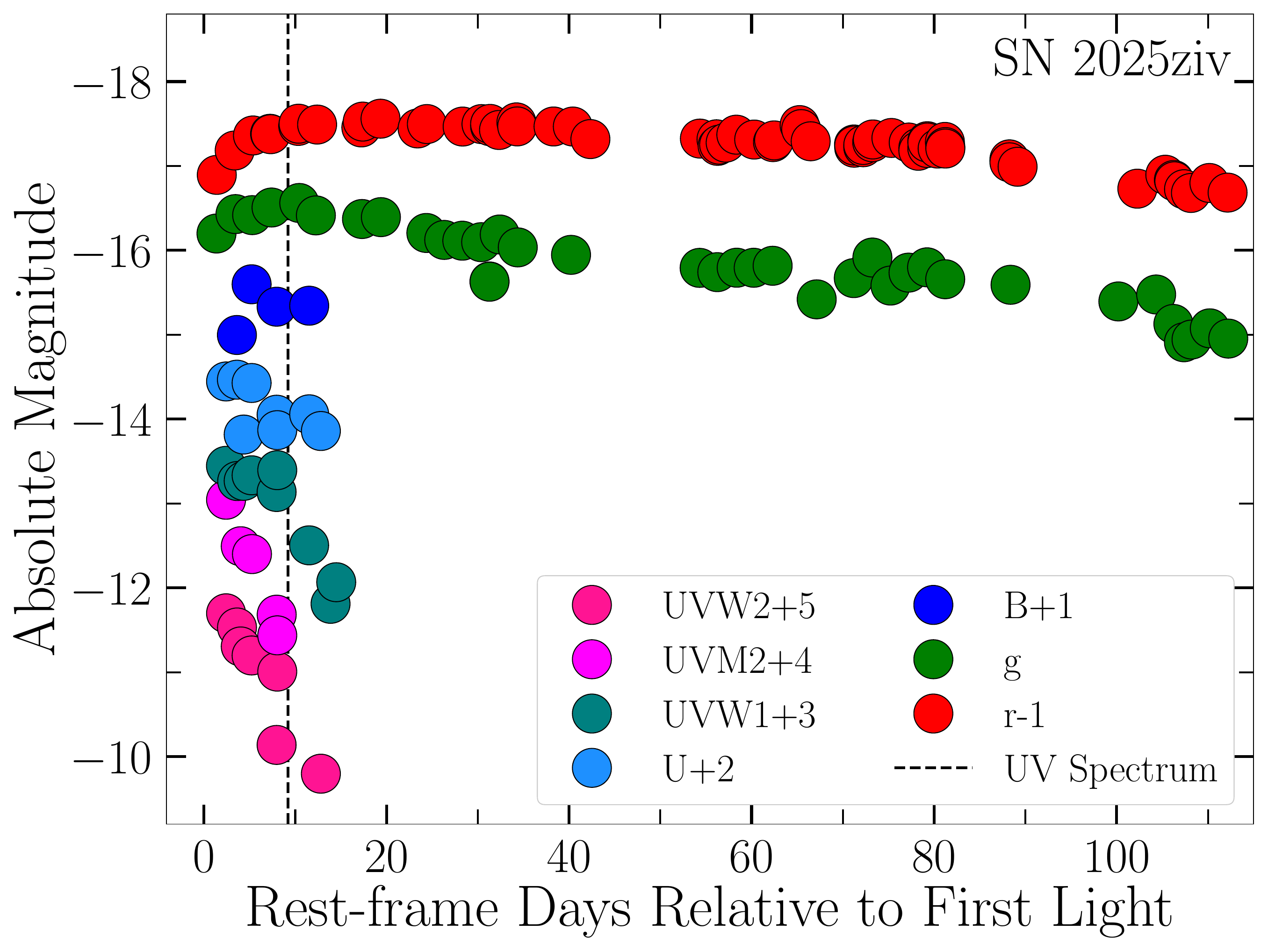}}
\subfigure{\includegraphics[width=0.47\textwidth]{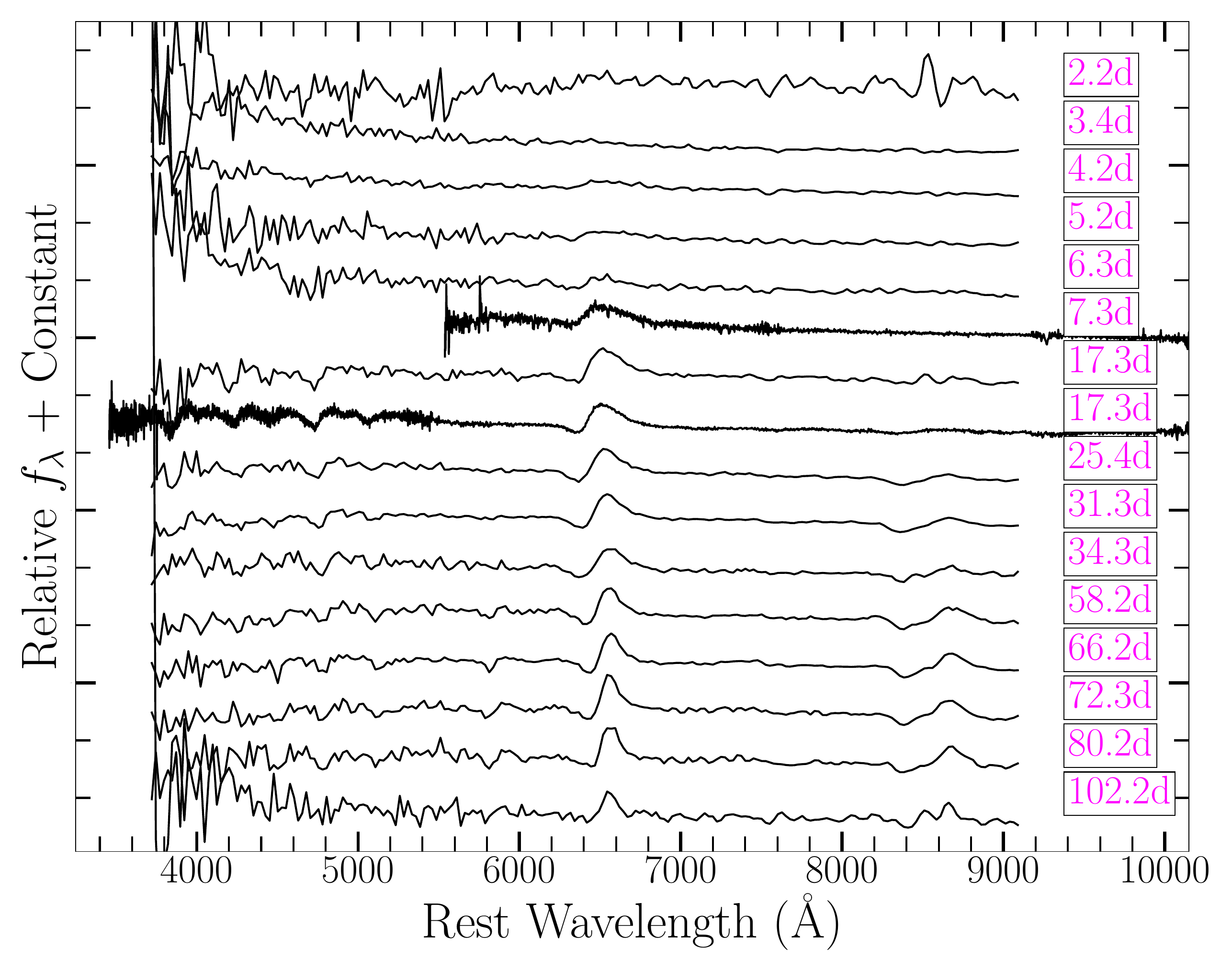}}\
\caption{Multi-color light curves (left) and optical spectral series (right) of SN~II 2025ziv. Phase of UV spectrum marked as vertical dashed black line in light curve plot.    \label{fig:25ziv} }
\end{figure*}

Here we present SN properties for all sample objects in Table \ref{tab:sample_specs}. Logs of UV/optical/NIR spectroscopic observations of all sample objects are presented in Table \ref{tab:spec_all}. Blackbody model fits to UVOIR SEDs are shown in Figures \ref{fig:BB1}--\ref{fig:BB2}. All multicolor light curves and spectral sequences for each previously unpublished sample object are shown in Figures \ref{fig:24faq} - \ref{fig:25ziv}.

\renewcommand\thetable{A\arabic{table}} 
\setcounter{table}{0}

\begin{deluxetable*}{ccccccccc}[h!]
\tablecaption{Sample Object Properties \label{tab:sample_specs}}
\tablecolumns{9}
\tablewidth{0pt}
\tablehead{\colhead{Name} & \colhead{Time of First Light} & \colhead{SN Type} & \colhead{Redshift} & \colhead{Distance} &  \colhead{$E(B-V)_{\textrm{MW}}$} &  \colhead{$E(B-V)_{\textrm{host}}$} & { UV Spectral Epochs} & {UV Instrument} \\
\colhead{} & \colhead{(MJD)} & \colhead{} & \colhead{$z$} & \colhead{(Mpc)} & \colhead{(mag)} & \colhead{(mag)} & \colhead{(days)} }
\startdata
2024faq & $60397.0 \pm 0.5$ & II & 0.026 & 114.7 & 0.012 & $<0.01$ & 7.7 & HST STIS \\%
2024iss & $60442.3 \pm 0.03$ & IIb & 0.003238 & 14.0 & 0.01 & $0.019 \pm 0.004$ & 7.0 & HST STIS\\
2024ryv & $60532.8 \pm 0.5$ & II & 0.0147 & 64.3 & 0.06 & $<$0.002 & 4.7 & HST STIS\\
2024tvi & $60545.8 \pm 1.5$ & IIn & 0.0296 & 131.0 & 0.02 & $<$0.006 & 13.7 & HST STIS\\
2024pso & $60509.0 \pm 0.4$ & IIb & 0.0183 & 80.3 & 0.05 & $0.19 \pm 0.046$ & 8.6 & HST STIS\\
2024aifs & $60789.5 \pm 3.0$ & IIn & 0.0263 & 116.1 & 0.11 & $<$0.007 & 15.0 & HST STIS\\
2025ryz & $60877.44 \pm 1.0$ & II & 0.0162 & 71.0 & 0.02 & $<$0.02 & 10.4 & HST STIS\\
2025bco & $60713.5 \pm 1.0$ & IIn & 0.03633 & 161.5 & 0.018 & $<$0.002 & 6.3 & HST STIS\\
2025adj & $60698.0 \pm 0.5$ & II & 0.011 & 48.0 & 0.02 & $<$0.005 & 8.0 & HST STIS\\
2025ziv & $60949.0 \pm 0.5$ & II & 0.0147 & 64.3 & 0.27 & $<$0.02 & 9.2 & HST STIS\\
2025whv & $60913.8 \pm 0.5$ & SLSN-I & 0.11 & 515.3 & 0.06 & $<$0.01 & 15.7 & HST STIS\\
\hline 
1987A & $46849.62 \pm 1.0$ & II & -- & 0.0513 & 0.0 & $0.19\pm0.2$ & 1.7, 2.9, 3.5, 4.6, 5.8 & IUE\\
1993J & $49073.5 \pm 0.8$ & IIb & -- & 3.63 & 0.067 & $0.10\pm0.11$ & 2.2, 2.6, 8.9 & IUE\\
1998S & $50871.2 \pm 3.5$ & II & 0.002795 & 17 & 0.02 & $0.20\pm0.02$ & 14 & HST STIS\\
1999em & $51475.4 \pm 2.0$ & II & 0.00239 & 11.7 & 0.035 & $0.10\pm0.05$ & 13 & HST STIS\\
2013df & $56449.5 \pm 0.9$ & IIb & 0.0024 & 16.6 & 0.017 & $0.08\pm0.02$ & 13 & HST STIS\\
2020fqv & $58938.9 \pm 0.2$ & II & 0.0075 & 17.3 & 0.029 & $0.81\pm0.18$ & 3.4 & HST STIS\\
2021yja & $59464.4 \pm 0.06$ & II & 0.00568 & 23.4 & 0.019 & $0.085\pm0.015$ & 10, 21 & HST STIS\\
2021yja & --& -- & -- & -- & -- & -- & 4.5, 7 & Swift-UVOT\\
2022wsp & $59854.58 \pm 0.05$ & II & 0.00932 & 25.0 & 0.05 & $0.43 \pm 0.09$ & 10, 20 & HST STIS\\
2022acko & $59918.17 \pm 0.4$ & II & 0.00526 & 19.0 & 0.024 & $0.03 \pm 0.01$ & 5.2,6.0,7.3 & HST STIS\\
2023ixf & $60082.83 \pm 0.02$ & II & 0.0008 & 6.9 & 0.008 & $0.033 \pm 0.01$ & 3.6, 4.7, 5.6, 8.7 & HST STIS\\
\enddata
\end{deluxetable*}

\startlongtable
\begin{deluxetable*}{cccccccc}
\tablecaption{Ultraviolet/Optical/Infrared Spectroscopy \label{tab:spec_all}}
\tablecolumns{8}
\tablewidth{0pt}
\tablehead{\colhead{Name} & \colhead{UT Date} & \colhead{MJD} &
\colhead{Phase\tablenotemark{a}} &
\colhead{Telescope} & \colhead{Instrument} & \colhead{Wavelength Range} & \colhead{Data Source}\\
\colhead{} & \colhead{} & \colhead{} & \colhead{[days]} & \colhead{} & \colhead{} & \colhead{[$\mu$m]} & \colhead{}
}
\startdata
    2024faq & 2024-03-29 08:30:36 & 60398.4 & 1.4 & P60 & SEDM & 0.38 - 0.92 & ZTF\\
    2024faq & 2024-04-04 15:43:12 & 60404.7 & 7.7 & HST & STIS & 0.17 - 1.03 & HST GO-17507 \\
    2024faq & 2024-04-06 00:00:00 & 60406.0 & 9.0 & UH88 & SNIFS & 0.33 - 0.97 & SCAT\\
    2024faq & 2024-04-12 11:22:01 & 60412.5 & 15.5 & Shane & Kast & 0.33 - 1.1 & UCSC\\
    2024faq & 2024-04-16 11:26:42 & 60416.5 & 19.5 & Shane & Kast & 0.36 - 1.03 & TReX\tablenotemark{b}\\
    2024faq & 2024-05-16 11:15:55 & 60446.5 & 49.5 & Shane & Kast & 0.36 - 1.03 & TReX\tablenotemark{b}\\
    2024faq & 2024-05-17 08:47:02 & 60447.4 & 50.4 & Shane & Kast & 0.33 - 1.1 & UCSC\\
    2024faq & 2024-05-30 07:55:24 & 60460.3 & 63.3 & Shane & Kast & 0.36 - 1.03 & TReX\tablenotemark{b}\\
    2024faq & 2024-06-14 08:58:16 & 60475.4 & 78.4 & Shane & Kast & 0.36 - 1.03 & TReX\tablenotemark{b}\\
    2024ryv & 2024-08-13 01:07:41 & 60535.0 & 2.2 & LT & SPRAT & 0.40 - 0.80 & ZTF\\
    2024ryv & 2024-08-14 05:44:01 & 60536.2 & 3.4 & P60 & SEDM & 0.38 - 0.92 & ZTF\\
    2024ryv & 2024-08-15 05:31:11 & 60537.2 & 4.4 & P60 & SEDM & 0.38 - 0.92 & ZTF\\
    2024ryv & 2024-08-15 10:42:58 & 60537.4 & 4.6 & Shane & Kast & 0.36 - 1.03 & TReX\tablenotemark{b}\\
    2024ryv & 2024-08-15 13:13:46 & 60537.6 & 4.7 & HST & STIS & 0.17 - 1.03 & HST GO-17507 \\
    2024ryv & 2024-08-16 09:01:09 & 60538.4 & 5.6 & P200 & DBSP & 0.34 - 1.00 & ZTF\\
    2024ryv & 2024-08-20 09:07:50 & 60542.4 & 9.6 & P60 & SEDM & 0.38 - 0.92 & ZTF\\
    2024ryv & 2024-08-21 09:47:04 & 60543.4 & 10.6 & P60 & SEDM & 0.38 - 0.92 & ZTF\\
    2024ryv & 2024-08-27 08:26:01 & 60549.4 & 16.6 & P200 & DBSP & 0.34 - 1.00 & ZTF\\
    2024ryv & 2024-08-28 05:59:33 & 60550.2 & 17.4 & P60 & SEDM & 0.38 - 0.92 & ZTF\\
    2024ryv & 2024-08-28 10:56:04 & 60550.5 & 17.7 & Shane & Kast & 0.36 - 1.03 & TReX\tablenotemark{b}\\
    2024ryv & 2024-09-25 11:46:45 & 60578.5 & 45.7 & Shane & Kast & 0.36 - 1.03 & TReX\tablenotemark{b}\\
    2024ryv & 2024-09-26 07:18:29 & 60579.3 & 46.5 & P200 & DBSP & 0.34 - 1.00 & ZTF\\
    2024ryv & 2024-10-25 05:50:35 & 60608.2 & 75.4 & Shane & Kast & 0.36 - 1.03 & TReX\tablenotemark{b}\\
    2024ryv & 2024-11-09 07:43:06 & 60623.3 & 90.5 & Shane & Kast & 0.36 - 1.03 & TReX\tablenotemark{b}\\
    2024tvi & 2024-09-06 12:31:20 & 60559.5 & 13.7 & HST & STIS & 0.17 - 1.03 & HST GO-17507 \\
    2024tvi & 2024-09-10 07:22:48 & 60563.3 & 17.5 & P200 & DBSP & 0.34 - 1.00 & ZTF\\
    2024tvi & 2024-09-12 01:38:14 & 60565.1 & 19.3 & NOT & ALFOSC & 0.39 - 0.95 & ZTF\\
    2024tvi & 2024-09-25 05:57:07 & 60578.2 & 32.4 & Shane & Kast & 0.36 - 1.03 & TReX\tablenotemark{b}\\
    2024tvi & 2024-10-11 06:50:44 & 60594.3 & 48.5 & Shane & Kast & 0.36 - 1.03 & TReX\tablenotemark{b}\\
    2024tvi & 2024-10-25 04:42:14 & 60608.2 & 62.4 & Shane & Kast & 0.36 - 1.03 & TReX\tablenotemark{b}\\
    2024pso & 2024-07-21 04:04:37 & 60512.2 & 3.2 & NOT & ALFOSC & 0.40 - 0.96 & ZTF\\
    2024pso & 2024-07-22 07:31:05 & 60513.3 & 4.3 & P60 & SEDM & 0.38 - 0.92 & ZTF\\
    2024pso & 2024-07-24 07:12:55 & 60515.3 & 6.3 & P60 & SEDM & 0.38 - 0.92 & ZTF\\
    2024pso & 2024-07-24 10:55:21 & 60515.5 & 6.5 & P200 & DBSP & 0.34 - 1.00 & ZTF\\
    2024pso & 2024-07-25 07:43:58 & 60516.3 & 7.3 & P200 & DBSP & 0.34 - 1.00 & ZTF\\
    2024pso & 2024-07-26 13:03:26 & 60517.5 & 8.6 & HST & STIS & 0.17 - 1.03 & HST GO-17507 \\
    2024pso & 2024-07-28 11:28:36 & 60519.5 & 10.5 & Shane & Kast & 0.36 - 1.03 & TReX\tablenotemark{b}\\
    2024pso & 2024-08-01 10:32:44 & 60523.4 & 14.4 & P200 & DBSP & 0.34 - 1.00 & ZTF\\
    2024pso & 2024-08-10 12:54:43 & 60532.5 & 23.5 & Keck II & KCWI & 0.37 - 0.88 & ZTF\\
    2024pso & 2024-08-15 10:12:26 & 60537.4 & 28.4 & Shane & Kast & 0.36 - 1.03 & TReX\tablenotemark{b}\\
    2024pso & 2024-08-27 08:12:29 & 60549.3 & 40.3 & P200 & DBSP & 0.34 - 1.00 & ZTF\\
    2024pso & 2024-08-28 10:21:12 & 60550.4 & 41.4 & Shane & Kast & 0.36 - 1.03 & TReX\tablenotemark{b}\\
    2024pso & 2024-09-25 10:48:04 & 60578.5 & 69.5 & Shane & Kast & 0.36 - 1.03 & TReX\tablenotemark{b}\\
    2024pso & 2024-10-11 07:48:52 & 60594.3 & 85.3 & Shane & Kast & 0.36 - 1.03 & TReX\tablenotemark{b}\\
    2024pso & 2024-11-09 07:09:52 & 60623.3 & 114.3 & Shane & Kast & 0.36 - 1.03 & TReX\tablenotemark{b}\\
    2024pso & 2025-01-01 01:44:31 & 60676.1 & 167.1 & Keck II & KCWI & 0.37 - 0.88 & ZTF\\
    2025ryz & 2025-07-25 12:41:20 & 60881.5 & 4.1 & Keck I & LRIS & 0.32 - 1.03 & ZTF\\
    2025ryz & 2025-07-29 09:00:59 & 60885.4 & 7.9 & Shane & Kast & 0.36 - 1.0 & ZTF\\
    2025ryz & 2025-07-31 19:52:40 & 60887.8 & 10.4 & HST & STIS & 0.17 - 1.03 & HST GO-17807 \\
    2025ryz & 2025-08-02 09:25:47 & 60889.4 & 12.0 & Shane & Kast & 0.36 - 1.03 & ZTF\\
    2025ryz & 2025-08-04 09:08:30 & 60891.4 & 13.9 & P200 & NGPS & 0.56 - 1.02 & ZTF\\
    2025ryz & 2025-08-16 04:00:03 & 60903.2 & 25.7 & SOAR & Goodman & 0.40 - 1.0 & ZTF\\
    2025whv & 2025-09-04 00:10:04 & 60922.0 & 7.4 & NOT & ALFOSC & 0.39 - 0.97 & ZTF\\
    2025whv & 2025-09-11 11:25:08 & 60929.5 & 14.1 & HST & STIS & 0.17 - 1.03 & HST GO-17807 \\
    2025whv & 2025-09-11 23:58:42 & 60930.0 & 14.6 & NOT & ALFOSC & 0.35 - 0.96 & ZTF\\
    2025whv & 2025-09-21 02:18:14 & 60939.1 & 22.8 & NOT & ALFOSC & 0.39 - 0.96 & ZTF\\
    2024aifs & 2025-05-02 11:21:59 & 60797.5 & 8.0 & Shane & Kast & 0.36 - 1.03 & ZTF\\
    2024aifs & 2025-05-09 10:30:47 & 60804.4 & 15.0 & HST & STIS & 0.17 - 1.03 & HST GO-17807 \\
    2024aifs & 2025-05-13 04:35:54 & 60808.2 & 18.7 & NOT & ALFOSC & 0.35 - 0.96 & ZTF\\
    2024aifs & 2025-05-25 10:12:08 & 60820.4 & 30.9 & P200 & NGPS & 0.56 - 1.02 & ZTF\\
    2024aifs & 2025-05-27 03:04:53 & 60822.1 & 32.6 & NOT & ALFOSC & 0.39 - 0.96 & ZTF\\
    2024aifs & 2025-06-20 03:41:59 & 60846.2 & 56.7 & NOT & ALFOSC & 0.35 - 0.96 & ZTF\\
    2024aifs & 2025-06-30 01:04:13 & 60856.0 & 66.5 & NOT & ALFOSC & 0.39 - 0.96 & ZTF\\
    2024aifs & 2025-07-18 03:40:45 & 60874.2 & 84.7 & LT & SPRAT & 0.40 - 0.80 & ZTF\\
    2024aifs & 2025-07-21 02:27:18 & 60877.1 & 87.6 & NOT & ALFOSC & 0.39 - 0.96 & ZTF\\
    2024aifs & 2025-07-24 23:41:48 & 60881.0 & 91.5 & NOT & ALFOSC & 0.39 - 0.96 & ZTF\\
    2025ziv & 2025-10-03 05:45:36 & 60951.2 & 2.2 & P60 & SEDM & 0.38 - 0.92 & ZTF\\
    2025ziv & 2025-10-04 08:33:51 & 60952.4 & 3.4 & P60 & SEDM & 0.38 - 0.92 & ZTF\\
    2025ziv & 2025-10-05 05:53:45 & 60953.2 & 4.2 & P60 & SEDM & 0.38 - 0.92 & ZTF\\
    2025ziv & 2025-10-06 05:24:41 & 60954.2 & 5.2 & P60 & SEDM & 0.38 - 0.92 & ZTF\\
    2025ziv & 2025-10-07 06:33:10 & 60955.3 & 6.3 & P60 & SEDM & 0.38 - 0.92 & ZTF\\
    2025ziv & 2025-10-08 08:17:39 & 60956.3 & 7.3 & P200 & NGPS & 0.56 - 1.02 & ZTF\\
    2025ziv & 2025-10-10 03:49:22 & 60958.2 & 9.2 & HST & STIS & 0.17 - 1.03 & HST GO-17807 \\
    2025ziv & 2025-10-18 06:22:57 & 60966.3 & 17.3 & P60 & SEDM & 0.38 - 0.92 & ZTF\\
    2025ziv & 2025-10-18 08:20:32 & 60966.3 & 17.3 & Shane & Kast & 0.36 - 1.03 & ZTF\\
    2025ziv & 2025-10-26 08:28:24 & 60974.4 & 25.4 & P60 & SEDM & 0.38 - 0.92 & ZTF\\
    2025ziv & 2025-11-01 06:38:51 & 60980.3 & 31.3 & P60 & SEDM & 0.38 - 0.92 & ZTF\\
    2025ziv & 2025-11-04 07:00:32 & 60983.3 & 34.3 & P60 & SEDM & 0.38 - 0.92 & ZTF\\
    2025ziv & 2025-11-28 05:37:34 & 61007.2 & 58.2 & P60 & SEDM & 0.38 - 0.92 & ZTF\\
    2025ziv & 2025-12-06 04:28:23 & 61015.2 & 66.2 & P60 & SEDM & 0.38 - 0.92 & ZTF\\
    2025ziv & 2025-12-12 06:48:58 & 61021.3 & 72.3 & P60 & SEDM & 0.38 - 0.92 & ZTF\\
    2025ziv & 2025-12-20 05:15:08 & 61029.2 & 80.2 & P60 & SEDM & 0.38 - 0.92 & ZTF\\
    2025ziv & 2026-01-11 05:12:10 & 61051.2 & 102.2 & P60 & SEDM & 0.38 - 0.92 & ZTF\\
    2025bco & 2025-02-13 19:46:48 & 60719.8 & 6.3 & HST & STIS & 0.17 - 0.30 & HST GO-17788 \\
    2025bco & 2025-02-14 07:04:04 & 60720.3 & 6.8 & SOAR & Goodman & 0.38 - 0.70 & ZTF\\
    2025bco & 2025-02-14 07:04:04 & 60720.3 & 6.8 & SOAR & Goodman & 0.49 - 0.90 & ZTF\\
    2025bco & 2025-02-15 00:00:00 & 60721.0 & 7.5 & Keck II & NIRES & 0.97 - 2.47 & ZTF\\
    2025bco & 2025-02-16 02:13:02 & 60722.1 & 8.6 & NOT & ALFOSC & 0.35 - 0.96 & ZTF\\
    2025bco & 2025-02-17 10:59:22 & 60723.5 & 10.0 & P200 & NGPS & 0.57 - 1.02 & ZTF\\
    2025bco & 2025-02-20 04:37:09 & 60726.2 & 12.7 & NOT & ALFOSC & 0.35 - 0.96 & ZTF\\
    2025bco & 2025-05-25 06:35:16 & 60820.3 & 106.8 & P200 & NGPS & 0.56 - 0.92 & ZTF\\
    2025bco & 2026-01-14 10:13:17 & 61054.4 & 340.9 & P200 & NGPS & 0.56 - 1.04 & ZTF\\
    2025adj & 2025-01-25 09:33:19 & 60700.4 & 2.4 & P60 & SEDM & 0.38 - 0.92 & ZTF\\
    2025adj & 2025-01-25 10:09:57 & 60700.4 & 2.4 & Shane & Kast & 0.35 - 1.05 & ZTF\\
    2025adj & 2025-01-26 03:36:52 & 60701.2 & 3.2 & LT & SPRAT & 0.40 - 0.81 & ZTF\\
    2025adj & 2025-01-26 16:25:17 & 60701.7 & 3.7 & Keck I & LRIS & 0.31 - 1.03 & ZTF\\
    2025adj & 2025-01-30 09:07:55 & 60705.4 & 7.4 & P60 & SEDM & 0.38 - 0.92 & ZTF\\
    2025adj & 2025-01-30 23:55:52 & 60706.0 & 8.0 & HST & STIS & 0.17 - 1.03 & HST GO-17788 \\
    2025adj & 2025-02-02 11:17:48 & 60708.5 & 10.5 & P200 & NGPS & 0.56 - 1.02 & ZTF\\
    2025adj & 2025-03-22 11:41:42 & 60756.5 & 58.5 & P200 & NGPS & 0.56 - 1.02 & ZTF\\
    2025adj & 2025-04-09 11:42:00 & 60774.5 & 76.5 & Shane & Kast & 0.33 - 1.09 & UCSC\\
    2025adj & 2025-05-26 07:44:15 & 60821.3 & 123.3 & P200 & NGPS & 0.56 - 1.02 & ZTF\\
    2025adj & 2025-06-06 07:00:00 & 60832.3 & 134.3 & Shane & Kast & 0.34 - 1.09 & UCSC\\
    2025adj & 2025-08-04 06:31:32 & 60891.3 & 193.3 & P200 & NGPS & 0.56 - 1.02 & ZTF\\
\enddata
\tablenotetext{a}{Relative to time of first light.}
\tablenotetext{b}{TRansient EXtragalactic team at UC Berkeley (PIs Margutti and Chornock)}
\end{deluxetable*}

\end{document}